\documentclass{aa}  

\usepackage{pdflscape}
\usepackage{graphicx}
\usepackage{changepage}
\usepackage{multirow}
\usepackage{subcaption}
\usepackage[labelfont=bf]{caption}
\usepackage{txfonts}
\usepackage{dcolumn}
\newcolumntype{d}[1]{D{.}{.}{#1}}
\newcolumntype{m}[1]{D{,}{\,\pm\,}{#1}}
\newcolumntype{s}[1]{D{,}{\,}{#1}}
\usepackage[colorlinks = true,
linkcolor= blue,
citecolor = blue,
filecolor = blue,
urlcolor = blue,
]{hyperref}
\usepackage{lscape}

\begin{document}

   \title{Testing the mass-radius relation of white dwarfs in common proper motion pairs} \subtitle{I. Hydrogen-dominated atmospheres}
\titlerunning{Testing the mass-radius relation of white dwarfs}

   \author{Roberto Raddi
\inst{1}\fnmsep\thanks{\email{roberto.raddi@upc.edu}}
\and Alberto Rebassa-Mansergas \inst{1,2}
\and Santiago Torres \inst{1,2}
\and Maria E. Camisassa \inst{1}
\and Ralf Napiwotzki \inst{3}
\and \\
Detlev Koester \inst{4}
\and Pier-Emanuel Tremblay \inst{5}
\and Ulrich Heber \inst{6}
\and Leandro Althaus \inst{7,8}
          }
   \institute{Universitat Polit\`ecnica de Catalunya, Departament de F\'isica, c/ Esteve Terrades 5, 08860 Castelldefels, Spain
   \and
      Institute for Space Studies of Catalonia (IEEC), c/Esteve Terradas,1, Edifici RDIT, Despatx 212, Campus del Baix Llobregat UPC
– Parc Mediterrani de la Tecnologia, 08860 Castelldefels, Spain
   \and
 Centre for Astrophysics Research, University of Hertfordshire, Hatfield, AL10 9AB, UK
\and  
   Institut f\"ur Theoretische Physik und Astrophysik, Christian-Albrechts-Universit\"at, Kiel 24118, Germany
   \and
    Department of Physics, University of Warwick, Coventry CV4 7AL, UK
    \and
   Dr. Karl Remeis-Observatory \& ECAP, Astronomical Institute, Friedrich-Alexander University Erlangen-Nuremberg (FAU), Sternwartstr. 7, 96049, Bamberg, Germany
   \and
Facultad de Ciencias Astron\'omicas y Geof\'isicas, Universidad Nacional de La Plata, Paseo del Bosque s/n, 1900 La Plata, Argentina
   \and
Instituto de Astrof\'isica La Plata, UNLP-CONICET, Paseo del Bosque s/n, 1900 La Plata, Argentina
}

   \date{Received 6 September 2024; accepted 30 January 2025}

% \abstract{}{}{}{}{} 
% 5 {} token are mandatory
 
  \abstract
   {The mass of white dwarfs is one of the most important properties that constrains their past and future evolution. Direct estimates of white dwarf masses are crucial for assessing the validity of theoretical evolutionary models and methods of analysis. }
   {The main goal of this work was to measure the masses and radii of white dwarfs that belong to widely separated, common proper motion binaries with non-degenerate companions. These can be assessed, independently from theoretical mass-radius relations, through measurements of gravitational redshifts and photometric radii.}
   {We studied 50 white dwarfs with hydrogen-dominated atmospheres, performing a detailed analysis of high-resolution ($R \approx 18\,500$) spectra by means of state-of-the-art grids of synthetic models and specialized software. Hence, we measured accurate radial velocities from the H$\alpha$ and H$\beta$ line-cores, thus obtaining the white dwarf gravitational redshifts. Jointly with a  photometric analysis that is formalized by a Bayesian inference method, we measured precise white dwarf radii that allowed us to directly measure the white dwarf masses from their gravitational redshifts.}
   {The distributions of measured  masses and radii agree within  6\,\% (at the 1-$\sigma$ level) from the theoretical mass-radius relation, thus delivering a much smaller scatter in comparison with previous analyses that used gravitational redshift measurements from low-resolution spectra. A comparison against model-dependent spectroscopic estimates produces a larger scatter of 15\,\% on the mass determinations. We find an agreement within $ \approx $10\,\% from previous model-based, photometric mass estimates from the literature.}
   {Combining gravitational redshift measurements and photometric analysis of white dwarfs delivers precise and accurate, empirical estimates of their masses and radii. This work confirms the reliability of the theoretical mass-radius relation from the lightest to the heaviest white dwarfs in our sample ($\approx 0.38$--$1.3$\,M$_\odot$).}


   \maketitle
%
%-------------------------------------------------------------------

\section{Introduction}

The mass of stars, along with their initial chemical composition and rotation, is a key ingredient of theoretical evolutionary models \citep{hurley2000,ekstrom2012}.\ 
Despite its importance, measuring precise stellar masses remains a difficult task  \citep{popper1980,serenelli2021} that
is made easier for visual binaries
and/or requires precise spectro-photometric characterization and trigonometric parallaxes. White dwarfs, which are the most common end product of stellar evolution, are not exempt from this problem. 

White dwarfs are fundamental for constraining important aspects of stellar evolutionary models \citep[e.g. the initial-to-final-mass relation linking white dwarf masses to those of their progenitors;][]{marigo2022} or global properties of the Milky Way, such as the star formation history \citep[e.g.][]{tremblay2014,torres2021,cukanovaite2023}, that are directly connected to the white dwarf mass distribution \citep[][]{tremblay2016,jimenez-esteban2023} and luminosity functions \citep[][]{garcia-berro2016}. White dwarfs do not sustain nuclear burning in their cores and their structure is supported by electron degeneracy pressure, thus their evolution is a cooling process that shapes their mass-radius relation as a function of time \citep{althaus2010}. Due to their Earth-sized radii, average white dwarfs are intrinsically faint stars that pose observational difficulties to their identification and accurate characterization.

Already before the publication of the {\em Gaia} Data Release 1 \citep{gaiadr1}, the mass-radius relation for a limited number of white dwarfs with measured parallaxes was tested via model-dependent spectroscopic determinations of surface gravities \citep[e.g.][]{provencal1998,holberg2012}, but the improvement provided by the {\em Gaia} parallaxes has been notable both in terms of precision and accuracy \citep{tremblay2017, bedard2017}. Nowadays, a statistically significant all-sky sample of white dwarfs, containing at least 350\,000 reliable candidates \citep{jimenez-esteban2018,gentilefusillo2019,gentilefusillo2021}, has been identified in the ESA {\em Gaia} Data Release 2 \citep[][]{gaiadr2} and the Early Data Release 3 \citep[EDR3;][]{gaiaedr3}. Hence, excellent comparisons have been achieved for about 10\,000 white dwarfs with existing low-resolution spectra  and broad-band photometry \citep{bergeron2019,tremblay2019}. 
On the other hand, very precise and accurate results are also achievable with complementary methods, including the analysis of white dwarfs in eclipsing binaries that have enabled a detailed characterization of low-mass helium (He) or carbon-oxygen (CO) core objects that are formed via binary interactions \citep{parsons2017}. The disadvantage of the eclipsing binaries is that the majority of white dwarfs have low masses due to their evolution through a common envelope phase, thus the mass-radius relation is not easily tested in the high white dwarf mass regime.

Before the {\em Gaia} era, several authors have attempted to derive white dwarf masses from the measure of their gravitational redshifts  in pioneering works \citep{adams1925,popper1954,greenstein1967,trimble1972,greenstein1972,wegner1974}. This method is suitable for all mass ranges, as the general theory of relativity predicts for the light emitted by a massive object to be systematically displaced by a gravitational-redshift factor, $\varv_{\rm gr} = G M / c R$; where $G$ is the gravitational constant, $c$ is the speed of light, $M$ and $R$ are the mass and radius of the considered object.

The lack of high-resolution spectroscopy of white dwarfs is what mostly plagues the measurements of gravitational redshifts. In fact, it is well-known that radial velocities measured from low-resolution spectra can introduce relatively large systematic uncertainties and scatter around the true value \citep{chandra2020,raddi2022,arsenau2024}. This issue is due to the Stark effect, because of which the pressure-broadened lines are characterized by asymmetries and pressure shifts. Such a phenomenon has been widely studied for the Balmer lines of hydrogen-atmosphere (DA spectral type) white dwarfs, which have their individual radial velocities to be discrepant by a few tens of km/s \citep{wiese1971,grabowski1987,halenka2015}. The narrow cores of the H$\alpha$ and H$\beta$ lines are almost unaffected by the pressure shifts \citep[$< 1$\,km/s systematic shifts;][]{napiwotzki2020}, because they form in the upper layers of stellar atmospheres in conditions of non-local thermodynamic equilibrium (NLTE) where Doppler broadening dominates, and thus, due to the typically slow rotation of white dwarfs they have narrow Gaussian profiles \citep{koester1988,heber1997}. These sharp line-cores can be used to measure accurate radial velocities from sufficiently high-resolution spectra \citep[$\sigma_\varv < 10$\,km/s;][]{napiwotzki2020}. Under the assumption that peculiar velocities of randomly selected white dwarfs would cancel out, \citet{falcon2010} showed that the mean mass-distribution of the white dwarf population can be retrieved from their distribution of gravitational redshifts measured from high-resolution spectra, thus confirming the value of such data.  More recent work by \citet{crumpler2024} used the same idea to derive an empirical mass-radius relation, by binning with a bootstrapping procedure the masses and radii obtained from $\approx 26\,000$ isolated white dwarfs with low-resolution spectra from SDSS-V \citep[][]{kollmeier2017}; their work  confirms the temperature dependence of the mass-radius relation that is predicted by theoretical relations.

White dwarfs in wide, common proper motion binaries with non-degenerate companions can be used effectively to measure their gravitational redshift.  By assuming a negligible contribution from the orbital motion of the companion star, typically of the order of 0.1 km/s \citep[cf][]{vonHippel1996,arsenau2024}, the gravitational redshift of a white dwarf can be simplified as  $\varv_{\rm gr} = \varv_{\rm WD} - \varv_{\rm sys}$, where $\varv_{\rm sys}$ is  the systemic radial velocity that is assumed to be equal to that of the white dwarf companion, when its gravitational redshift is considered to be negligible. This method can be rather accurate if the white dwarf distance is known, enabling to directly estimate both its radius and mass \citep[e.g.][]{shipman1997,joyce2018}. Typically good results that are in agreement with spectroscopic estimates, but often characterized by large uncertainties, have been obtained in the past \citep{koester1987,bergeron1995, reid1996, holberg2012,silvestri2001}.  White dwarfs in open clusters are also used to measure gravitational-redshifts, by taking advantage of a larger number of co-moving objects. However, the nearby clusters where high-resolution spectroscopy is achievable, thus far, only include the Praesepe \citep{casewell2009} and Hyades \citep{pasquini2019,pasquini2023} clusters. 

The identification of thousand such white dwarf in common proper motion pairs \citep{elbadry2021,rebassa2021,raddi2022} offers an excellent opportunity to empirically investigate the mass-radius relation of white dwarf. Indeed, \citet{arsenau2024} analyzed 135 white dwarfs from the Sloan Digital Sky Survey \citep{sdss} that belong to wide binaries. By measuring radial velocities from their low-resolution spectra, they obtained masses and radii broadly in agreement with theoretical predictions although characterized by a large dispersion.

\begin{figure*}[th]
    \centering
\begin{subfigure}[t]{0.49\textwidth}
\includegraphics[width=\textwidth]{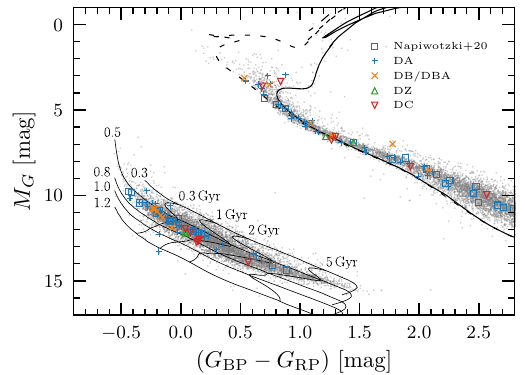}
\caption{}
    \label{fig:selection_a}
    \end{subfigure}
   \begin{subfigure}[t]{0.49\textwidth}
    \includegraphics[width=\textwidth]{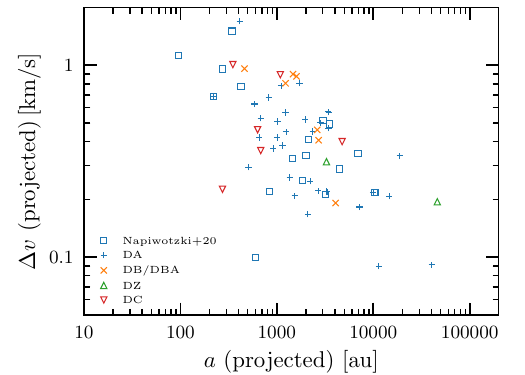}
             \caption{}
         \label{fig:selection_b}
        \end{subfigure}
 \caption{Properties of the selected targets. Symbols and colors are explained in the legends. {\bf (a)}: color-magnitude diagram of the studied sample.  For reference, we plot the white dwarf cooling tracks for He-, CO-, and oxygen-neon (ONe)-core white dwarfs  \citep{althaus2013,camisassa2016,camisassa2019} and the isochrones, which we converted to the {\em Gaia} magnitudes by means of synthetic spectra \citep{koester2010}; the corresponding masses, in Solar units, and cooling ages are labeled. The solar metallicity BaSTI isochrones with [Fe/H]~$= +0.06$ \citep[][]{hidalgo2018}  are also plotted at 1 and 10~Gyr (dashed and solid curves, respectively). {\bf (b)}: projected velocity difference plotted against projected orbital semi-major axis of the studied binaries. }
    \label{fig:selection}
\end{figure*}

In the present work, we analyze a sample of 50 DA white dwarfs with high-resolution spectroscopy. These white dwarfs belong to  widely separated, common proper motion binaries, whose non-degenerate companions have published radial velocities in the {\em Gaia} Data Release 3 \citep[DR3;][]{gaiadr3}. Thanks to the accurate {\em Gaia} parallaxes, we directly measure the white dwarf radii, which combined to sufficiently precise and accurate gravitational redshifts allow us to measure empirical masses. Our results are compared to a theoretical mass-radius relation as well as to model-dependent estimates.

\section{Sample selection}

In \citet{raddi2022}, we presented a sample of 7256 reliable white dwarfs in common proper motion pairs with non-degenerate companions, by means of which we studied the age -- velocity dispersion relation. Those stars were identified by relying on improvements of the selection criteria introduced by \citet{gentilefusillo2021} and \citet{elbadry2021} for white dwarfs and common proper-motion pairs, respectively. The parallaxes of the identified pairs agreed within 3\,$\sigma$ uncertainties. Our list of wide binaries included 1092 systems with accurate radial velocities that were either published in the {\em Gaia} Data Release 2 \citep[DR2;][]{katz2019} or in other spectroscopic surveys, that is RAVE\,DR5 \citep{kunder2017}, LAMOST\,DR5 \citep{luo2019,xiang2019}, APOGEE\,DR16 \citep{jonsson2020}, and GALAH+\,DR3 \citep{buder2021}. Moreover, the common proper motion pairs were chosen not to have nearby brighter stars at less than 10-arcsec separation from the non-degenerate stars, in order to safely avoid contaminated radial velocity measurements \citep[see tests on {\em Gaia} DR2][]{boubert2019}. With the publication of new radial velocities in the {\em Gaia} DR3 \citep[][]{katz2023}, a total of 3224 such binaries possess radial velocity measurements for the non-degenerate members.

Among the sample of wide binaries, we found 28 DA white dwarfs that were observed by the Supernova Ia Progenitor surveY \citep[SPY;][]{napiwotzki2001,napiwotzki2020}, hence possessing high-resolution spectra ($R \approx 18\,500$; 0.36\,\AA\ at H$\alpha$) taken with the UV-visual echelle spectrograph \citep[UVES;][]{dekker2000} at the European Southern Observatory (ESO) Very Large Telescope (VLT) in Cerro Paranal (Chile). The SPY project aimed at identifying close double-degenerate binaries, by measuring radial velocity variations or resolving spectroscopic-binaries among a sample of bright white dwarfs that were observed at different epochs. These high-resolution spectra are suitable for measuring accurate radial velocities of white dwarfs as shown by various authors \citep{napiwotzki2001,napiwotzki2020,pauli2006,falcon2010}. Out of the 28 white dwarfs with SPY spectra, we selected 17 objects that were reliably classified as not having unresolved stellar components according to the  criteria given by \citet{napiwotzki2020} and that  have wide common proper motion companions with {\em Gaia} radial velocity measurements. The published atmospheric parameters and radial velocities of the considered SPY white dwarfs are listed in Table\,\ref{tab:spy}; the 11 stars excluded from the following analysis, listed in Table\,\ref{tab:wd_excluded}, do not have radial velocity measurements in {\em Gaia}.  

\begin{figure}
    \centering
    \includegraphics{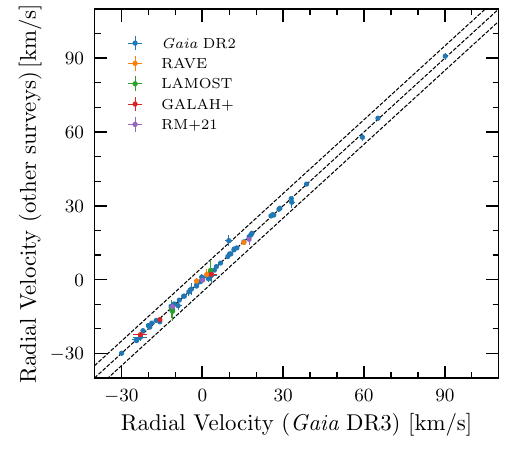}
    \caption{Comparison among radial velocity measurements for the non-degenerate companions that have both radial velocities in {\em Gaia} DR3 and other surveys. The color scheme is shown in the legend. Uncertainties are typically of 1--2\,km/s and the error bars have almost the size of the used symbols. The dashed lines represent a difference of $\pm 5$\,km/s from the equality.}
    \label{fig:rv_comparison}
\end{figure}

In addition, we selected 48 new white dwarfs to be observed with UVES, whose companions had existing radial velocity measurements. The adopted selection criteria were their visibility from Paranal, projected separations below $a = 50\,000$\,au so to exclude most of the chance alignment systems \citep{elbadry2021}, and being sufficiently bright to be observable with UVES.  The {\em Gaia} color-magnitude diagram of the 48 observed targets and 17 SPY white dwarfs is shown in Fig.\,\ref{fig:selection_a}, while their projected separations and tangential velocity differences are displayed in Fig.\,\ref{fig:selection_b}. We note the diagonal trend in the latter figure that limits the maximum separation and projected velocity for a bound binary whose total mass is below $\sim$8.5\,M$_\odot$ \citep[see][and references therein for more details]{raddi2022}.  

Figure\,\ref{fig:rv_comparison} shows the comparison among those non-degenerate companion stars that have radial velocity measurements in {\em Gaia} DR3  as well as in {\em Gaia} DR2 or other spectroscopic surveys. The agreement is much better than $\pm 5$\,km/s. We note that one object, 0207+0355, has the {\em Gaia} DR2 and DR3 measurements that are more discrepant than 5\,km/s, but it is not listed as radial-velocity variable in {\em Gaia} DR3. All studied targets have more than six {\em Gaia} DR3 radial velocity measurements, but we did not exclude those that do not have  deblended observations. However, we flag them in Table\,\ref{tab:nwd_gaia}. Neither we did limit our targets based on their \verb|flags_gspspec| that are more relevant for chemical analysis \citep{recio-blanco2023}.

Relevant {\em Gaia} parameters of the studied white dwarfs and non-degenerate stars are listed in Table\,\ref{tab:wd_gaia} and \ref{tab:nwd_gaia}, respectively.

\section{Observations}
We were awarded with 39\,hr of observing time at the ESO VLT with UVES, which was mounted at the Nasmyth B focus of the 8.2-m UT2 (Kueyen). We employed a standard configuration, also adopted by the SPY survey \citep{napiwotzki2020}, and unless otherwise indicated we used the dichroic 1 that has the central wavelengths at 3900\,\AA\ and 5640\,\AA\ in the blue and red arms, respectively. This setup allowed us to cover all the Balmer series down to the Balmer jump. We adopted a 2.1-arcsec wide slit that delivers a resolving power of $R\approx 18\,500$, corresponding to a spectral resolution of about 0.36\,\AA\ at the H$\alpha$, as for the SPY project. The spectra are oversampled, with dispersions of 0.03 and 0.045\,\AA\ per pixel, respectively in the red and blue arms. The observations were executed in service mode; the seeing was on average 1-arcsec and roughly covered a 0.4--2 arcsec interval. The typical signal-to-noise ratio per pixel is ${\rm SNR} = 10$, per exposure. All in all, we observed 51 objects; these data include 48 new targets, plus the SPY white dwarf WD\,1544$-$377 for comparison purposes, and two objects with featureless spectra (DC type), LAWD\,1 and LAWD\,57, to assess the flux calibration.  The observing logs, presented in Table\,\ref{tab:logs}, list all the individual exposure times and relevant information for the observed science targets.

We reduced and calibrated the raw data and extracted the 1D spectra by means of the ESO Reflex pipeline \citep{freudling2013}, using the standard set of UVES recipes. The final products are very mildly affected by the quasi-periodic ripple pattern that is typically observed in UVES spectra, which is due to a small misalignment between the spectral and flat-field traces. Nevertheless the results are comparable to those obtained for the data reduction of SPY targets \citep{napiwotzki2020}. The telluric correction was applied by scaling the  average template provided by \citet{napiwotzki2020}, which is made available along with the individual SPY spectra at the VizieR catalog access tool \citep{ochsenbein20000}.

\section{Spectroscopic analysis}
\label{sec:spectroscopic_analysis}
Of the 48 new observed white dwarfs, 33 have hydrogen dominated atmospheres (DA spectral type) that include a DAZ showing Ca\,{\sc ii} H\&K absorption lines and a magnetic white dwarf (DAH), nine turned out to possess helium-dominated atmosphere (seven DB/DBA/DBZ spectra, and two DZ that show Ca\,{\sc ii} H\&K absorption), and six are featureless DC white dwarfs. Our spectral classification is given in Table\,\ref{tab:wd_gaia}.

In the present work, we focus on the analysis of 50 hydrogen-dominated white dwarfs that include the 33 new observations and the 17 objects with SPY spectra, while the non-DA spectra will be analyzed in another publication.

The spectroscopic analysis detailed here is crucial for measuring the radial velocities of the sample. The spectroscopic parameters will be used to derive model-dependent masses and radii,  as it is typically done in the literature, which will be compared to the independent measurements of photometric radii, and masses derived from the gravitational redshift.

\subsection{Spectroscopic parameters}

We analyzed the new UVES spectra and the old ones from the SPY survey with two grids of synthetic spectra for DA white dwarfs. The first grid consists of 1D LTE pure-hydrogen models with the mixing-length parameter ML2/$\alpha = 0.8$ that are computed following the prescriptions presented by \citet{koester2010} and updated with more recent physics \citep{koester2020}; their effective temperature and surface gravity range between $T_{\rm eff} = 3000$ -- 80\,000\,K and $\log{g} = 6$ -- 9.5\,dex. The second grid consists of 3D LTE pure-hydrogen models \citep[][and references therein for details on convection, neutral hydrogen broadening, and H$_2$ collision induced absorption; and line-profiles from \citet{tremblay2009}]{tremblay2013,tremblay2015}, that cover ranges of $T_{\rm eff} =  3000$ -- 40\,000\,K and $\log{g} = 7$ -- 9\,dex.  

For our spectroscopic analysis we employed the program {\sc fitsb2} \citep[based on {\sc fitprof};][]{napiwotzki1999,napiwotzki2020}, which measures $T_{\rm eff}$, $\log{g}$, and radial velocity ($\varv_{\rm WD}$) via $\chi^2$ minimization with the {\sc amoeba} routine \citep{press1992}. The atmospheric parameters, $T_{\rm eff}$ and $\log{g}$, of DA white dwarfs are known to be degenerate in the range of $T_{\rm eff} = 13\,000$ -- 15\,000 for $\log{g} = 8$ dex, at which temperature the equivalent widths of Balmer lines reach a maximum. Thus, we adopted initial $T_{\rm eff}$ values obtained from photometry \citep{gentilefusillo2021} as they help to discriminate for the correct spectroscopic parameters. We fitted the first six Balmer lines, reaching a compromise between line visibility due to the energy level dissolution and spectral noise, because the latest Balmer lines are surface gravity indicators. We included the H$\alpha$ line although its red wing is not entirely covered by the spectral range, but this is not expected to have a strong impact on the results. The statistical uncertainties are determined via a Monte Carlo (MC) method, which determines a number of realizations of the observed spectrum based on the variance of the data. For these high-resolution spectra, systematic uncertainties are known to be much larger \citep{koester2009}. In addition, we estimated systematic uncertainties on a spectrum-by-spectrum basis by fitting individual sub-exposures of the observed white dwarfs and measuring their scatter with respect to the average spectrum. A more realistic assessment of the overall systematic uncertainties is discussed next in this section. 

The atmospheric parameters measured with the two grids are listed in Table\,\ref{tab:spectra_params}, where the errors account for both the statistical and systematic uncertainty estimates. The best-fits results obtained with the Koester grid of synthetic spectra are displayed in Fig.\,\ref{fig:fig01}  and \ref{fig:fig02}, while those obtained from the Tremblay grid are shown in Fig.\,\ref{fig:fig03}  and \ref{fig:fig04}. We refer to Appendix\,\ref{app:spy} for a comparison with the spectroscopic measurements published by \citet{napiwotzki2020}.

\begin{figure}[t]
    \centering
    \includegraphics[width=\linewidth]{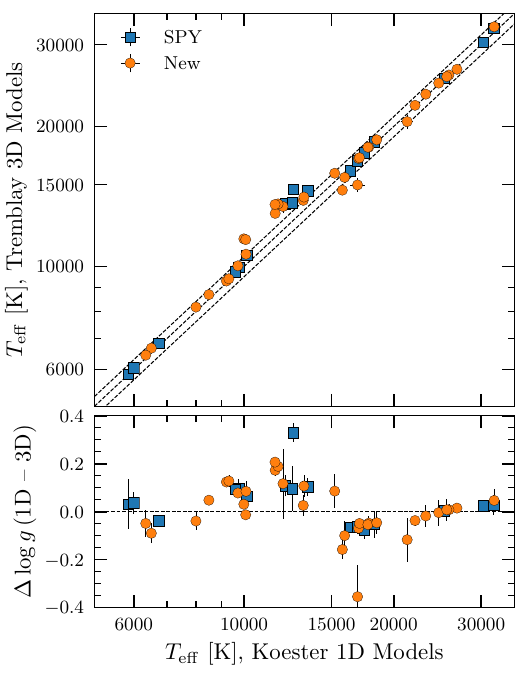}
    \caption{Comparison between spectroscopic parameters measured with two set of synthetic spectra. The dashed lines represent the equality and a $5$\,\% temperature difference in the upper panel, and the equality in the bottom panel.}
    \label{fig:comparison_spectro}
\end{figure}
\begin{figure*}[t!]
    \centering
    \includegraphics[width=\linewidth]{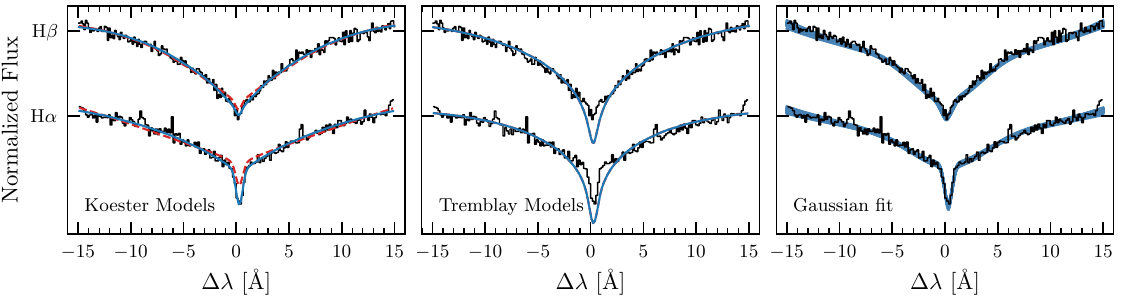}
    \includegraphics[width=\linewidth]{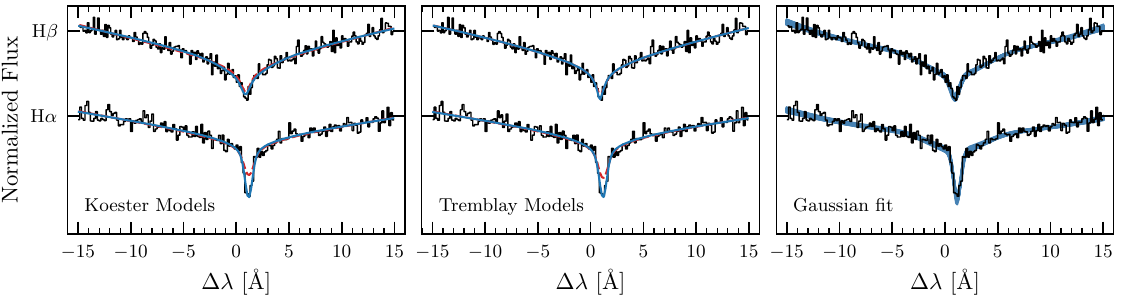}    
    \caption{Examples of the radial-velocity fitting for the white dwarfs 0209$-$0140 (top panels) and 0807$-$3622 (bottom panels), respectively of $T_{\rm eff} \approx 9500$\,K and $T_{\rm eff} \approx 17\,000$\,K. The red-dashed curves represent the model-fit without the additional Gaussian functions, which are included in the blue-solid curves. The shaded region in the Gaussian fit panels represent 50 random draws of the best-fit distributions.}
    \label{fig:fit_rv}
\end{figure*}

In Fig.\,\ref{fig:comparison_spectro}, we present the results of the spectroscopic analysis, comparing the two estimates of $T_{\rm eff}$ and $\log{g}$. We corrected the atmospheric parameters obtained by means of the \citet{koester2010} 1D models with the 3D corrections proposed by \citet{tremblay2013}, when the $T_{\rm eff}$ is in the 6000 -- 14\,500\,K range. These corrections are also listed in Table\,\ref{tab:spectra_params}, but we stress that caution should be taken due to the fact that they were derived by \citet{tremblay2013} via the analysis of low-resolution spectra with their 1D and 3D models. The studied sample approximately spans over $T_{\rm eff} = 6000$ -- 30\,000\,K. The measured $T_{\rm eff}$ and the $\log{g}$ broadly agree within $\pm 5$\,\% and $\pm 0.15$\,dex, respectively, that imply an estimate of the overall systematic uncertainties. Such differences are more marked between 10\,000 -- 15\,000\,K, where the treatment of atmospheric convection is relevant. We note that the atmospheric parameters of the object WD\,1544$-$377, which we re-observed for comparison purposes, are in excellent  1-$\sigma$ agreement with the parameters measured from the SPY spectra (cf Table\,\ref{tab:spectra_params}), thus suggesting a good stability of the instrument over a 20-years period.

In Table\,\ref{tab:spectra_params}, we include model-dependent estimates of white dwarf masses that are derived via interpolation of $T_{\rm eff}$ and $\log{g}$ onto the white dwarf evolutionary tracks computed by the La Plata group \citep[][]{althaus2013,camisassa2016,camisassa2019}. These tracks account for different white dwarf core-compositions (He, CO, and ONe) and thickness of the hydrogen layer, which depend on the detailed evolution of the white dwarf progenitor.

\subsection{Radial velocities}

\citet{napiwotzki2020} extensively discussed the accuracy of radial velocities for white dwarfs, analyzing the impact of pressure shifts on the measured values and emphasizing the importance of using only the closest spectral range around the NLTE cores of the H$\alpha$ and H$\beta$ lines. While those authors were mostly concerned with relative radial-velocity shifts, they estimated that measuring the combined radial velocity from H$\alpha$+H$\beta$ would provide the best absolute estimate.

We measured radial velocities with two different methods, and compared the results in order to quantify the systematic uncertainties in terms of white dwarf masses. In the first one, we adopted a procedure that is similar to that adopted by \citet{napiwotzki2020}. We measured the radial velocity shifts by fitting the H$\alpha$+H$\beta$ line-cores  on a line-by-line basis and averaging the result as well as the H$\alpha$ core alone, in both cases limiting the spectral range to $\pm 15$\,\AA, and using the spectroscopic best fit to model the line wings. We note that the exclusion of the reddest part of the H$\alpha$ wings from the UVES spectra does not impact the radial velocity measurement. Frequently the line cores are not well modeled by neither of the employed synthetic spectral grids, due to the exclusion of NLTE effects in the models, thus it is necessary to add a Gaussian and/or a Lorentzian to improve the fit \citep{napiwotzki2020}. Thus we followed their iterative procedure by using the best-fit spectroscopic measurement as starting point, then adding the Gaussian/Lorentzian when necessary, and iterating between small $T_{\rm eff}$, $\log{g}$, and $\varv_{\rm WD}$ adjustments until the solution converges in a few iterations. The radial velocity uncertainties are estimated via a Monte Carlo method as in the previous section. The measured values are listed in Table\,\ref{tab:spectra_params}, in which we indicate whether a Gaussian and/or Lorentzian were also used. 

In the second method, we fitted the H$\alpha$  and H$\beta$ line cores by modeling them and the neighboring $\pm15$\,\AA\ region with two Gaussian functions with the same central wavelength plus a quadratic polynomial \citep[see][for a similar approach]{maoz2017}. The purpose of this  alternative method is that of assessing possible systematic shifts in the  radial-velocity measurements from the first method. The fit is performed with the {\sc python} library {\sc lmfit} \citep{lmfit}, using the Nelder-Mead (i.e. {\sc amoeba}) algorithm to find the minimum $\chi^2$, and estimating the uncertainties with the Markov-Chain Monte Carlo sampler of the {\sc emcee} module \citep{foreman-mackey2013}. The results  for the measurements from the average of H$\alpha$+H$\beta$ and H$\alpha$ alone are also listed in Table\,\ref{tab:spectra_params}.

As an example, Figure\,\ref{fig:fit_rv} shows a comparison of the best-fitting models, identified with the different methods for two white dwarfs. In the upper panels, we note that the 3D models overestimate the core depth. This issue was discussed by \citet{tremblay2013}, who noted that the 3D structure deviates from the 1D equivalents in the upper atmosphere due to the cooling effect of convective overshoot, thus  affecting the overall quality of the fits. Radial velocity measurements for such peculiar cases do not show significant differences.  The same comparison is shown for all the studied stars in Fig.\,\ref{fig:fig01rv} through Fig.\,\ref{fig:fig10rv}.

The average differences among the two  radial-velocity fitting methods are listed in Table\,\ref{tab:rv_comparison}. In order to understand the impact of measuring radial velocities with different methods, we converted the differences to Solar-mass units by using the gravitational redshift formula and considering a typical white dwarf of $T_{\rm eff} = 10\,000$\,K and 0.6\,M$_{\odot}$. The systematic differences in radial velocities would result in uncertainties of the order of 0.01 -- 0.02\,M$_{\odot}$, with a maximum scatter of 0.07\,M$_{\odot}$. The listed results are compatible with those presented by \citet{napiwotzki2020}, who measured the H$\beta$ to be on average blueshifted with respect to the H$\alpha$.

As an additional test, we downgraded the spectral resolution of our sample down to 2\,\AA\ as it is typical for most of the follow-up observations or legacy spectroscopic surveys. In this case, we measured the radial velocities by fitting all five visible Balmer lines with the \citet{koester2010} models, including the entire line profile. The comparison with the 
radial velocities measured in the first method delivers an average difference of $\Delta \varv_{\rm WD} = 2.85 \pm 11.84$\,km/s that corresponds to a striking systematic uncertainty of $M = 0.06 \pm 0.23$\,M$_{\odot}$. This result is not surprising, and it explains the large scatter that is frequently observed for empirical or semi-empirical determinations of the mass-radius relation; typically using low-resolution spectra of white dwarfs and measuring radial velocities by means of the cross-correlation method would tend to overestimate masses and introduce random noise.

\begin{figure*}[t!]
    \centering
    \includegraphics[width=\linewidth]{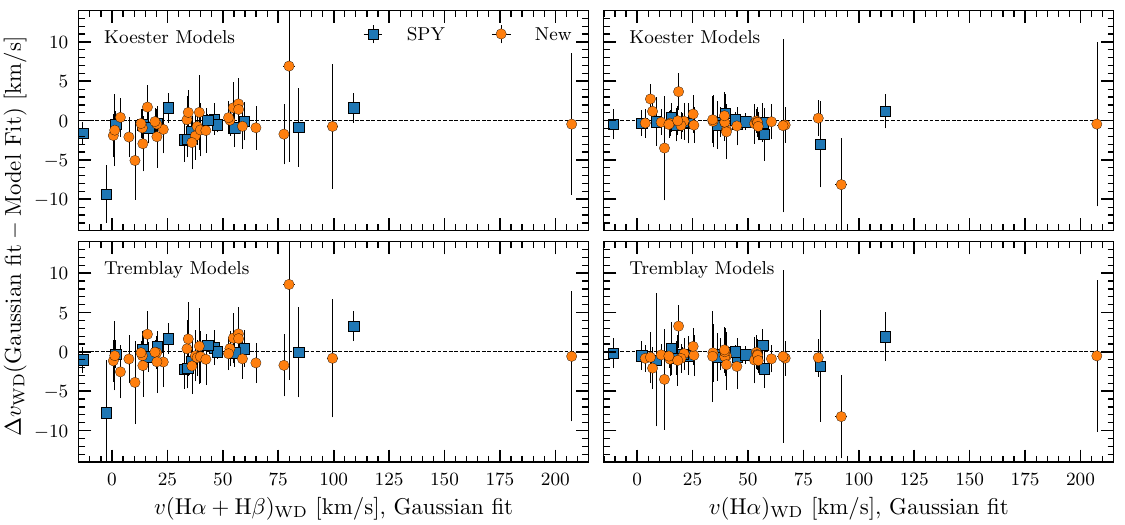}
    \caption{Comparison among the radial velocity estimates. 
    On the $x$-axes we plot the radial velocities measured with the second method, which fits two Gaussian functions and a quadratic polynomial to the  line cores and wings.  The $y$-axes display the difference between the Gaussian fit and the first method that uses the spectral templates to fit the line wings, respectively for the Koester and Tremblay models on the left and right panels. The left panels compare the H$\alpha$+H$\beta$ measurements, while the  right panels show the same comparison for the H$\alpha$ line. The colors and shapes of the symbols are the same as in Fig.\,\ref{fig:comparison_spectro}.}
    \label{fig:comparison_wd-rv}
\end{figure*}

We plot some of the comparisons of Table\,\ref{tab:rv_comparison} in Fig.\,\ref{fig:comparison_wd-rv}, which shows no significant trends as a function of radial velocity. In this figure, we note that the scatter is reduced when only the H$\alpha$ core is used for the comparisons. When discussing the results in Sect.\,\ref{sec:results}, we anticipate a better agreement between observed and theoretical masses and radii when  using the  radial velocities measured from the combined H$\alpha$+H$\beta$ .

\begin{table}
    \centering
        \caption{Radial velocity differences from various methods.}
    \label{tab:rv_comparison}
        \small
    \begin{tabular}{@{}lc m{2.2} m{2.2}@{}}
  \hline
  \hline
   \noalign{\smallskip}
Comparison   & Lines    &  \multicolumn{1}{c}{$\Delta \varv_{\rm WD}$ (km/s)} &  \multicolumn{1}{c}{$\Delta M$ (M$_{\odot}$)} \\
 \noalign{\smallskip}
  \hline
   \noalign{\smallskip}
 K1D      &  (H$\alpha$+H$\beta$) -- (H$\alpha$) & -0.79 , 1.42 & -0.02 , 0.03\\
Gfit --K1D  & H$\alpha$+H$\beta$     &  -0.63 ,  1.54 & -0.01,  0.04\\
 Gfit --K1D  & H$\alpha$                         &  -0.35 ,  1.54 & -0.01,  0.03\\
\noalign{\smallskip}
  \hline
   \noalign{\smallskip}
 T3D      & (H$\alpha$+H$\beta$) -- (H$\alpha$)  & -1.52 , 1.70 & -0.03 , 0.03\\
 Gfit -- T3D &  H$\alpha$+H$\beta$     &  -0.23 ,  2.10 &  0.00,  0.04\\
Gfit -- T3D & H$\alpha$                          &  -0.68 ,  1.45 & -0.01, 0.03\\
 \noalign{\smallskip}
  \hline
   \noalign{\smallskip}
K1D -- T3D & (H$\alpha$+H$\beta$)  & 0.41 , 0.71 & 0.01, 0.01\\
K1D -- T3D & H$\alpha$             & -0.32 , 0.77 & -0.01, 0.01\\
 Gfit &  (H$\alpha$+H$\beta$) -- (H$\alpha$)  &  -1.07,  -3.48 &  -0.02,  0.07\\
 \noalign{\smallskip}
  \hline
   \noalign{\smallskip}
  LR -- K1D & (H$\alpha$...H$\zeta$) -- (H$\alpha$+H$\beta$)  & 2.85 , 11.84 & 0.06, 0.23\\
   \noalign{\smallskip}
    \hline
       \noalign{\smallskip}
\multicolumn{4}{p{3.2in}}{{\bf Notes.} K1D and T3D refer to the Koester and Tremblay models, respectively; Gfit corresponds to a Gaussian fit of the H$\alpha$ or H$\alpha$+H$\beta$ performed independently; LR refers to the low-resolution  analysis that is performed with the Koester models;  the corresponding $\Delta M$ are  estimated for a typical white dwarf of $T_{\rm eff} = 10\,000$\,K and $M = 0.61$\,M$_\odot$. }
    \end{tabular}
\end{table}
\subsection{Peculiar objects}
\subsubsection{A DAZ white dwarf} 
In the studied sample, the object 0209$-$0140 shows the thin resonance lines of Ca\,{\sc ii} H\&K  that could be due to the accretion of planetary material. We measured the radial velocity of the bluest of the two lines, at 3933.66\,\AA, that is $\varv($Ca\,{\sc ii}\,H$) = 12.6 \pm 0.5$\,km/s. This line is known to be blue-shifted due to the quadratic Stark effect \citep[cf][]{dimitrijevic1992,vennes2011}; the comparison with the velocity obtained via the Gaussian-fit of the H$\alpha$ line results in a difference of $\Delta \varv = -1.8 \pm 0.7 $\,km/s (cf $\Delta \varv = -0.8 \pm 1.5$\,km/s with respect to the radial velocity from the fit of H$\alpha$+H$\beta$), which is in 3--$\sigma$ agreement with theoretical predictions for a blueshift of $-3.7$\,km/s , likely suggesting a photospheric nature for this line.

\subsubsection{A magnetic white dwarf}
\begin{figure}
    \centering
\includegraphics[width=\linewidth]{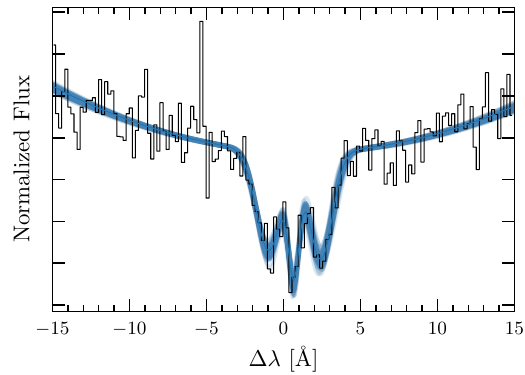}
    \caption{Triple Gaussian fit of the H$\alpha$ core of the magnetic white dwarf 1426$-$5716. The shaded curve represent random draws from the best-fit distribution.}
    \label{fig:dah}
\end{figure}

The white dwarf 1426$-$5716 shows Zeeman splitting of the H$\alpha$, H$\beta$, and H$\gamma$ lines, due to the presence of a surface magnetic field. For this object, we only measured the redshift of the H$\alpha$ core and the average intensity of the magnetic field. 
We fitted the wings of the H$\alpha$ line within $\pm 15$\,\AA\ from the center with second-order polynomial, while we performed a three Gaussian fit of the core by varying the central wavelength of the $\pi$ component  and the separation, $\Delta \lambda_{\rm Z}$, of the two $\sigma$ components  that are assumed to be symmetric with respect to the center and identical. The best-fit result is shown in Fig.\,\ref{fig:dah}, from which we obtained a redshift for the line core of $\Delta \lambda({\rm H}\alpha) = 0.66 \pm 0.04 $\,\AA\ that corresponds to a radial velocity of $\varv({\rm H}\alpha)_{\rm WD} = 30.3 \pm 1.7$\,km/s. 
The magnetic field of this white dwarf is rather weak, causing just a line split of $\Delta \lambda_{\rm Z} = 1.7 \pm 0.1$\,\AA\ that implies a magnetic field intensity of  $<|B|>$\,$ = 84 \pm 5 $\,kG \citep[cf eq.\,1 in][]{landstreet2016}.
\section{Photometric analysis}
\label{sec:phot_analysis}

In order to determine the masses and radii of the studied white dwarfs without relying on a theoretical mass-radius relation, we performed a photometric analysis that consists of a fit of the observed spectral energy distribution (SED$_{\rm \boldsymbol{o}}$)  with a theoretical representation of it (SED$_{\rm \boldsymbol{\theta}}$). The observed SED$_{\rm \boldsymbol{o}}$ is obtained by the externally calibrated BP/RP low-resolution spectra published in the {\em Gaia} DR3 \citep{de-angeli2023,montegrifo2023}, while the theoretical SED$_{\rm \boldsymbol{\theta}}$ is a function of the stellar radiance, radius, distance, and interstellar extinction, that is:
\begin{equation}
  \mathrm{SED}_{\boldsymbol{\theta}} \equiv f_\lambda = \left(\frac{R}{d} \right)^2 F_\lambda (T_{\rm eff},\,g) \times 10^{-0.4 A_\lambda}, 
  \label{eq:sed}
\end{equation}
where $f_\lambda$ and $F_\lambda$ are the apparent and intrinsic flux of a white dwarf, $R$ is the radius, $d$ is the distance, and $A_\lambda$ is the interstellar extinction.   
The stellar flux is a function of T$_{\rm eff}$ and $\log{g}$, and it is represented by the previously described grids of synthetic spectra for DA white dwarfs for which we have that the surface gravity is linked to the white dwarf mass and radius via the Newton's gravitational law: $g = G M / R^2$. 

We define a parameterized model, in which $\boldsymbol{\theta_i} = \{T_{{\rm eff},\,i},\, R_i,\, d_i,\, A(55)_i,\, M_i \}$ are the set of unknown parameters (temperature, radius, distance, extinction, and mass) that we choose to represent the SED$_{\boldsymbol{\theta}}$ of the $i$-th white dwarf in our sample. We use the monochromatic extinction at 5500\,\AA\ as unknown parameter, which is tied to the \citet{fitzpatrick2019} parametrization of the Milky Way's reddening law as a function of wavelength; hence, we adopt the 
total-to-selective extinction parameter $R(55)$ = 3.02 that is equivalent to the standard $R(V) = 3.1$ for the average Galactic extinction in the Johnson $V$-band.

Our goal is to determine the set of unknown parameters that best represent the observations for each white dwarf, $\boldsymbol{o}_i = \{{\rm SED}_{\boldsymbol{o},\,i},\, \varpi_i,\, \varv_{{\rm gr},\,i},\, \ell_i,\, b_i\}$, where $\ell_i$ and $b_i$ are the Galactic coordinates of a given star from which the measured interstellar extinction depends. This problem can be formalized by a probabilistic approach via the Bayes theorem:
\begin{equation}
    P(\boldsymbol{\theta} | \boldsymbol{o}) = \frac{P(\boldsymbol{o} | \boldsymbol{\theta}) \times P(\boldsymbol{\theta})}{P(\boldsymbol{o})}  \propto P(\boldsymbol{o} | \boldsymbol{\theta}) \times P(\boldsymbol{\theta}),
    \label{eq:bayes}
\end{equation}
where, omitting the $i$ subscript, $P(\boldsymbol{\theta})$ represents our prior knowledge on the model parameters,  $P(\boldsymbol{o})$ is to be considered as a normalization constant that can be ignored, and $P(\boldsymbol{o} | \boldsymbol{\theta})$ is the joint probability distribution relating the model and the data or, in other words, the likelihood, $\mathcal{L}(\boldsymbol{\theta} | \boldsymbol{o})$, of estimating an unknown physical parameter $\boldsymbol{\theta}$ given a measured quantity $\boldsymbol{o}$. We can break down Eq.\,\ref{eq:bayes} further, by rearranging the individual variables and observed quantities, thus expressing the joint probability distribution as:
\begin{equation}
\begin{aligned}
 P(\boldsymbol{\theta} | \boldsymbol{o}) &= P({\rm SED}_{\boldsymbol{o}} | {\rm SED}_{\boldsymbol{\theta}}) \times P(\varpi | d) 
 \times P(\varv_{\rm gr} | R,\, M) \\ 
&  \times P(\varpi,\,\ell,\,b | A(55)) \times P(T_{\rm eff}) \times P(R) \\
& \times P(d) \times P(A(55)  \times P(M).
\label{eq:bayes2}
 \end{aligned}
\end{equation}
The first joint probability term, on the right hand side, is equivalent to a standard photometric fit and summarizes the probability of observing a white dwarf spectrum given a set of physical parameters, the second one relates the probability of measuring a given parallax based on the real white dwarf distance, the third one corresponds to the probability of measuring the gravitational redshift based on the corresponding  white dwarf radii and masses, and finally the last joint probability term represent the relation between extinction and location within the Galaxy. The remaining five probabilities summarize our prior knowledge on the adopted model.

\subsection{Likelihoods}
\label{sec:likelihood}
Assuming normally distributed, uncorrelated uncertainties, we can write the joint probabilities (likelihoods) of Eq.\,\ref{eq:bayes2} as Gaussian distributions, $\mathcal{N} (\chi,\,\sigma_{\boldsymbol{o}})$, where we have that:
\begin{equation}
\chi^2 = \left(\frac{\mu_{\boldsymbol{\theta}} - \mu_{\boldsymbol{o}}}{\sigma_{\boldsymbol{o}}}
\right)^2 = 
\begin{cases}
\begin{aligned}
&\sum^{j = n}_{j = 1} \frac{[m_{\boldsymbol{\theta},\,j} - m_{\boldsymbol{o},\,j}]^2}{\sigma_{m_j}^2}\\
&\frac{[(d - (1/\varpi)]^2}{\sigma_{1/\varpi}^2} \\
&\frac{[(G M/R c) - \varv_{\rm gr}]^2}{\sigma_{\varv_{\rm gr}}^2} \\
&\frac{[A(55) - A_0]^2}{\sigma_{A_0}^2}
\label{eq:likelihood}
\end{aligned}
\end{cases}
\end{equation}
In our analysis, we express the observed {\em Gaia} BP/RP spectrum as ${\rm SED}_{\boldsymbol{o}} = \sum_{j =1}^{j=24} m_j$, where $m_j$ are fictitious magnitudes computed in 24 rectangular pass-bands with a 300\,\AA\ width (noting that the {\em Gaia} spectra cover the 3360 -- 10\,200\,\AA\ range). By applying the same parametrization to SED$_{\boldsymbol{\theta}}$, the first line of Eq.\,\ref{eq:likelihood} corresponds to the traditional formulation of photometric fitting. Both the apparent magnitude of the model and the observed magnitude in the $j$-th band are defined as:
\begin{equation}
m_{j} =  \frac{\int_{\lambda_j}^{\lambda_{j+1}} f_\lambda S_\lambda  \lambda \mathrm{d}\lambda}{\int_{\lambda_j}^{\lambda_{j+1}} S_\lambda  \lambda \mathrm{d}\lambda}, 
\label{eq:magnitude}
\end{equation}
where $f_\lambda$ is the apparent flux of the observed spectrum or that of the model  (Eq.\ref{eq:sed}) and $S_\lambda$ is the rectangular band-pass. We note that in our model we adopt a monochromatic parametrization of the interstellar extinction, meaning that we apply the interstellar extinction to our models before computing the integrated magnitudes. 

The other likelihoods of Eq.\,\ref{eq:likelihood} use the mathematical relations among parallax and distance, as well as the definition of gravitational redshift. The likelihood function applied to the interstellar extinction includes a mean value, $A_0$, that is extracted from the 3D maps of interstellar dust \citep{lallement2022}; the extinction at 5500\,\AA\ is given as a function of distance and Galactic coordinates (assumed to have infinite precision), $A_0 = \mathcal{F}(\varpi,\,\ell,\,b)$, to which we associate a conservative 10\,\% uncertainty of $\sigma_{A_0} = 0.1 \times \mathcal{F}(\varpi,\, \ell,\,b)$\,mag. 

We took into account two additional caveats. First, we apply a flux-correction to the BP/RP spectra, of the order of  $-1\pm1$\,\%, as we normalize them to the measured {\em Gaia} G magnitudes. Secondly, we account for systematic effects in the radial velocities of the non-degenerate companions that are assumed to be equal to the systemic velocity in the equation: $\varv_{\rm gr} = \varv_{\rm WD} - \varv_{\rm sys}$. We apply the suggested correction for {\em Gaia} radial velocities \citep{katz2023} and correct for the companion's gravitational redshift given in the {\em Gaia} archive or interpolated by us, which are respectively in the region of 0.2 and 0.6\,km/s. 

Finally, we noted that {\em Gaia} DR3 provided estimates of individual zero-point corrections for the parallaxes of each detected source and, moreover, the selected white dwarfs belong to binary systems where the companions have typically even more precise parallaxes. In Sect.\,\ref{sec:results}, we discuss the impact of different assumptions on the distance-based parallaxes or assuming no fixed extinction-distance relation. 
\subsection{Priors}
We adopted the following priors on the physical parameters to measure:
\begin{description}
    \item[$P(T_{\rm eff})$:] Gaussian prior  $\mathcal{N}(T_{\rm best}, \sigma_T)$ such that $T_{\rm best}$ is the best-fit result inferred by fixing the values of $d = 1/\varpi$ and $A(55) = A_0$; the $\sigma_{T}$ accounts for the measured spectroscopic uncertainties and an additional systematic temperature uncertainty of 5\,\%. 
    \item[$P(R)$:] Gaussian prior $\mathcal{N}(R_{\rm best}, \sigma_R)$ centered on the best-fit result inferred by fixing the values of $d = 1/\varpi$ and $A(55) = A_0$; a reasonable $\sigma_R$ is inferred from the spectroscopic estimates and adding a 0.05\,dex to the $\log{g}$ uncertainty.
    
    \item[$P(d)$:] A prior that is proportional to the volume of a spherical shell containing the studied star, i.e. proportional to $d^{2}$.
    
    \item[$P(A(55))$:] Uniform prior, $\mathcal{U}(0.0001\,{\rm mag},\,A_{\rm max}$), with the maximum value defined as $A_{\rm max} = \mathcal{F}(350\,{\rm pc},\,\ell_i,\,b_i)$.
    \item[$P(M)$:] Uniform prior, $\mathcal{U}(0.1\,M_\odot,\,1.4\,M_\odot)$.
\end{description}
We anticipate here that for the star WD\,1147+255, we measure a physically unrealistic mass (Sect.\,\ref{sec:results}), which required a modified $P(M) = \mathcal{U}(0.1\,M_\odot,\,3\,M_\odot)$ in order to estimate it without incurring into numerical errors.

\begin{figure*}[t!]
    \centering
\begin{subfigure}[t]{0.49\textwidth}
\includegraphics[width=\textwidth]{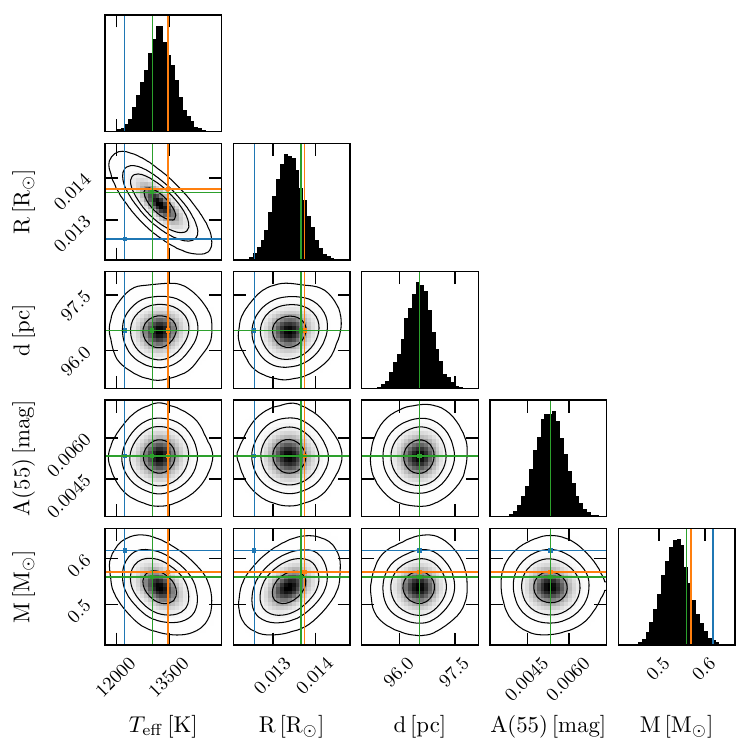}
\caption{Corner polot}
    \label{fig:corner}
    \end{subfigure}
   \begin{subfigure}[t]{0.49\textwidth}
    \includegraphics[width=\textwidth]{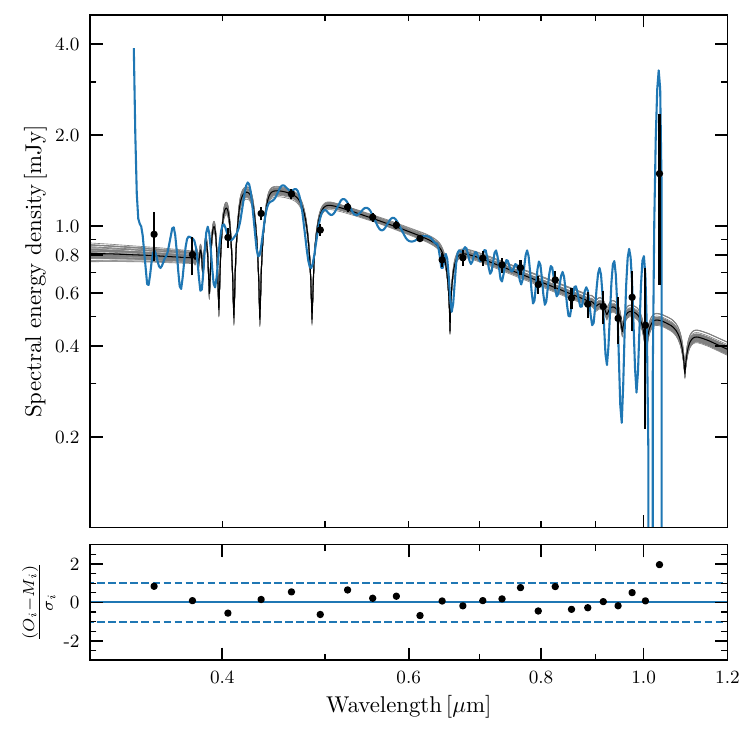}
             \caption{SED fit}
         \label{fig:sed}
        \end{subfigure}
 \caption{Results of the photometric analysis for the star 0319$-$7254. (\textbf{a}): Corner plot  \citep{corner} showing the correlations among the measured physical parameters. The colored lines respectively mark the estimated parameters obtained from the spectral analysis (blue: Koester models; orange: Tremblay models) or other works (green: the \citet{gentilefusillo2021} parameters as well as the parallax-based distance and the interstellar extinction from the 3D maps). (\textbf{b}): the best-fitting SED (black line) is compared to the {\em Gaia} BP/RP spectrum (cyan). The gray shaded line represents fifty random draws from the parameters distributions (using the Koester models). The bottom-panel shows the residuals between the model and the observed data.}
\end{figure*}
\subsection{Sampling the joint distribution}

Having set the probabilistic model, we sample the posterior $P(\boldsymbol{\theta} | \boldsymbol{o})$ as the product of priors and likelihoods (cf Eq.\,\ref{eq:bayes2}). Thus the problem is equivalent to maximizing the overall logarithmic likelihood, which can be written as the sum of individual logarithmic likelihoods and priors:

\begin{equation}
\begin{aligned}
\ln{\mathcal{L}} & = \ln{\mathcal{L}({\rm SED}_{\boldsymbol{\theta}} | {\rm SED}_{\boldsymbol{o}})} +  \ln{\mathcal{L}(d | \varpi)} + \ln{\mathcal{L}(R,\,M | \varv_{\rm GR})} \\
& + \ln{\mathcal{L}(A(55) | \,\varpi,\,\ell,\,b)}  + \ln{P(T_{\rm eff})} + \ln{P(R)} \\
& + \ln{P(d)} + \ln{P(A(55))} + \ln{P(M)}  = \\
& = -\frac{1}{2} \sum_{k=1}^{k=4} \left[\chi^2_{\boldsymbol{\theta},\,k}  + \ln{(2\pi \sigma_{\boldsymbol{o},\,k}^2)}\right]  - \frac{1}{2}  \left[\chi^2_{T} + \ln{(2\pi \sigma_T^2)}\right] \\
& - \frac{1}{2}  \left[\chi^2_{R} + \ln{(2\pi \sigma_R^2)}\right] + 2\log{d},
\end{aligned}
\end{equation}
where $\chi^2_{\boldsymbol{\theta},\,k}$ and $\sigma_{\boldsymbol{o},\,k}$ are the four likelihoods defined in Eq.\,\ref{eq:likelihood}; the logarithms of uniform priors are equal to zero  within the specified limits, while the distance prior is also included.

We first use the \verb|brute| function of the {\sc python} module {\sc lmfit}, to explore the parameter space. Then, we find the best-fit values via the \verb|minimize| function, using the Nelder-Mead algorithm. At this point, we sample the probability distributions with the {\sc python} library {\sc emcee}. The routine converges within a few-thousand steps in less than two minutes per star using 42 walkers, by running it on a single core of a cluster with two 32-cores AMD EPYC 9354P 3,25 GHz units with 64\,GB of RAM each. An example of the correlations among each physical parameter is shown in Fig.\,\ref{fig:corner} along with the corresponding best-fitting model in Fig.\,\ref{fig:sed}. 

In order to mitigate for differences in the chosen synthetic spectra, we performed distinct photometric fits using the two different spectral libraries (i.e., the Koester and Tremblay models) to represent the SED$_{\boldsymbol{\theta}}$. Moreover, for each fit we determined the physical parameters using both the radial velocity measurements, $\varv({\rm H}\alpha+{\rm H}\beta)_{\rm WD}$ and $\varv({\rm H}\alpha)_{\rm WD}$, which we averaged among the different methods.

We list the  photometric estimates of physical parameters in Table\,\ref{tab:phot_params}. We note that $T_{\rm eff}$, $R$, $d$, and $A(55)$ are either averaged values for the two spectral libraries, or fixed by their likelihoods; the masses and the other parameters are listed separately for the two radial velocity estimates, that is $\varv$\,(H$\alpha$+H$\beta$)$_{\rm WD}$ and $\varv$\,(H$\alpha$)$_{\rm WD}$, combining the results from the model and the Gaussian fits. This table also lists the correlation among mass and radius, $\rho(M,\,R)$, that arises from the posterior distributions (Fig.\,\ref{fig:corner} for an example), as well as the differences $\Delta R$ and $\Delta M$ with respect to the theoretical evolutionary tracks, which are discussed in Section \ref{sec:results}.

\section{Results}
\label{sec:results}
The main goal of this work was to measure masses and radii of white dwarfs in widely separated, common proper motion binaries that are independent from theoretical mass-radius relations. In this section, we will discuss the comparison with theoretical predictions and the impact of different radial velocity measurements, interstellar extinction, and parallax-based distances.

\subsection{Mass-radius relation}
\begin{figure*}
    \centering
\begin{subfigure}[t]{0.49\textwidth}
\includegraphics[width=\textwidth]{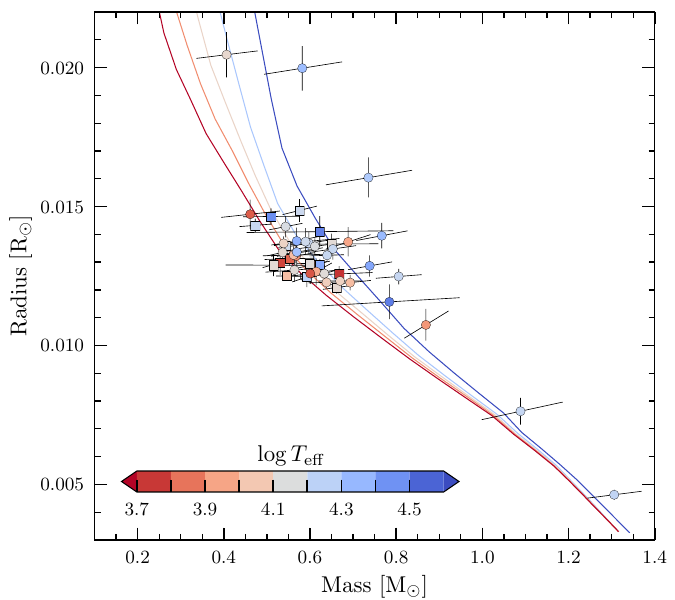}
\caption{$\varv_{\rm WD}$ measured from the H$\alpha$+H$\beta$ lines.}
    \label{fig:mr_hahb}
    \end{subfigure}
   \begin{subfigure}[t]{0.49\textwidth}
    \includegraphics[width=\textwidth]{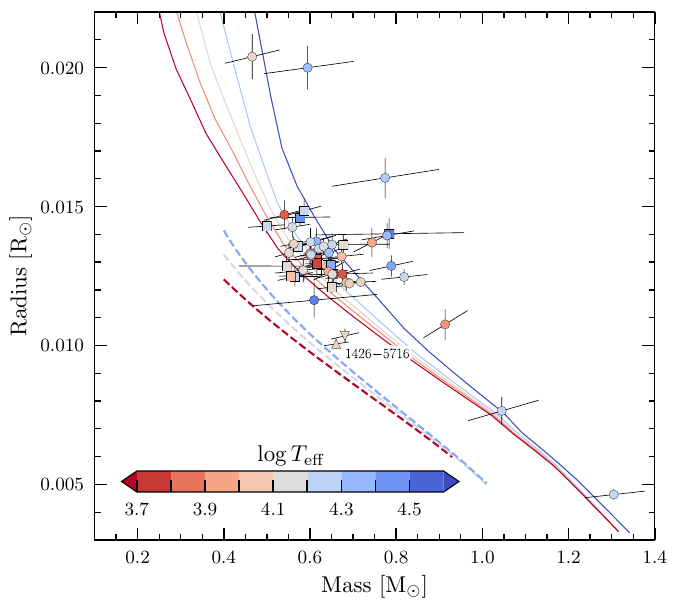}
             \caption{$\varv_{\rm WD}$ measured from the H$\alpha$ line.}
         \label{fig:mr_ha}
        \end{subfigure}
    \caption{Empirical masses and radii, computed adopting the radial velocity measured from the H$\alpha$+H$\beta$ lines together (left panel) and H$\alpha$ line only (right panel), are overplotted onto theoretical relations. The color-scheme maps the measured $T_{\rm eff}$, while the inclined error bars account for the mass/radius correlation. Symbols represent the SPY sample (squares) and new UVES data (circles);  up-/down-pointing triangles are used in the right panel for the DAH white dwarf, when it is analyzed with convective \citep{tremblay2011} or radiative models \citep{tremblay2015a}, respectively. The theoretical curves account for different core-compositions \citep[He, CO, and ONe; see Fig.\,\ref{fig:vgrav} for comparison and][for more details]{althaus2013,camisassa2016,camisassa2019}. The mass-radius relation for Fe-core white dwarfs \citep{panei2000} is represented by dashed lines.}
    \label{fig:mr}
\end{figure*}

We display the results of our photometric analysis in Fig.\,\ref{fig:mr}, where we compare the empirical measurements to the theoretical mass-radius relation. In the two panels we show the results obtained by adopting the average of the $\varv({\rm H}\alpha+{\rm H}\beta)_{\rm WD}$ measurements (Fig.\,\ref{fig:mr_hahb}) and those for the average of the $\varv({\rm H}\alpha)_{\rm WD}$ values (Fig.\,\ref{fig:mr_ha}).
A  visual inspection of this figure suggests a rather good agreement between theoretical predictions and measured values across the full range of masses. Moreover, we note a sensible reduction of uncertainties and scatter of the data points with respect to previous analyses that use the same method applied to low-resolution spectra \citep{arsenau2024}, or equivalent ones \citep{reid1996,tremblay2017}. 

Despite the overall good agreement, we note a group of six outliers on the right side of the theoretical curves, which are more evident in Fig.\,\ref{fig:mr_ha} (0115$-$1534, 0806$-$0006, 1054$-$4123, 1126$-$7631, 1524$-$2318, and WD\,1015$-$216) and have masses clustering around $0.8$\,M$_\odot$. There are other two less obvious outliers, 1140$-$3142 and 2245$-$1102, with masses of $\approx0.6$ and 0.9\,M$_\odot$, respectively. The 
disagreement with the theoretical curves  reduces when the $\varv({\rm H}\alpha+{\rm H}\beta)_{\rm WD}$ is used, but for the other six objects we may still be overestimating both the radius and the mass. The most extreme outlier, WD\,1147+255, is not shown in Fig.\,\ref{fig:mr}; for this object we measure an unrealistically large mass of $\approx 2.5$\,M$_\odot$. This star was also analyzed by \citet{reid1996}, who reported inconsistencies in the available radial velocity measurements of the companion. We note that {\em Gaia} DR3 detects a much-fainter $G \approx 20$\,mag source at 5-arcsec separation from the companion that, nevertheless, should not have an effect on its radial velocity measurement. Instead, we speculate that the non-degenerate companion of WD\,1147+255 is likely an unresolved binary or an active star, whose radial velocity is not sufficiently reliable. 

We note that a poor calibration of the {\it Gaia} data for the outliers, with the exception of WD\,1147+255, should be excluded, given the stringent quality cuts we applied for their selection. Instead, a small systematic offset in the  noisier spectra could be caused by a less effective telluric removal; for instance, inspection of the spectral regions used for the radial velocity fits (Fig.\,\ref{fig:fig01rv}--\ref{fig:fig10rv}) shows that 1054$-$4123, 1126$-$7631, 1140$-$3142,  and 1524$-$2318 have noisier cores or small dips that, however, do not cause large discrepancies among radial velocity measurements (Table\,\ref{tab:spectra_params}). For the star 0806$-$0006 the 3D models are not able to reproduce the H$\alpha$ core, but its radial velocity measurements are still consistent. On the other hand, the radial velocity measurements of WD\,1015$-$216 that is one of the hottest stars are rather discrepant, due to the weakness of its H$\alpha$ and H$\beta$ cores, which leads to very different positions in Fig.\,\ref{fig:mr}; a similar issue is noted for the star 0818+1211.

Although white dwarfs are known to be slow rotators \citep{koester1988}, the NLTE core of the H$\alpha$ is sensitive to rotational broadening or to the presence of unresolved Zeeman splitting due to very weak magnetic fields at their surface \citep{heber1997,koester1998,karl2005}. Both these effect could also contribute to offsetting the gravitational redshift.  While white dwarf rotation remains complicated to be extracted, in particular from noisy data, the high resolution provides a tool for detecting week magnetic fields down to a few kG such as that of 1426$-$5716, discussed further below. Here, we note that also the massive white dwarf 0207+0335 (Fig.\,\ref{fig:fig02rv}) possesses unusual, although noisy,  H$\alpha$ and H$\beta$ cores that may hint at the presence of another weak magnetic field. Concerning both the two most massive white dwarfs ($M > 1.0$\,M$_\odot$), we note that their measurements agree with the tracks within the error bars. It may seem that the radial velocity measured from the H$\alpha$ gives a better result for 0207+0335, while the H$\alpha$ core of 0608$-$0059 is poorly fitted by the model atmospheres. However, we note that in this mass regime the pressure shifts would have a larger impact, and a larger sample of objects is necessary to test the mass-radius relation in this parameter space.

Another possibility for the observed outliers can also be the presence of an unseen third companion, which could both affect the radius and the radial-velocity measurements.  Given the short baseline of our observations it is challenging to infer the presence of another unresolved object, unless the spectrum clearly appears as that of a double white dwarf. A longer baseline, like that adopted by the SPY project, is more effective at detecting line-core variations that can be associated to unresolved binaries \citep[e.g. fig.\,12 in][]{napiwotzki2020}. However, we note that our lowest-mass, He-core white dwarf, 1011+0536, might have a split H$\alpha$ core (Fig.\,\ref{fig:fig04rv}) that needs further time-resolved follow-up. Another star, 1434$-$3256, that is not an outlier, also shows an unusual H$\alpha$ core that may be altered by the presence of an unresolved object.

The magnetic white dwarf, 1426$-$5716,  that was mentioned above, also appears as an outlier  falling on the left side of the mass-radius relation (up-pointing triangle in Fig.\,\ref{fig:mr_ha}). For comparison with this object, we added the theoretical mass-radius relation of iron white dwarfs \citep[Fe-core;][]{panei2000} that would appear as more compact than standard CO- and ONe-core white dwarfs. A few well-known white dwarfs, such as Procyon\,B, GD\,140, and EG\,50, were previously questioned to possess Fe-cores due to their smaller-radii \citep[][and references therein]{panei2000} and later reassessed as standard white dwarfs \citep{provencal2002}. The existence of Fe-core white dwarfs would be intriguing, as it is linked to the supernova explosion of ONe-core white dwarfs \citep{isern1991,jones2019}; however, those uncommon objects are predicted to be hydrogen-depleted, thus they would be unlikely to appear as a normal -- although magnetic -- DA white dwarf such as 1426$-$5716.

On the other hand, weak magnetic fields of more than 50\,kG are predicted to suppress convection in cool white dwarf atmospheres, however, without having an impact on the mass-radius relation \citep{tremblay2015a}. Nevertheless, such magnetic fields are observed to alter the ultraviolet flux \citep{gentilefusillo2018}. We performed an additional photometric fit as in Sect.\,\ref{sec:phot_analysis}, but employing radiative synthetic  spectra \citep{tremblay2015a,gentilefusillo2018}. This test confirmed a rather small effect, increasing the measured mass and radius of 1426$-$5716 by $\approx 3$\,\% as shown in Fig.\,\ref{fig:mr_ha} by the additional down-pointing triangle. Such an increase moves the white dwarf closer to theoretical predictions,  but it is not sufficient to significantly improve the comparison. Hence, we speculate that a slightly noisy spectrum combined with a complex line-profile may have negatively impacted its radial velocity measurement. Including the H$\beta$ into the fit would add more noise, due to the lower SNR of the blue arm.

\begin{figure}[th!]
    \centering    
    \includegraphics[width=\linewidth]{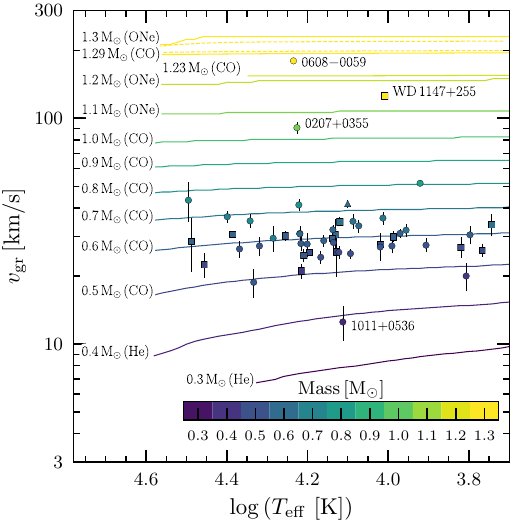}
    \caption{Distribution of gravitational redshifts as a function of $\log{T_{\rm eff}}$. The theoretical curves are plotted as solid lines, which are labeled by the corresponding mass and core compositions \citep[He-, CO-, and ONe-cores][]{althaus2013,camisassa2016,camisassa2019}. For reference, we also plot the models for two ultra-massive CO-core white dwarfs of 1.23 and 1.29\,M$_\odot$ \citep{camisassa2022} and the relativistic models for the 1.3\,M$_\odot$ ONe- and the 1.29\,M$_\odot$ CO-core white dwarfs \citep[dashed curves;][]{althaus2022,althaus2023}. The observed stars are color coded accordingly to their measured masses; the adopted symbols are the same as in Fig.\,\ref{fig:mr}. The four labeled stars are discussed in the text.}
    \label{fig:vgrav}
\end{figure}

\subsection{Physical properties of the sample}
\label{sec:physical_properties}

In Fig.\,\ref{fig:vgrav}, we display the distribution of gravitational redshifts as a function of $T_{\rm eff}$. The plotted values are those obtained by adopting the average $\varv_{\rm WD}$ of the H$\alpha$+H$\beta$ fits from Table\,\ref{tab:spectra_params}. The adopted theoretical mass-radius relation \citep{althaus2013,camisassa2016,camisassa2019} transitions from He- to CO-core and from CO to ONe-core at specific white dwarf masses that depend on the progenitor evolution. By interpolating the measured  gravitational redshifts onto the theoretical curves of Fig.\,\ref{fig:vgrav}, we determine the expected values of mass and radius for our sample. The percentage differences with respect to the measured masses and radii are listed as $\Delta R$ and $\Delta M$ in Table\,\ref{tab:phot_params}. These values are typically very similar, for each star, due to the direct proportionality between mass and radius in the gravitational-redshift formula, implying that an overestimate/underestimate of the radius corresponds to a larger/smaller mass. The agreement with the theoretical estimates is confirmed to be very good for the results employing the $\varv({\rm H}\alpha+{\rm H}\beta)_{\rm WD}$, leading to an average  $<$\,$\Delta M$\,$>$\,$ = 0.5^{+5.6}_{-4.0}$\,\% and $<$\,$\Delta R$\,$>$\,$= 0.5^{+6.0}_{-4.0}$\,\%. 
On the other hand, when using the $\varv({\rm H}\alpha)_{\rm WD}$ measurements, we obtain on average slightly larger differences of $<$\,$\Delta M$\,$>$\,$=  2.6^{+9.5}_{-2.3}$\,\% and $<$\,$\Delta R$\,$>$\,$=  2.5^{+9.5}_{-2.4}$\,\%.

The visual comparison with the overplotted theoretical curves of Fig.\,\ref{fig:vgrav}, as expected, shows that the studied sample is mostly dominated by CO-core white dwarfs. The white dwarf with the lowest mass, 1011+0536, is the only one that could harbor a He-core, based on our results, and observations of eclipsing binaries that evolved through binary interactions \citep{parsons2017}. However, at the level of precision of our measurements it is not possible to exclude a low-mass CO-core, if episodes of intense mass-loss took place during the evolution of this object \citep[e.g.][]{prada-moroni2009}. The two most massive white dwarfs, 0207+355 and 0608--0059, are at the CO-/ONe-core boundary, where it is also difficult to establish with accuracy better than a few percent what is the most likely core composition. The most massive of the two, 0608$-$0059, with a mass of $\approx 1.3$\,M$_\odot$ is expected to have an ONe-core. Interestingly, we measure $\Delta M \approx  2.6$\,\% with respect to ultramassive CO-core white dwarf models \citep{camisassa2022}, suggesting a slightly better agreement for the mass estimate in contrast to $ \Delta M \approx 4.4$\,\% with respect to ONe-core models. The mass of this white dwarf is also approaching a regime where general relativity effects should be taken into account \citep{althaus2022,althaus2023}, but the relatively large error on the gravitational redshift would not allow sufficient precision to distinguish between relativistic and non-relativistic models.

The outlier WD\,1147+255, labeled in Fig.\,\ref{fig:vgrav}, which was already discussed in the previous section, possesses a gravitational redshift that implies a mass of $\approx 1.15$\,M$_\odot$,  confirming an unreliable measurement that could arise from the radial velocity variability of its wide common proper-motion companion.

\begin{figure}[t]
    \centering
    \includegraphics[width=\linewidth]{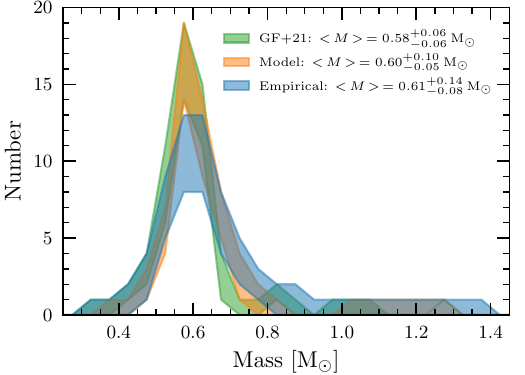}
    \caption{Mass distribution of the studied sample. Our direct measurements (blue curve) are compared to the model-dependent estimates obtained via interpolation of the gravitational redshift curves of Fig.\,\ref{fig:vgrav} (orange curve), and the photometric estimates by \citet{gentilefusillo2021} (green curves). The width of the colored bands represents the scatter due to the $1$\,$\sigma$ uncertainties.}
    \label{fig:mass-distribution}
\end{figure}

In Fig.\,\ref{fig:mass-distribution}, we plot the mass-distribution of our sample. The average mass distribution of the studied sample is compatible with that of a random selection of field white dwarfs. Accounting for the measured uncertainties, we find a relatively good agreement between the peak of our mass distribution and that obtained from the model-dependent analysis of \citet{gentilefusillo2021} for the same sample of white dwarfs.
Model-dependent masses interpolated from the gravitational-redshift curves of Fig.\,\ref{fig:vgrav} compare slightly better with the results of \citet{gentilefusillo2021}.
As discussed before, we measure an excess of white dwarfs with $\approx 0.8$\,M$_\odot$ that should mostly be attributed systematic uncertainties in the radial velocity measurements. Employing radial velocity measurements from the H$\alpha$ line would shift our results towards a slightly larger peak mass.

\subsection{Comparison with other methods}
\begin{figure*}
    \centering
    \includegraphics[width=\linewidth]{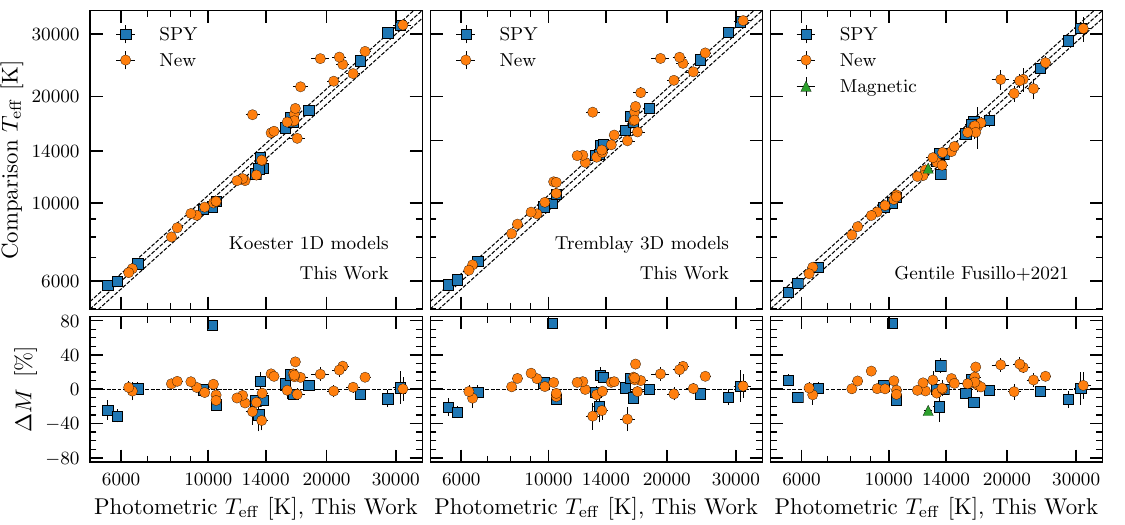}
    \caption{Comparison among different estimates of $T_{\rm eff}$ and masses. From left to right, we compare our photometric results to those obtained from the spectroscopic analyses of Sect.\,\ref{sec:spectroscopic_analysis} and the published values from \citet{gentilefusillo2021}. Symbols and colors are shown in the legend. The most discrepant object in the lower panels is WD\,1147+255.}
    \label{fig:comparison_all}
\end{figure*}

Figure\,\ref{fig:comparison_all} gives an overview of the
comparisons among $T_{\rm eff}$ and mass estimates, obtained from our spectroscopic (Sect.\,\ref{sec:spectroscopic_analysis}) and photometric analyses (Sect.\,\ref{sec:phot_analysis}), as well as a comparison with the measurements provided by \citet{gentilefusillo2021}. Our spectroscopic measures are derived via interpolation of $T_{\rm eff}$ and $\log{g}$ onto the reference mass-radius relation \citep[][]{althaus2013,camisassa2016,camisassa2019}. On the other hand, \citet{gentilefusillo2021} performed a photometric fit of the three {\em Gaia} bands by fixing the theoretical mass-radius relation, that is adopting those computed by \citet{bedard2020} for CO-core white dwarfs and \citet{serenelli2001} for the He-core white dwarfs. They also used the 3D extinction-distance maps \citep{lallement2022}, although in their analysis the effect of the extinction is modeled as band-integrated values.

We observe an overall good agreement among our spectroscopic and photometric $T_{\rm eff}$, except for some hotter white dwarfs that are typically found at larger distances where the interstellar extinction may have an impact in the photometric measurement. The photometric masses are on average slightly larger (1--2\,\%) with respect to the spectroscopic estimates, and the mass-differences present a scatter of about 15\,\% around the mean value that is seen to vary as a function of the 
white dwarf temperature. The comparison with the $T_{\rm eff}$ published by \citet{gentilefusillo2021} is much better, due to the similar  photometric approach. On the other hand, the mass comparison confirms the average systematic difference of about 1.2\,\% with respect to their results. The scatter around this average is of about 12\,\%. 

\begin{figure}
    \centering
    \includegraphics[width=\linewidth]{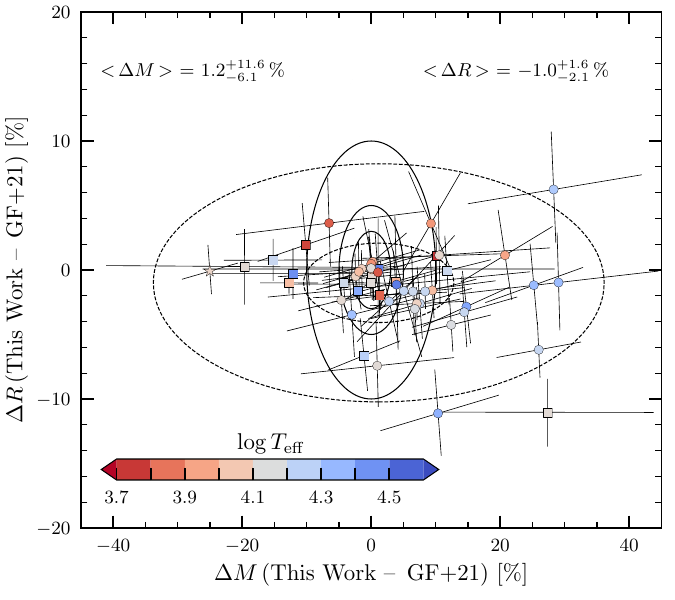}
    \caption{Comparison of measured masses and radii against the model-dependent photometric estimates by \citet{gentilefusillo2021} The color-scheme and symbols are the same as in Fig.\,\ref{fig:mr}. The solid curves represent the 3, 5, and 10\,\% contours with respect to the equality, while the dashed curves represent the 1--3\,$\sigma$ levels around the average differences (listed at the top corners of the figure).}
         \label{fig:mr_comparison}
\end{figure}

A more insightful comparison with the work of \citet{gentilefusillo2021} is presented in Fig.\,\ref{fig:mr_comparison}, where the relative differences of mass and radius measurements highlight the slight overestimate of masses ($<$\,$\Delta M$\,$>$\,$ = 2.5^{+11.2}_{-7.4}$\,\%) in contrast to the photometric radii that are slightly underestimated ($<$\,$\Delta R$\,$>$\,$ = -1 \pm 2$\,\%). The negative value of $\Delta R$ is likely due to our renormalization of the BP/RP spectra and to a different approach in measuring the interstellar extinction. 

\subsection{Parallax and extinction biases}

The approach employed for our photometric analysis allowed us to test a few different assumptions. Along with the {\em Gaia} EDR3, a zero-point correction for each source was provided \citep{lindegren2021}. This correction is roughly equivalent or much smaller than the parallax errors for the white dwarfs ($\lesssim 5\,$\,\% for our sample). Additionally, as members of binary systems, we note that the non-degenerate companions have either equivalent or much smaller parallax measurement errors. A change in the adopted parallax-based distance likelihood would imply a consequent change in the extinction derived from the 3D maps \citep{lallement2022}.

Numerical test showed that accounting for the parallax zero-point in Eq.\,\ref{eq:likelihood}, produces changes of the order of  $\pm 1$\,\% on the white dwarf masses, and at most 2\,\% for the two hottest, most distant white dwarfs. Adopting the companion's parallaxes to constrain the distance, again produces masses that are typically different by $\pm 1$\,\%; however, we note larger changes (as large as 6)\,\%  in excess for 0207+0355 and in defect for 1011+0536 that are the second most massive and the least massive white dwarfs in our sample, respectively. These two stars also are among the most distant objects in our sample.

We also considered the possibility of keeping the extinction as a free parameter, only limiting its maximum value to that provided by the 3D maps at a distance of 350\,pc along the corresponding sightlines. In this case, the scatter is again  below $2$\,\% and it does not  strongly depend on distance,  but a few objects beyond 100\,pc experience differences between 2--7\,\%. It is worth pointing out that, when assuming the prior on the interstellar extinction, all white dwarfs but 1524--2318 have values of $A(55)$ below 0.025\,mag.

These tests suggest that, inevitably, a direct measurement of white dwarf masses is affected by additional uncertainties, which may depend on the relevant assumptions on distance and extinction priors.

\section{Summary and conclusions}
In this work, we have presented the high-resolution spectroscopic follow-up of 48  white dwarfs with VLT/UVES. The observed objects belong to widely separated, common proper motion pairs with non-degenerate stars. We focused our analysis on the 33 objects classified as DA white dwarfs, to which we added 17 more objects of the same spectral class that are drawn from the SPY survey \citep{napiwotzki2020}, also observed with VLT/UVES and belonging to common proper motion pairs.

We performed a spectroscopic analysis with two sets of synthetic spectra, which parametrize the convection in 1D \citep{koester2010} or treat it with 3D hydrodynamics \citep{tremblay2013}, and we derived precise and accurate radial velocities by fitting a narrow spectral region of $\pm 15$\,\AA\ centered on the NLTE cores of the H$\alpha$ and H$\beta$ lines. Testing the simultaneous fitting of both lines and 
the H$\alpha$ line alone confirms small systematic differences that are known as pressure shifts in the Stark broadened wings of the Balmer lines of white dwarfs.

By taking advantage of the  radial velocity of the companions, which is a suitable proxy for the systemic radial velocity, we have directly measured the gravitational redshifts of the white dwarfs. Thus, we modeled the spectral energy distributions of the white dwarfs employing the {\em Gaia} BP/RP spectra, from which we directly measured the white dwarfs radii by means of a formalized Bayesian approach that uses the {\em Gaia} parallaxes and 3D extinction maps as likelihoods for the white dwarf distances and extinction. Finally, we obtained model-independent  masses estimates using the gravitational redshifts as external constraints.

Our results confirm a very good agreement with theoretical mass-radius relations \citep{althaus2013,camisassa2016,camisassa2019}, with a dispersion of 6\,\% around the equality (at the 1-$\sigma$ level). We also find that measuring white dwarf radial velocities from the H$\alpha$+H$\beta$ together delivers better agreement with the theoretical mass-radius relations, with respect to radial velocities of the H$\alpha$ only.

In our sample of binaries, we found a low-mass white dwarf of  0.41\,M$_\odot$, which could either possess a He- or a CO-core, and two massive white dwarfs of $M > 1.05$\,M$_{\odot}$ that could either possess ONe-cores or ultramassive CO-cores. Despite the good precision of the results, it is not sufficient to clearly disentangle the degeneracy among core compositions. The majority of white dwarfs have an average mass of $\sim 0.60$\,M$_\odot$ and CO-cores, as it is expected for a random selection. There is a number of outliers, concentrated around $\sim 0.8$\,M$_\odot$, which we speculate to either have biased radial velocities due to their  noisier line cores spectra or that could belong to unresolved binary systems (hence forming triple hierarchical systems)  that can be investigated through time-resolved spectroscopy. The most extreme outlier, WD\,1147+255, has an unrealistic mass of 2.5\,M$_\odot$, but its companion is known to have incompatible radial velocity measurements in the literature, thus it might be confirmed to hide another unresolved companion. We found two more peculiar white dwarfs in our sample. A DAZ white dwarf, whose radial velocity measured from the Ca\,{\sc ii}\,H line seems to confirm an externally polluted atmosphere. A weakly magnetic white dwarf, whose mass does not match the theoretical mass-radius relation, but we noted that its radial velocity may be slightly offset due to noise in the spectrum.  Inspection of the H$\alpha$ cores also hints at another magnetic white dwarf among the most massive objects in our sample, and two possible spectroscopic binaries visible from unusual line-cores.

A comparison between our model-independent mass determinations against the spectroscopic masses confirms a rather good agreement, but with a  scatter of the order of 15\,\% around the equality. Additionally, comparing our results with the photometric estimates provided by \citet{gentilefusillo2021}, indicates a slight offset of a few percent with respect to their physical parameters. 

We also tested that additional uncertainties may arise from different assumptions on distances, for instance by accounting for the {\em Gaia} parallax zero-point or adopting the companion's parallax to constrain the distance. Different assumptions on the interstellar extinction could affect the final results too.

To conclude, we stress that high-resolution spectroscopy of white dwarfs is confirmed to provide the only reliable way of measuring their radial velocities, as it mitigates the impact of pressure shift due to their Stark broadened lines, thus enabling a precise and accurate comparison with theoretical mass-radius relations on a star-by-star basis. Moreover, we stress that common proper motion pairs are favorable for analyzing the most massive white dwarfs that are more difficult to detect in eclipsing binaries. Nevertheless, a signal-to-noise ratio of at least 10 per pixel is recommended to achieve the best results, as well as observations at different epochs are also required to verify dubious outliers. The main limitations remain the intrinsic low-luminosity of white dwarfs, which decreases with their masses and age, hence it is implied that 8-m class or larger telescopes need to be used. 

Future multi-fiber facilities, like the 4-metre Multi-Object Spectroscopic Telescope \citep[4MOST;][]{4most} and existing ones, like the  WHT Enhanced Area Velocity Explorer \citep[WEAVE;][]{weave},  SDSS-V \citep{kollmeier2017}, and the Dark Energy Spectroscopic Instrument \citep[DESI;][]{desi2022}, will provide large numbers of white dwarf companions \citep[e.g.][]{s11} with well characterized low- and  medium-resolution ($R = 2000$--$5000$) spectra, unresolved binary fractions, and very reliable radial velocities that will complement those provided by the upcoming {\em Gaia} data releases.  These projects are also providing large unbiased samples of white dwarfs \citep{manser2024,crumpler2024} that enable a powerful statistical tool to investigate further the mass-radius relation. Of these upcoming projects, only WEAVE will deliver high-resolution ($R \approx 20,000$) spectra of white dwarfs, but the 4-m William Herschel telescope can likely provide reliable data down to $\approx 16$-mag white dwarfs, and below that magnitude range the results need to be verified. However, looking ahead on a 15--20 yr timescale, other survey facilities under development, like the planned 12-m Wide Field Spectroscopic Telescope \citep[WST;][]{wst} could provide large numbers of fainter white dwarfs with high-resolution spectroscopy that can be used for precise studies.

\section{Data availability}
\begin{acknowledgements}
We thank the anonymous referee for their constructive comments.  This work has made use of data from the European Space Agency (ESA) mission
{\it Gaia} (\url{https://www.cosmos.esa.int/gaia}), processed by the {\it Gaia}
Data Processing and Analysis Consortium (DPAC,
\url{https://www.cosmos.esa.int/web/gaia/dpac/consortium}). Funding for the DPAC
has been provided by national institutions, in particular the institutions
participating in the {\it Gaia} Multilateral Agreement.
This research has made use of the SIMBAD database \citep{wenger2000}, operated at CDS, Strasbourg, France. This research has made use of the VizieR catalogue access tool, CDS, Strasbourg, France. The original description of the VizieR service was published in \citet{ochsenbein20000}. Based on observations collected at the European Organisation for Astronomical Research in the Southern Hemisphere under ESO programmes 0108.D-0328(A),  0108.D-0328(B), 165.H-0588, 167.D-0407, 71.D-0383, 72.D-0487. This work made extensive use of the \verb|python| libraries, \verb|numpy| \citep{numpy} and \verb|matplotlib| \citep{matplotlib}, and \verb|lmfit| \citep{lmfit}.

RR acknowledges support from Grant RYC2021-030837-I and MEC acknowledges grant RYC2021-032721-I, both funded by MCIN/AEI/ 10.13039/501100011033 and by “European Union NextGeneration EU/PRTR”. This research was supported in part by the National Science Foundation under Grant No. NSF PHY-1748958. This work was partially supported by the AGAUR/Generalitat de Catalunya grant SGR-386/2021 and by the Spanish MINECO grant PID2020-117252GB-I00.

 \end{acknowledgements}


\bibliographystyle{aa}
\bibliography{bibliography}

\begin{thebibliography}{121}
\expandafter\ifx\csname natexlab\endcsname\relax\def\natexlab#1{#1}\fi

\bibitem[{{Abdurro'uf} {et~al.}(2022){Abdurro'uf}, {Accetta}, {Aerts}, {Silva
  Aguirre}, {Ahumada}, {Ajgaonkar}, {Filiz Ak}, {Alam}, {Allende Prieto},
  {Almeida}, {Anders}, {Anderson}, {Andrews}, {Anguiano}, {Aquino-Ort{\'\i}z},
  {Arag{\'o}n-Salamanca}, {Argudo-Fern{\'a}ndez}, {Ata}, {Aubert},
  {Avila-Reese}, {Badenes}, {Barb{\'a}}, {Barger}, {Barrera-Ballesteros},
  {Beaton}, {Beers}, {Belfiore}, {Bender}, {Bernardi}, {Bershady}, {Beutler},
  {Bidin}, {Bird}, {Bizyaev}, {Blanc}, {Blanton}, {Boardman}, {Bolton},
  {Boquien}, {Borissova}, {Bovy}, {Brandt}, {Brown}, {Brownstein}, {Brusa},
  {Buchner}, {Bundy}, {Burchett}, {Bureau}, {Burgasser}, {Cabang}, {Campbell},
  {Cappellari}, {Carlberg}, {Wanderley}, {Carrera}, {Cash}, {Chen}, {Chen},
  {Cherinka}, {Chiappini}, {Choi}, {Chojnowski}, {Chung}, {Clerc}, {Cohen},
  {Comerford}, {Comparat}, {da Costa}, {Covey}, {Crane}, {Cruz-Gonzalez},
  {Culhane}, {Cunha}, {Dai}, {Damke}, {Darling}, {Davidson}, {Davies},
  {Dawson}, {De Lee}, {Diamond-Stanic}, {Cano-D{\'\i}az}, {S{\'a}nchez},
  {Donor}, {Duckworth}, {Dwelly}, {Eisenstein}, {Elsworth}, {Emsellem},
  {Eracleous}, {Escoffier}, {Fan}, {Farr}, {Feng}, {Fern{\'a}ndez-Trincado},
  {Feuillet}, {Filipp}, {Fillingham}, {Frinchaboy}, {Fromenteau}, {Galbany},
  {Garc{\'\i}a}, {Garc{\'\i}a-Hern{\'a}ndez}, {Ge}, {Geisler}, {Gelfand},
  {G{\'e}ron}, {Gibson}, {Goddy}, {Godoy-Rivera}, {Grabowski}, {Green},
  {Greener}, {Grier}, {Griffith}, {Guo}, {Guy}, {Hadjara}, {Harding},
  {Hasselquist}, {Hayes}, {Hearty}, {Hern{\'a}ndez}, {Hill}, {Hogg},
  {Holtzman}, {Horta}, {Hsieh}, {Hsu}, {Hsu}, {Huber}, {Huertas-Company},
  {Hutchinson}, {Hwang}, {Ibarra-Medel}, {Chitham}, {Ilha}, {Imig}, {Jaekle},
  {Jayasinghe}, {Ji}, {Johnson}, {Jones}, {J{\"o}nsson}, {Katkov}, {Khalatyan},
  {Kinemuchi}, {Kisku}, {Knapen}, {Kneib}, {Kollmeier}, {Kong}, {Kounkel},
  {Kreckel}, {Krishnarao}, {Lacerna}, {Lane}, {Langgin}, {Lavender}, {Law},
  {Lazarz}, {Leung}, {Leung}, {Lewis}, {Li}, {Li}, {Lian}, {Liang}, {Lin},
  {Lin}, {Lin}, {Lintott}, {Long}, {Longa-Pe{\~n}a}, {L{\'o}pez-Cob{\'a}},
  {Lu}, {Lundgren}, {Luo}, {Mackereth}, {de la Macorra}, {Mahadevan},
  {Majewski}, {Manchado}, {Mandeville}, {Maraston}, {Margalef-Bentabol},
  {Masseron}, {Masters}, {Mathur}, {McDermid}, {Mckay}, {Merloni},
  {Merrifield}, {Meszaros}, {Miglio}, {Di Mille}, {Minniti}, {Minsley},
  {Monachesi}, {Moon}, {Mosser}, {Mulchaey}, {Muna}, {Mu{\~n}oz}, {Myers},
  {Myers}, {Nadathur}, {Nair}, {Nandra}, {Neumann}, {Newman}, {Nidever},
  {Nikakhtar}, {Nitschelm}, {O'Connell}, {Garma-Oehmichen}, {Luan Souza de
  Oliveira}, {Olney}, {Oravetz}, {Ortigoza-Urdaneta}, {Osorio}, {Otter},
  {Pace}, {Padilla}, {Pan}, {Pan}, {Parikh}, {Parker}, {Peirani}, {Pe{\~n}a
  Ram{\'\i}rez}, {Penny}, {Percival}, {Perez-Fournon}, {Pinsonneault},
  {Poidevin}, {Poovelil}, {Price-Whelan}, {B{\'a}rbara de Andrade Queiroz},
  {Raddick}, {Ray}, {Rembold}, {Riddle}, {Riffel}, {Riffel}, {Rix}, {Robin},
  {Rodr{\'\i}guez-Puebla}, {Roman-Lopes}, {Rom{\'a}n-Z{\'u}{\~n}iga}, {Rose},
  {Ross}, {Rossi}, {Rubin}, {Salvato}, {S{\'a}nchez}, {S{\'a}nchez-Gallego},
  {Sanderson}, {Santana Rojas}, {Sarceno}, {Sarmiento}, {Sayres}, {Sazonova},
  {Schaefer}, {Schiavon}, {Schlegel}, {Schneider}, {Schultheis}, {Schwope},
  {Serenelli}, {Serna}, {Shao}, {Shapiro}, {Sharma}, {Shen}, {Shetrone}, {Shu},
  {Simon}, {Skrutskie}, {Smethurst}, {Smith}, {Sobeck}, {Spoo}, {Sprague},
  {Stark}, {Stassun}, {Steinmetz}, {Stello}, {Stone-Martinez},
  {Storchi-Bergmann}, {Stringfellow}, {Stutz}, {Su}, {Taghizadeh-Popp},
  {Talbot}, {Tayar}, {Telles}, {Teske}, {Thakar}, {Theissen}, {Tkachenko},
  {Thomas}, {Tojeiro}, {Hernandez Toledo}, {Troup}, {Trump}, {Trussler},
  {Turner}, {Tuttle}, {Unda-Sanzana}, {V{\'a}zquez-Mata}, {Valentini},
  {Valenzuela}, {Vargas-Gonz{\'a}lez}, {Vargas-Maga{\~n}a}, {Alfaro},
  {Villanova}, {Vincenzo}, {Wake}, {Warfield}, {Washington}, {Weaver},
  {Weijmans}, {Weinberg}, {Weiss}, {Westfall}, {Wild}, {Wilde}, {Wilson},
  {Wilson}, {Wilson}, {Wolf}, {Wood-Vasey}, {Yan}, {Zamora}, {Zasowski},
  {Zhang}, {Zhao}, {Zheng}, {Zheng}, \& {Zhu}}]{sdss}
{Abdurro'uf}, {Accetta}, K., {Aerts}, C., {et~al.} 2022, \apjs, 259, 35

\bibitem[{{Adams}(1925)}]{adams1925}
{Adams}, W.~S. 1925, The Observatory, 48, 337

\bibitem[{{Althaus} {et~al.}(2022){Althaus}, {Camisassa}, {Torres}, {Battich},
  {C{\'o}rsico}, {Rebassa-Mansergas}, \& {Raddi}}]{althaus2022}
{Althaus}, L.~G., {Camisassa}, M.~E., {Torres}, S., {et~al.} 2022, \aap, 668,
  A58

\bibitem[{{Althaus} {et~al.}(2023){Althaus}, {C{\'o}rsico}, {Camisassa},
  {Torres}, {Gil-Pons}, {Rebassa-Mansergas}, \& {Raddi}}]{althaus2023}
{Althaus}, L.~G., {C{\'o}rsico}, A.~H., {Camisassa}, M.~E., {et~al.} 2023,
  \mnras, 523, 4492

\bibitem[{{Althaus} {et~al.}(2010){Althaus}, {C{\'o}rsico}, {Isern}, \&
  {Garc{\'\i}a-Berro}}]{althaus2010}
{Althaus}, L.~G., {C{\'o}rsico}, A.~H., {Isern}, J., \& {Garc{\'\i}a-Berro}, E.
  2010, \aapr, 18, 471

\bibitem[{{Althaus} {et~al.}(2013){Althaus}, {Miller Bertolami}, \&
  {C{\'o}rsico}}]{althaus2013}
{Althaus}, L.~G., {Miller Bertolami}, M.~M., \& {C{\'o}rsico}, A.~H. 2013,
  \aap, 557, A19

\bibitem[{{Arseneau} {et~al.}(2024){Arseneau}, {Chandra}, {Hwang}, {Zakamska},
  {Pallathadka}, {Crumpler}, {Hermes}, {El-Badry}, {Rix}, {Stassun},
  {G{\"a}nsicke}, {Brownstein}, \& {Morrison}}]{arsenau2024}
{Arseneau}, S., {Chandra}, V., {Hwang}, H.-C., {et~al.} 2024, \apj, 963, 17

\bibitem[{{B{\'e}dard} {et~al.}(2020){B{\'e}dard}, {Bergeron}, {Brassard}, \&
  {Fontaine}}]{bedard2020}
{B{\'e}dard}, A., {Bergeron}, P., {Brassard}, P., \& {Fontaine}, G. 2020, \apj,
  901, 93

\bibitem[{{B{\'e}dard} {et~al.}(2017){B{\'e}dard}, {Bergeron}, \&
  {Fontaine}}]{bedard2017}
{B{\'e}dard}, A., {Bergeron}, P., \& {Fontaine}, G. 2017, \apj, 848, 11

\bibitem[{{Bergeron} {et~al.}(2019){Bergeron}, {Dufour}, {Fontaine}, {Coutu},
  {Blouin}, {Genest-Beaulieu}, {B{\'e}dard}, \& {Rolland}}]{bergeron2019}
{Bergeron}, P., {Dufour}, P., {Fontaine}, G., {et~al.} 2019, \apj, 876, 67

\bibitem[{{Bergeron} {et~al.}(1995){Bergeron}, {Liebert}, \&
  {Fulbright}}]{bergeron1995}
{Bergeron}, P., {Liebert}, J., \& {Fulbright}, M.~S. 1995, \apj, 444, 810

\bibitem[{{Boubert} {et~al.}(2019){Boubert}, {Strader}, {Aguado}, {Seabroke},
  {Koposov}, {Sanders}, {Swihart}, {Chomiuk}, \& {Evans}}]{boubert2019}
{Boubert}, D., {Strader}, J., {Aguado}, D., {et~al.} 2019, \mnras, 486, 2618

\bibitem[{{Buder} {et~al.}(2020){Buder}, {Sharma}, {Kos}, {Amarsi},
  {Nordlander}, {Lind}, {Martell}, {Asplund}, {Bland-Hawthorn}, {Casey}, {De
  Silva}, {D'Orazi}, {Freeman}, {Hayden}, {Lewis}, {Lin}, {Schlesinger},
  {Simpson}, {Stello}, {Zucker}, {Zwitter}, {Beeson}, {Buck}, {Casagrande},
  {Clark}, {Cotar}, {Da Costa}, {de Grijs}, {Feuillet}, {Horner}, {Khanna},
  {Kafle}, {Liu}, {Montet}, {Nandakumar}, {Nataf}, {Ness}, {Spina}, {Traven},
  {Tepper-Garcia}, {Ting}, {Vogrincic}, {Wittenmyer}, {Zerjal}, \& {the GALAH
  collaboration}}]{buder2021}
{Buder}, S., {Sharma}, S., {Kos}, J., {et~al.} 2020, arXiv e-prints,
  arXiv:2011.02505

\bibitem[{{Camisassa} {et~al.}(2019){Camisassa}, {Althaus}, {C{\'o}rsico}, {De
  Ger{\'o}nimo}, {Miller Bertolami}, {Novarino}, {Rohrmann}, {Wachlin}, \&
  {Garc{\'\i}a-Berro}}]{camisassa2019}
{Camisassa}, M.~E., {Althaus}, L.~G., {C{\'o}rsico}, A.~H., {et~al.} 2019,
  \aap, 625, A87

\bibitem[{{Camisassa} {et~al.}(2016){Camisassa}, {Althaus}, {C{\'o}rsico},
  {Vinyoles}, {Serenelli}, {Isern}, {Miller Bertolami}, \&
  {Garc{\'\i}a-Berro}}]{camisassa2016}
{Camisassa}, M.~E., {Althaus}, L.~G., {C{\'o}rsico}, A.~H., {et~al.} 2016,
  \apj, 823, 158

\bibitem[{{Camisassa} {et~al.}(2022){Camisassa}, {Althaus}, {Koester},
  {Torres}, {Gil-Pons}, \& {C{\'o}rsico}}]{camisassa2022}
{Camisassa}, M.~E., {Althaus}, L.~G., {Koester}, D., {et~al.} 2022, \mnras,
  511, 5198

\bibitem[{{Casewell} {et~al.}(2009){Casewell}, {Dobbie}, {Napiwotzki},
  {Burleigh}, {Barstow}, \& {Jameson}}]{casewell2009}
{Casewell}, S.~L., {Dobbie}, P.~D., {Napiwotzki}, R., {et~al.} 2009, \mnras,
  395, 1795

\bibitem[{{Chandra} {et~al.}(2020){Chandra}, {Hwang}, {Zakamska}, \&
  {Cheng}}]{chandra2020}
{Chandra}, V., {Hwang}, H.-C., {Zakamska}, N.~L., \& {Cheng}, S. 2020, \apj,
  899, 146

\bibitem[{{Crumpler} {et~al.}(2024){Crumpler}, {Chandra}, {Zakamska},
  {Pallathadka}, {Arseneau}, {Fusillo}, {Hermes}, {Badenes}, {Chakraborty},
  {G{\"a}nsicke}, \& {Schmidt}}]{crumpler2024}
{Crumpler}, N.~R., {Chandra}, V., {Zakamska}, N.~L., {et~al.} 2024, \apj, 977,
  237

\bibitem[{{Cukanovaite} {et~al.}(2023){Cukanovaite}, {Tremblay}, {Toonen},
  {Temmink}, {Manser}, {O'Brien}, \& {McCleery}}]{cukanovaite2023}
{Cukanovaite}, E., {Tremblay}, P.~E., {Toonen}, S., {et~al.} 2023, \mnras, 522,
  1643

\bibitem[{{De Angeli} {et~al.}(2023){De Angeli}, {Weiler}, {Montegriffo},
  {Evans}, {Riello}, {Andrae}, {Carrasco}, {Busso}, {Burgess}, {Cacciari},
  {Davidson}, {Harrison}, {Hodgkin}, {Jordi}, {Osborne}, {Pancino},
  {Altavilla}, {Barstow}, {Bailer-Jones}, {Bellazzini}, {Brown}, {Castellani},
  {Cowell}, {Delchambre}, {De Luise}, {Diener}, {Fabricius}, {Fouesneau},
  {Fr{\'e}mat}, {Gilmore}, {Giuffrida}, {Hambly}, {Hidalgo}, {Holland},
  {Kostrzewa-Rutkowska}, {van Leeuwen}, {Lobel}, {Marinoni}, {Miller},
  {Pagani}, {Palaversa}, {Piersimoni}, {Pulone}, {Ragaini}, {Rainer},
  {Richards}, {Rixon}, {Ruz-Mieres}, {Sanna}, {Sarro}, {Rowell}, {Sordo},
  {Walton}, \& {Yoldas}}]{de-angeli2023}
{De Angeli}, F., {Weiler}, M., {Montegriffo}, P., {et~al.} 2023, \aap, 674, A2

\bibitem[{{de Jong} {et~al.}(2012){de Jong}, {Bellido-Tirado}, {Chiappini},
  {Depagne}, {Haynes}, {Johl}, {Schnurr}, {Schwope}, {Walcher}, {Dionies},
  {Haynes}, {Kelz}, {Kitaura}, {Lamer}, {Minchev}, {M{\"u}ller}, {Nuza},
  {Olaya}, {Piffl}, {Popow}, {Steinmetz}, {Ural}, {Williams}, {Winkler},
  {Wisotzki}, {Ansorge}, {Banerji}, {Gonzalez Solares}, {Irwin}, {Kennicutt},
  {King}, {McMahon}, {Koposov}, {Parry}, {Sun}, {Walton}, {Finger}, {Iwert},
  {Krumpe}, {Lizon}, {Vincenzo}, {Amans}, {Bonifacio}, {Cohen}, {Francois},
  {Jagourel}, {Mignot}, {Royer}, {Sartoretti}, {Bender}, {Grupp}, {Hess},
  {Lang-Bardl}, {Muschielok}, {B{\"o}hringer}, {Boller}, {Bongiorno}, {Brusa},
  {Dwelly}, {Merloni}, {Nandra}, {Salvato}, {Pragt}, {Navarro}, {Gerlofsma},
  {Roelfsema}, {Dalton}, {Middleton}, {Tosh}, {Boeche}, {Caffau}, {Christlieb},
  {Grebel}, {Hansen}, {Koch}, {Ludwig}, {Quirrenbach}, {Sbordone}, {Seifert},
  {Thimm}, {Trifonov}, {Helmi}, {Trager}, {Feltzing}, {Korn}, \&
  {Boland}}]{4most}
{de Jong}, R.~S., {Bellido-Tirado}, O., {Chiappini}, C., {et~al.} 2012, in
  Society of Photo-Optical Instrumentation Engineers (SPIE) Conference Series,
  Vol. 8446, Ground-based and Airborne Instrumentation for Astronomy IV, ed.
  I.~S. {McLean}, S.~K. {Ramsay}, \& H.~{Takami}, 84460T

\bibitem[{{Dekker} {et~al.}(2000){Dekker}, {D'Odorico}, {Kaufer}, {Delabre}, \&
  {Kotzlowski}}]{dekker2000}
{Dekker}, H., {D'Odorico}, S., {Kaufer}, A., {Delabre}, B., \& {Kotzlowski}, H.
  2000, in Society of Photo-Optical Instrumentation Engineers (SPIE) Conference
  Series, Vol. 4008, Optical and IR Telescope Instrumentation and Detectors,
  ed. M.~{Iye} \& A.~F. {Moorwood}, 534--545

\bibitem[{{DESI Collaboration} {et~al.}(2022){DESI Collaboration}, {Abareshi},
  {Aguilar}, {Ahlen}, {Alam}, {Alexander}, {Alfarsy}, {Allen}, {Allende
  Prieto}, {Alves}, {Ameel}, {Armengaud}, {Asorey}, {Aviles}, {Bailey},
  {Balaguera-Antol{\'\i}nez}, {Ballester}, {Baltay}, {Bault}, {Beltran},
  {Benavides}, {BenZvi}, {Berti}, {Besuner}, {Beutler}, {Bianchi}, {Blake},
  {Blanc}, {Blum}, {Bolton}, {Bose}, {Bramall}, {Brieden}, {Brodzeller},
  {Brooks}, {Brownewell}, {Buckley-Geer}, {Cahn}, {Cai}, {Canning}, {Capasso},
  {Carnero Rosell}, {Carton}, {Casas}, {Castander}, {Cervantes-Cota},
  {Chabanier}, {Chaussidon}, {Chuang}, {Circosta}, {Cole}, {Cooper}, {da
  Costa}, {Cousinou}, {Cuceu}, {Davis}, {Dawson}, {de la Cruz-Noriega}, {de la
  Macorra}, {de Mattia}, {Della Costa}, {Demmer}, {Derwent}, {Dey}, {Dey},
  {Dhungana}, {Ding}, {Dobson}, {Doel}, {Donald-McCann}, {Donaldson},
  {Douglass}, {Duan}, {Dunlop}, {Edelstein}, {Eftekharzadeh}, {Eisenstein},
  {Enriquez-Vargas}, {Escoffier}, {Evatt}, {Fagrelius}, {Fan}, {Fanning},
  {Fawcett}, {Ferraro}, {Ereza}, {Flaugher}, {Font-Ribera}, {Forero-Romero},
  {Frenk}, {Fromenteau}, {G{\"a}nsicke}, {Garcia-Quintero}, {Garrison},
  {Gazta{\~n}aga}, {Gerardi}, {Gil-Mar{\'\i}n}, {Gontcho A Gontcho},
  {Gonzalez-Morales}, {Gonzalez-de-Rivera}, {Gonzalez-Perez}, {Gordon},
  {Graur}, {Green}, {Grove}, {Gruen}, {Gutierrez}, {Guy}, {Hahn}, {Harris},
  {Herrera}, {Herrera-Alcantar}, {Honscheid}, {Howlett}, {Huterer},
  {Ir{\v{s}}i{\v{c}}}, {Ishak}, {Jelinsky}, {Jiang}, {Jimenez}, {Jing},
  {Joyce}, {Jullo}, {Juneau}, {Kara{\c{c}}ayl{\i}}, {Karamanis}, {Karcher},
  {Karim}, {Kehoe}, {Kent}, {Kirkby}, {Kisner}, {Kitaura}, {Koposov},
  {Kov{\'a}cs}, {Kremin}, {Krolewski}, {L'Huillier}, {Lahav}, {Lambert},
  {Lamman}, {Lan}, {Landriau}, {Lane}, {Lang}, {Lange}, {Lasker}, {Le Guillou},
  {Leauthaud}, {Le Van Suu}, {Levi}, {Li}, {Magneville}, {Manera}, {Manser},
  {Marshall}, {Martini}, {McCollam}, {McDonald}, {Meisner},
  {Mena-Fern{\'a}ndez}, {Meneses-Rizo}, {Mezcua}, {Miller}, {Miquel},
  {Montero-Camacho}, {Moon}, {Moustakas}, {Mueller}, {Mu{\~n}oz-Guti{\'e}rrez},
  {Myers}, {Nadathur}, {Najita}, {Napolitano}, {Neilsen}, {Newman}, {Nie},
  {Ning}, {Niz}, {Norberg}, {Noriega}, {O'Brien}, {Obuljen},
  {Palanque-Delabrouille}, {Palmese}, {Zhiwei}, {Pappalardo}, {PENG},
  {Percival}, {Perruchot}, {Pogge}, {Poppett}, {Porredon}, {Prada},
  {Prochaska}, {Pucha}, {P{\'e}rez-Fern{\'a}ndez}, {P{\'e}rez-R{\`a}fols},
  {Rabinowitz}, \& {Raichoor}}]{desi2022}
{DESI Collaboration}, {Abareshi}, B., {Aguilar}, J., {et~al.} 2022, \aj, 164,
  207

\bibitem[{{Dimitrijevi{\'c}} \& {Sahal-Br{\'e}chot}(1992)}]{dimitrijevic1992}
{Dimitrijevi{\'c}}, M.~S. \& {Sahal-Br{\'e}chot}, S. 1992, Bulletin
  Astronomique de Belgrade, 145, 83

\bibitem[{{Ekstr{\"o}m} {et~al.}(2012){Ekstr{\"o}m}, {Georgy}, {Eggenberger},
  {Meynet}, {Mowlavi}, {Wyttenbach}, {Granada}, {Decressin}, {Hirschi},
  {Frischknecht}, {Charbonnel}, \& {Maeder}}]{ekstrom2012}
{Ekstr{\"o}m}, S., {Georgy}, C., {Eggenberger}, P., {et~al.} 2012, \aap, 537,
  A146

\bibitem[{{El-Badry} {et~al.}(2021){El-Badry}, {Rix}, \&
  {Heintz}}]{elbadry2021}
{El-Badry}, K., {Rix}, H.-W., \& {Heintz}, T.~M. 2021, \mnras, 506, 2269

\bibitem[{{Falcon} {et~al.}(2010){Falcon}, {Winget}, {Montgomery}, \&
  {Williams}}]{falcon2010}
{Falcon}, R.~E., {Winget}, D.~E., {Montgomery}, M.~H., \& {Williams}, K.~A.
  2010, \apj, 712, 585

\bibitem[{{Fitzpatrick} {et~al.}(2019){Fitzpatrick}, {Massa}, {Gordon},
  {Bohlin}, \& {Clayton}}]{fitzpatrick2019}
{Fitzpatrick}, E.~L., {Massa}, D., {Gordon}, K.~D., {Bohlin}, R., \& {Clayton},
  G.~C. 2019, \apj, 886, 108

\bibitem[{Foreman-Mackey(2016)}]{corner}
Foreman-Mackey, D. 2016, The Journal of Open Source Software, 1, 24

\bibitem[{{Foreman-Mackey} {et~al.}(2013){Foreman-Mackey}, {Hogg}, {Lang}, \&
  {Goodman}}]{foreman-mackey2013}
{Foreman-Mackey}, D., {Hogg}, D.~W., {Lang}, D., \& {Goodman}, J. 2013, \pasp,
  125, 306

\bibitem[{{Freudling} {et~al.}(2013){Freudling}, {Romaniello}, {Bramich},
  {Ballester}, {Forchi}, {Garc{\'{\i}}a-Dabl{\'o}}, {Moehler}, \&
  {Neeser}}]{freudling2013}
{Freudling}, W., {Romaniello}, M., {Bramich}, D.~M., {et~al.} 2013, \aap, 559,
  A96

\bibitem[{{Gaia Collaboration} {et~al.}(2018){Gaia Collaboration}, {Brown},
  {Vallenari}, {Prusti}, {de Bruijne}, {Babusiaux}, {Bailer-Jones}, {Biermann},
  {Evans}, {Eyer}, {Jansen}, {Jordi}, {Klioner}, {Lammers}, {Lindegren},
  {Luri}, {Mignard}, {Panem}, {Pourbaix}, {Randich}, {Sartoretti}, {Siddiqui},
  {Soubiran}, {van Leeuwen}, {Walton}, {Arenou}, {Bastian}, {Cropper},
  {Drimmel}, {Katz}, {Lattanzi}, {Bakker}, {Cacciari}, {Casta{\~n}eda},
  {Chaoul}, {Cheek}, {De Angeli}, {Fabricius}, {Guerra}, {Holl}, {Masana},
  {Messineo}, {Mowlavi}, {Nienartowicz}, {Panuzzo}, {Portell}, {Riello},
  {Seabroke}, {Tanga}, {Th{\'e}venin}, {Gracia-Abril}, {Comoretto},
  {Garcia-Reinaldos}, {Teyssier}, {Altmann}, {Andrae}, {Audard},
  {Bellas-Velidis}, {Benson}, {Berthier}, {Blomme}, {Burgess}, {Busso},
  {Carry}, {Cellino}, {Clementini}, {Clotet}, {Creevey}, {Davidson}, {De
  Ridder}, {Delchambre}, {Dell'Oro}, {Ducourant},
  {Fern{\'a}ndez-Hern{\'a}ndez}, {Fouesneau}, {Fr{\'e}mat}, {Galluccio},
  {Garc{\'\i}a-Torres}, {Gonz{\'a}lez-N{\'u}{\~n}ez}, {Gonz{\'a}lez-Vidal},
  {Gosset}, {Guy}, {Halbwachs}, {Hambly}, {Harrison}, {Hern{\'a}ndez},
  {Hestroffer}, {Hodgkin}, {Hutton}, {Jasniewicz}, {Jean-Antoine-Piccolo},
  {Jordan}, {Korn}, {Krone-Martins}, {Lanzafame}, {Lebzelter}, {L{\"o}ffler},
  {Manteiga}, {Marrese}, {Mart{\'\i}n-Fleitas}, {Moitinho}, {Mora}, {Muinonen},
  {Osinde}, {Pancino}, {Pauwels}, {Petit}, {Recio-Blanco}, {Richards},
  {Rimoldini}, {Robin}, {Sarro}, {Siopis}, {Smith}, {Sozzetti}, {S{\"u}veges},
  {Torra}, {van Reeven}, {Abbas}, {Abreu Aramburu}, {Accart}, {Aerts},
  {Altavilla}, {{\'A}lvarez}, {Alvarez}, {Alves}, {Anderson}, {Andrei},
  {Anglada Varela}, {Antiche}, {Antoja}, {Arcay}, {Astraatmadja}, {Bach},
  {Baker}, {Balaguer-N{\'u}{\~n}ez}, {Balm}, {Barache}, {Barata}, {Barbato},
  {Barblan}, {Barklem}, {Barrado}, {Barros}, {Barstow}, {Bartholom{\'e}
  Mu{\~n}oz}, {Bassilana}, {Becciani}, {Bellazzini}, {Berihuete}, {Bertone},
  {Bianchi}, {Bienaym{\'e}}, {Blanco-Cuaresma}, {Boch}, {Boeche}, {Bombrun},
  {Borrachero}, {Bossini}, {Bouquillon}, {Bourda}, {Bragaglia}, {Bramante},
  {Breddels}, {Bressan}, {Brouillet}, {Br{\"u}semeister}, {Brugaletta},
  {Bucciarelli}, {Burlacu}, {Busonero}, {Butkevich}, {Buzzi}, {Caffau},
  {Cancelliere}, {Cannizzaro}, {Cantat-Gaudin}, {Carballo}, {Carlucci},
  {Carrasco}, {Casamiquela}, {Castellani}, {Castro-Ginard}, {Charlot},
  {Chemin}, {Chiavassa}, {Cocozza}, {Costigan}, {Cowell}, {Crifo}, {Crosta},
  {Crowley}, {Cuypers}, {Dafonte}, {Damerdji}, {Dapergolas}, {David}, {David},
  {de Laverny}, {De Luise}, {De March}, {de Martino}, {de Souza}, {de Torres},
  {Debosscher}, {del Pozo}, {Delbo}, {Delgado}, {Delgado}, {Di Matteo},
  {Diakite}, {Diener}, {Distefano}, {Dolding}, {Drazinos}, {Dur{\'a}n},
  {Edvardsson}, {Enke}, {Eriksson}, {Esquej}, {Eynard Bontemps}, {Fabre},
  {Fabrizio}, {Faigler}, {Falc{\~a}o}, {Farr{\`a}s Casas}, {Federici},
  {Fedorets}, {Fernique}, {Figueras}, {Filippi}, {Findeisen}, {Fonti},
  {Fraile}, {Fraser}, {Fr{\'e}zouls}, {Gai}, {Galleti}, {Garabato},
  {Garc{\'\i}a-Sedano}, {Garofalo}, {Garralda}, {Gavel}, {Gavras}, {Gerssen},
  {Geyer}, {Giacobbe}, {Gilmore}, {Girona}, {Giuffrida}, {Glass}, {Gomes},
  {Granvik}, {Gueguen}, {Guerrier}, {Guiraud}, {Guti{\'e}rrez-S{\'a}nchez},
  {Haigron}, {Hatzidimitriou}, {Hauser}, {Haywood}, {Heiter}, {Helmi}, {Heu},
  {Hilger}, {Hobbs}, {Hofmann}, {Holland}, {Huckle}, {Hypki}, {Icardi},
  {Jan{\ss}en}, {Jevardat de Fombelle}, {Jonker}, {Juh{\'a}sz}, {Julbe},
  {Karampelas}, {Kewley}, {Klar}, {Kochoska}, {Kohley}, {Kolenberg},
  {Kontizas}, {Kontizas}, {Koposov}, {Kordopatis}, {Kostrzewa-Rutkowska},
  {Koubsky}, {Lambert}, {Lanza}, {Lasne}, {Lavigne}, {Le Fustec}, {Le
  Poncin-Lafitte}, {Lebreton}, {Leccia}, {Leclerc}, {Lecoeur-Taibi},
  {Lenhardt}, {Leroux}, {Liao}, {Licata}, {Lindstr{\o}m}, {Lister}, {Livanou},
  {Lobel}, {L{\'o}pez}, {Managau}, {Mann}, {Mantelet}, {Marchal}, {Marchant},
  {Marconi}, {Marinoni}, {Marschalk{\'o}}, {Marshall}, {Martino}, {Marton},
  {Mary}, {Massari}, {Matijevi{\v{c}}}, {Mazeh}, {McMillan}, {Messina},
  {Michalik}, {Millar}, {Molina}, {Molinaro}, {Moln{\'a}r}, {Montegriffo},
  {Mor}, {Morbidelli}, {Morel}, {Morris}, {Mulone}, {Muraveva}, {Musella},
  {Nelemans}, {Nicastro}, {Noval}, {O'Mullane}, {Ord{\'e}novic},
  {Ord{\'o}{\~n}ez-Blanco}, {Osborne}, {Pagani}, {Pagano}, {Pailler},
  {Palacin}, {Palaversa}, {Panahi}, {Pawlak}, {Piersimoni}, {Pineau}, {Plachy},
  {Plum}, {Poggio}, {Poujoulet}, {Pr{\v{s}}a}, {Pulone}, {Racero}, {Ragaini},
  {Rambaux}, {Ramos-Lerate}, {Regibo}, {Reyl{\'e}}, {Riclet}, {Ripepi}, {Riva},
  {Rivard}, {Rixon}, {Roegiers}, {Roelens}, {Romero-G{\'o}mez}, {Rowell},
  {Royer}, {Ruiz-Dern}, {Sadowski}, {Sagrist{\`a} Sell{\'e}s}, {Sahlmann},
  {Salgado}, {Salguero}, {Sanna}, {Santana-Ros}, {Sarasso}, {Savietto},
  {Schultheis}, {Sciacca}, {Segol}, {Segovia}, {S{\'e}gransan}, {Shih},
  {Siltala}, {Silva}, {Smart}, {Smith}, {Solano}, {Solitro}, {Sordo}, {Soria
  Nieto}, {Souchay}, {Spagna}, {Spoto}, {Stampa}, {Steele},
  {Steidelm{\"u}ller}, {Stephenson}, {Stoev}, {Suess}, {Surdej}, {Szabados},
  {Szegedi-Elek}, {Tapiador}, {Taris}, {Tauran}, {Taylor}, {Teixeira},
  {Terrett}, {Teyssandier}, {Thuillot}, {Titarenko}, {Torra Clotet}, {Turon},
  {Ulla}, {Utrilla}, {Uzzi}, {Vaillant}, {Valentini}, {Valette}, {van Elteren},
  {Van Hemelryck}, {van Leeuwen}, {Vaschetto}, {Vecchiato}, {Veljanoski},
  {Viala}, {Vicente}, {Vogt}, {von Essen}, {Voss}, {Votruba}, {Voutsinas},
  {Walmsley}, {Weiler}, {Wertz}, {Wevers}, {Wyrzykowski}, {Yoldas},
  {{\v{Z}}erjal}, {Ziaeepour}, {Zorec}, {Zschocke}, {Zucker}, {Zurbach}, \&
  {Zwitter}}]{gaiadr2}
{Gaia Collaboration}, {Brown}, A.~G.~A., {Vallenari}, A., {et~al.} 2018, \aap,
  616, A1

\bibitem[{{Gaia Collaboration} {et~al.}(2021){Gaia Collaboration}, {Brown},
  {Vallenari}, {Prusti}, {de Bruijne}, {Babusiaux}, {Biermann}, {Creevey},
  {Evans}, {Eyer}, {Hutton}, {Jansen}, {Jordi}, {Klioner}, {Lammers},
  {Lindegren}, {Luri}, {Mignard}, {Panem}, {Pourbaix}, {Randich}, {Sartoretti},
  {Soubiran}, {Walton}, {Arenou}, {Bailer-Jones}, {Bastian}, {Cropper},
  {Drimmel}, {Katz}, {Lattanzi}, {van Leeuwen}, {Bakker}, {Cacciari},
  {Casta{\~n}eda}, {De Angeli}, {Ducourant}, {Fabricius}, {Fouesneau},
  {Fr{\'e}mat}, {Guerra}, {Guerrier}, {Guiraud}, {Jean-Antoine Piccolo},
  {Masana}, {Messineo}, {Mowlavi}, {Nicolas}, {Nienartowicz}, {Pailler},
  {Panuzzo}, {Riclet}, {Roux}, {Seabroke}, {Sordo}, {Tanga}, {Th{\'e}venin},
  {Gracia-Abril}, {Portell}, {Teyssier}, {Altmann}, {Andrae}, {Bellas-Velidis},
  {Benson}, {Berthier}, {Blomme}, {Brugaletta}, {Burgess}, {Busso}, {Carry},
  {Cellino}, {Cheek}, {Clementini}, {Damerdji}, {Davidson}, {Delchambre},
  {Dell'Oro}, {Fern{\'a}ndez-Hern{\'a}ndez}, {Galluccio}, {Garc{\'\i}a-Lario},
  {Garcia-Reinaldos}, {Gonz{\'a}lez-N{\'u}{\~n}ez}, {Gosset}, {Haigron},
  {Halbwachs}, {Hambly}, {Harrison}, {Hatzidimitriou}, {Heiter},
  {Hern{\'a}ndez}, {Hestroffer}, {Hodgkin}, {Holl}, {Jan{\ss}en}, {Jevardat de
  Fombelle}, {Jordan}, {Krone-Martins}, {Lanzafame}, {L{\"o}ffler}, {Lorca},
  {Manteiga}, {Marchal}, {Marrese}, {Moitinho}, {Mora}, {Muinonen}, {Osborne},
  {Pancino}, {Pauwels}, {Petit}, {Recio-Blanco}, {Richards}, {Riello},
  {Rimoldini}, {Robin}, {Roegiers}, {Rybizki}, {Sarro}, {Siopis}, {Smith},
  {Sozzetti}, {Ulla}, {Utrilla}, {van Leeuwen}, {van Reeven}, {Abbas}, {Abreu
  Aramburu}, {Accart}, {Aerts}, {Aguado}, {Ajaj}, {Altavilla}, {{\'A}lvarez},
  {{\'A}lvarez Cid-Fuentes}, {Alves}, {Anderson}, {Anglada Varela}, {Antoja},
  {Audard}, {Baines}, {Baker}, {Balaguer-N{\'u}{\~n}ez}, {Balbinot}, {Balog},
  {Barache}, {Barbato}, {Barros}, {Barstow}, {Bartolom{\'e}}, {Bassilana},
  {Bauchet}, {Baudesson-Stella}, {Becciani}, {Bellazzini}, {Bernet}, {Bertone},
  {Bianchi}, {Blanco-Cuaresma}, {Boch}, {Bombrun}, {Bossini}, {Bouquillon},
  {Bragaglia}, {Bramante}, {Breedt}, {Bressan}, {Brouillet}, {Bucciarelli},
  {Burlacu}, {Busonero}, {Butkevich}, {Buzzi}, {Caffau}, {Cancelliere},
  {C{\'a}novas}, {Cantat-Gaudin}, {Carballo}, {Carlucci}, {Carnerero},
  {Carrasco}, {Casamiquela}, {Castellani}, {Castro-Ginard}, {Castro Sampol},
  {Chaoul}, {Charlot}, {Chemin}, {Chiavassa}, {Cioni}, {Comoretto}, {Cooper},
  {Cornez}, {Cowell}, {Crifo}, {Crosta}, {Crowley}, {Dafonte}, {Dapergolas},
  {David}, {David}, {de Laverny}, {De Luise}, {De March}, {De Ridder}, {de
  Souza}, {de Teodoro}, {de Torres}, {del Peloso}, {del Pozo}, {Delbo},
  {Delgado}, {Delgado}, {Delisle}, {Di Matteo}, {Diakite}, {Diener},
  {Distefano}, {Dolding}, {Eappachen}, {Edvardsson}, {Enke}, {Esquej}, {Fabre},
  {Fabrizio}, {Faigler}, {Fedorets}, {Fernique}, {Fienga}, {Figueras},
  {Fouron}, {Fragkoudi}, {Fraile}, {Franke}, {Gai}, {Garabato},
  {Garcia-Gutierrez}, {Garc{\'\i}a-Torres}, {Garofalo}, {Gavras}, {Gerlach},
  {Geyer}, {Giacobbe}, {Gilmore}, {Girona}, {Giuffrida}, {Gomel}, {Gomez},
  {Gonzalez-Santamaria}, {Gonz{\'a}lez-Vidal}, {Granvik},
  {Guti{\'e}rrez-S{\'a}nchez}, {Guy}, {Hauser}, {Haywood}, {Helmi}, {Hidalgo},
  {Hilger}, {H{\l}adczuk}, {Hobbs}, {Holland}, {Huckle}, {Jasniewicz},
  {Jonker}, {Juaristi Campillo}, {Julbe}, {Karbevska}, {Kervella}, {Khanna},
  {Kochoska}, {Kontizas}, {Kordopatis}, {Korn}, {Kostrzewa-Rutkowska},
  {Kruszy{\'n}ska}, {Lambert}, {Lanza}, {Lasne}, {Le Campion}, {Le Fustec},
  {Lebreton}, {Lebzelter}, {Leccia}, {Leclerc}, {Lecoeur-Taibi}, {Liao},
  {Licata}, {Lindstr{\o}m}, {Lister}, {Livanou}, {Lobel}, {Madrero Pardo},
  {Managau}, {Mann}, {Marchant}, {Marconi}, {Marcos Santos}, {Marinoni},
  {Marocco}, {Marshall}, {Martin Polo}, {Mart{\'\i}n-Fleitas}, {Masip},
  {Massari}, {Mastrobuono-Battisti}, {Mazeh}, {McMillan}, {Messina},
  {Michalik}, {Millar}, {Mints}, {Molina}, {Molinaro}, {Moln{\'a}r},
  {Montegriffo}, {Mor}, {Morbidelli}, {Morel}, {Morris}, {Mulone}, {Munoz},
  {Muraveva}, {Murphy}, {Musella}, {Noval}, {Ord{\'e}novic}, {Orr{\`u}},
  {Osinde}, {Pagani}, {Pagano}, {Palaversa}, {Palicio}, {Panahi}, {Pawlak},
  {Pe{\~n}alosa Esteller}, {Penttil{\"a}}, {Piersimoni}, {Pineau}, {Plachy},
  {Plum}, {Poggio}, {Poretti}, {Poujoulet}, {Pr{\v{s}}a}, {Pulone}, {Racero},
  {Ragaini}, {Rainer}, {Raiteri}, {Rambaux}, {Ramos}, {Ramos-Lerate}, {Re
  Fiorentin}, {Regibo}, {Reyl{\'e}}, {Ripepi}, {Riva}, {Rixon}, {Robichon},
  {Robin}, {Roelens}, {Rohrbasser}, {Romero-G{\'o}mez}, {Rowell}, {Royer},
  {Rybicki}, {Sadowski}, {Sagrist{\`a} Sell{\'e}s}, {Sahlmann}, {Salgado},
  {Salguero}, {Samaras}, {Sanchez Gimenez}, {Sanna}, {Santove{\~n}a},
  {Sarasso}, {Schultheis}, {Sciacca}, {Segol}, {Segovia}, {S{\'e}gransan},
  {Semeux}, {Shahaf}, {Siddiqui}, {Siebert}, {Siltala}, {Slezak}, {Smart},
  {Solano}, {Solitro}, {Souami}, {Souchay}, {Spagna}, {Spoto}, {Steele},
  {Steidelm{\"u}ller}, {Stephenson}, {S{\"u}veges}, {Szabados}, {Szegedi-Elek},
  {Taris}, {Tauran}, {Taylor}, {Teixeira}, {Thuillot}, {Tonello}, {Torra},
  {Torra}, {Turon}, {Unger}, {Vaillant}, {van Dillen}, {Vanel}, {Vecchiato},
  {Viala}, {Vicente}, {Voutsinas}, {Weiler}, {Wevers}, {Wyrzykowski}, {Yoldas},
  {Yvard}, {Zhao}, {Zorec}, {Zucker}, {Zurbach}, \& {Zwitter}}]{gaiaedr3}
{Gaia Collaboration}, {Brown}, A.~G.~A., {Vallenari}, A., {et~al.} 2021, \aap,
  649, A1

\bibitem[{{Gaia Collaboration} {et~al.}(2016){Gaia Collaboration}, {Brown},
  {Vallenari}, {Prusti}, {de Bruijne}, {Mignard}, {Drimmel}, {Babusiaux},
  {Bailer-Jones}, {Bastian}, {Biermann}, {Evans}, {Eyer}, {Jansen}, {Jordi},
  {Katz}, {Klioner}, {Lammers}, {Lindegren}, {Luri}, {O'Mullane}, {Panem},
  {Pourbaix}, {Randich}, {Sartoretti}, {Siddiqui}, {Soubiran}, {Valette}, {van
  Leeuwen}, {Walton}, {Aerts}, {Arenou}, {Cropper}, {H{\o}g}, {Lattanzi},
  {Grebel}, {Holland}, {Huc}, {Passot}, {Perryman}, {Bramante}, {Cacciari},
  {Casta{\~n}eda}, {Chaoul}, {Cheek}, {De Angeli}, {Fabricius}, {Guerra},
  {Hern{\'a}ndez}, {Jean-Antoine-Piccolo}, {Masana}, {Messineo}, {Mowlavi},
  {Nienartowicz}, {Ord{\'o}{\~n}ez-Blanco}, {Panuzzo}, {Portell}, {Richards},
  {Riello}, {Seabroke}, {Tanga}, {Th{\'e}venin}, {Torra}, {Els},
  {Gracia-Abril}, {Comoretto}, {Garcia-Reinaldos}, {Lock}, {Mercier},
  {Altmann}, {Andrae}, {Astraatmadja}, {Bellas-Velidis}, {Benson}, {Berthier},
  {Blomme}, {Busso}, {Carry}, {Cellino}, {Clementini}, {Cowell}, {Creevey},
  {Cuypers}, {Davidson}, {De Ridder}, {de Torres}, {Delchambre}, {Dell'Oro},
  {Ducourant}, {Fr{\'e}mat}, {Garc{\'\i}a-Torres}, {Gosset}, {Halbwachs},
  {Hambly}, {Harrison}, {Hauser}, {Hestroffer}, {Hodgkin}, {Huckle}, {Hutton},
  {Jasniewicz}, {Jordan}, {Kontizas}, {Korn}, {Lanzafame}, {Manteiga},
  {Moitinho}, {Muinonen}, {Osinde}, {Pancino}, {Pauwels}, {Petit},
  {Recio-Blanco}, {Robin}, {Sarro}, {Siopis}, {Smith}, {Smith}, {Sozzetti},
  {Thuillot}, {van Reeven}, {Viala}, {Abbas}, {Abreu Aramburu}, {Accart},
  {Aguado}, {Allan}, {Allasia}, {Altavilla}, {{\'A}lvarez}, {Alves},
  {Anderson}, {Andrei}, {Anglada Varela}, {Antiche}, {Antoja}, {Ant{\'o}n},
  {Arcay}, {Bach}, {Baker}, {Balaguer-N{\'u}{\~n}ez}, {Barache}, {Barata},
  {Barbier}, {Barblan}, {Barrado y Navascu{\'e}s}, {Barros}, {Barstow},
  {Becciani}, {Bellazzini}, {Bello Garc{\'\i}a}, {Belokurov}, {Bendjoya},
  {Berihuete}, {Bianchi}, {Bienaym{\'e}}, {Billebaud}, {Blagorodnova},
  {Blanco-Cuaresma}, {Boch}, {Bombrun}, {Borrachero}, {Bouquillon}, {Bourda},
  {Bouy}, {Bragaglia}, {Breddels}, {Brouillet}, {Br{\"u}semeister},
  {Bucciarelli}, {Burgess}, {Burgon}, {Burlacu}, {Busonero}, {Buzzi}, {Caffau},
  {Cambras}, {Campbell}, {Cancelliere}, {Cantat-Gaudin}, {Carlucci},
  {Carrasco}, {Castellani}, {Charlot}, {Charnas}, {Chiavassa}, {Clotet},
  {Cocozza}, {Collins}, {Costigan}, {Crifo}, {Cross}, {Crosta}, {Crowley},
  {Dafonte}, {Damerdji}, {Dapergolas}, {David}, {David}, {De Cat}, {de Felice},
  {de Laverny}, {De Luise}, {De March}, {de Martino}, {de Souza}, {Debosscher},
  {del Pozo}, {Delbo}, {Delgado}, {Delgado}, {Di Matteo}, {Diakite},
  {Distefano}, {Dolding}, {Dos Anjos}, {Drazinos}, {Duran}, {Dzigan},
  {Edvardsson}, {Enke}, {Evans}, {Eynard Bontemps}, {Fabre}, {Fabrizio},
  {Faigler}, {Falc{\~a}o}, {Farr{\`a}s Casas}, {Federici}, {Fedorets},
  {Fern{\'a}ndez-Hern{\'a}ndez}, {Fernique}, {Fienga}, {Figueras}, {Filippi},
  {Findeisen}, {Fonti}, {Fouesneau}, {Fraile}, {Fraser}, {Fuchs}, {Gai},
  {Galleti}, {Galluccio}, {Garabato}, {Garc{\'\i}a-Sedano}, {Garofalo},
  {Garralda}, {Gavras}, {Gerssen}, {Geyer}, {Gilmore}, {Girona}, {Giuffrida},
  {Gomes}, {Gonz{\'a}lez-Marcos}, {Gonz{\'a}lez-N{\'u}{\~n}ez},
  {Gonz{\'a}lez-Vidal}, {Granvik}, {Guerrier}, {Guillout}, {Guiraud},
  {G{\'u}rpide}, {Guti{\'e}rrez-S{\'a}nchez}, {Guy}, {Haigron},
  {Hatzidimitriou}, {Haywood}, {Heiter}, {Helmi}, {Hobbs}, {Hofmann}, {Holl},
  {Holland}, {Hunt}, {Hypki}, {Icardi}, {Irwin}, {Jevardat de Fombelle},
  {Jofr{\'e}}, {Jonker}, {Jorissen}, {Julbe}, {Karampelas}, {Kochoska},
  {Kohley}, {Kolenberg}, {Kontizas}, {Koposov}, {Kordopatis}, {Koubsky},
  {Krone-Martins}, {Kudryashova}, {Kull}, {Bachchan}, {Lacoste-Seris}, {Lanza},
  {Lavigne}, {Le Poncin-Lafitte}, {Lebreton}, {Lebzelter}, {Leccia}, {Leclerc},
  {Lecoeur-Taibi}, {Lemaitre}, {Lenhardt}, {Leroux}, {Liao}, {Licata},
  {Lindstr{\o}m}, {Lister}, {Livanou}, {Lobel}, {L{\"o}ffler}, {L{\'o}pez},
  {Lorenz}, {MacDonald}, {Magalh{\~a}es Fernandes}, {Managau}, {Mann},
  {Mantelet}, {Marchal}, {Marchant}, {Marconi}, {Marinoni}, {Marrese},
  {Marschalk{\'o}}, {Marshall}, {Mart{\'\i}n-Fleitas}, {Martino}, {Mary},
  {Matijevi{\v{c}}}, {Mazeh}, {McMillan}, {Messina}, {Michalik}, {Millar},
  {Miranda}, {Molina}, {Molinaro}, {Molinaro}, {Moln{\'a}r}, {Moniez},
  {Montegriffo}, {Mor}, {Mora}, {Morbidelli}, {Morel}, {Morgenthaler},
  {Morris}, {Mulone}, {Muraveva}, {Musella}, {Narbonne}, {Nelemans},
  {Nicastro}, {Noval}, {Ord{\'e}novic}, {Ordieres-Mer{\'e}}, {Osborne},
  {Pagani}, {Pagano}, {Pailler}, {Palacin}, {Palaversa}, {Parsons}, {Pecoraro},
  {Pedrosa}, {Pentik{\"a}inen}, {Pichon}, {Piersimoni}, {Pineau}, {Plachy},
  {Plum}, {Poujoulet}, {Pr{\v{s}}a}, {Pulone}, {Ragaini}, {Rago}, {Rambaux},
  {Ramos-Lerate}, {Ranalli}, {Rauw}, {Read}, {Regibo}, {Reyl{\'e}}, {Ribeiro},
  {Rimoldini}, {Ripepi}, {Riva}, {Rixon}, {Roelens}, {Romero-G{\'o}mez},
  {Rowell}, {Royer}, {Ruiz-Dern}, {Sadowski}, {Sagrist{\`a} Sell{\'e}s},
  {Sahlmann}, {Salgado}, {Salguero}, {Sarasso}, {Savietto}, {Schultheis},
  {Sciacca}, {Segol}, {Segovia}, {Segransan}, {Shih}, {Smareglia}, {Smart},
  {Solano}, {Solitro}, {Sordo}, {Soria Nieto}, {Souchay}, {Spagna}, {Spoto},
  {Stampa}, {Steele}, {Steidelm{\"u}ller}, {Stephenson}, {Stoev}, {Suess},
  {S{\"u}veges}, {Surdej}, {Szabados}, {Szegedi-Elek}, {Tapiador}, {Taris},
  {Tauran}, {Taylor}, {Teixeira}, {Terrett}, {Tingley}, {Trager}, {Turon},
  {Ulla}, {Utrilla}, {Valentini}, {van Elteren}, {Van Hemelryck}, {van
  Leeuwen}, {Varadi}, {Vecchiato}, {Veljanoski}, {Via}, {Vicente}, {Vogt},
  {Voss}, {Votruba}, {Voutsinas}, {Walmsley}, {Weiler}, {Weingrill}, {Wevers},
  {Wyrzykowski}, {Yoldas}, {{\v{Z}}erjal}, {Zucker}, {Zurbach}, {Zwitter},
  {Alecu}, {Allen}, {Allende Prieto}, {Amorim}, {Anglada-Escud{\'e}},
  {Arsenijevic}, {Azaz}, {Balm}, {Beck}, {Bernstein}, {Bigot}, {Bijaoui},
  {Blasco}, {Bonfigli}, {Bono}, {Boudreault}, {Bressan}, {Brown}, {Brunet},
  {Bunclark}, {Buonanno}, {Butkevich}, {Carret}, {Carrion}, {Chemin},
  {Ch{\'e}reau}, {Corcione}, {Darmigny}, {de Boer}, {de Teodoro}, {de Zeeuw},
  {Delle Luche}, {Domingues}, {Dubath}, {Fodor}, {Fr{\'e}zouls}, {Fries},
  {Fustes}, {Fyfe}, {Gallardo}, {Gallegos}, {Gardiol}, {Gebran}, {Gomboc},
  {G{\'o}mez}, {Grux}, {Gueguen}, {Heyrovsky}, {Hoar}, {Iannicola}, {Isasi
  Parache}, {Janotto}, {Joliet}, {Jonckheere}, {Keil}, {Kim}, {Klagyivik},
  {Klar}, {Knude}, {Kochukhov}, {Kolka}, {Kos}, {Kutka}, {Lainey}, {LeBouquin},
  {Liu}, {Loreggia}, {Makarov}, {Marseille}, {Martayan}, {Martinez-Rubi},
  {Massart}, {Meynadier}, {Mignot}, {Munari}, {Nguyen}, {Nordlander}, {Ocvirk},
  {O'Flaherty}, {Olias Sanz}, {Ortiz}, {Osorio}, {Oszkiewicz}, {Ouzounis},
  {Palmer}, {Park}, {Pasquato}, {Peltzer}, {Peralta}, {P{\'e}turaud},
  {Pieniluoma}, {Pigozzi}, {Poels}, {Prat}, {Prod'homme}, {Raison}, {Rebordao},
  {Risquez}, {Rocca-Volmerange}, {Rosen}, {Ruiz-Fuertes}, {Russo}, {Sembay},
  {Serraller Vizcaino}, {Short}, {Siebert}, {Silva}, {Sinachopoulos}, {Slezak},
  {Soffel}, {Sosnowska}, {Strai{\v{z}}ys}, {ter Linden}, {Terrell}, {Theil},
  {Tiede}, {Troisi}, {Tsalmantza}, {Tur}, {Vaccari}, {Vachier}, {Valles}, {Van
  Hamme}, {Veltz}, {Virtanen}, {Wallut}, {Wichmann}, {Wilkinson}, {Ziaeepour},
  \& {Zschocke}}]{gaiadr1}
{Gaia Collaboration}, {Brown}, A.~G.~A., {Vallenari}, A., {et~al.} 2016, \aap,
  595, A2

\bibitem[{{Gaia Collaboration} {et~al.}(2023){Gaia Collaboration}, {Vallenari},
  {Brown}, {Prusti}, {de Bruijne}, {Arenou}, {Babusiaux}, {Biermann},
  {Creevey}, {Ducourant}, {Evans}, {Eyer}, {Guerra}, {Hutton}, {Jordi},
  {Klioner}, {Lammers}, {Lindegren}, {Luri}, {Mignard}, {Panem}, {Pourbaix},
  {Randich}, {Sartoretti}, {Soubiran}, {Tanga}, {Walton}, {Bailer-Jones},
  {Bastian}, {Drimmel}, {Jansen}, {Katz}, {Lattanzi}, {van Leeuwen}, {Bakker},
  {Cacciari}, {Casta{\~n}eda}, {De Angeli}, {Fabricius}, {Fouesneau},
  {Fr{\'e}mat}, {Galluccio}, {Guerrier}, {Heiter}, {Masana}, {Messineo},
  {Mowlavi}, {Nicolas}, {Nienartowicz}, {Pailler}, {Panuzzo}, {Riclet}, {Roux},
  {Seabroke}, {Sordo}, {Th{\'e}venin}, {Gracia-Abril}, {Portell}, {Teyssier},
  {Altmann}, {Andrae}, {Audard}, {Bellas-Velidis}, {Benson}, {Berthier},
  {Blomme}, {Burgess}, {Busonero}, {Busso}, {C{\'a}novas}, {Carry}, {Cellino},
  {Cheek}, {Clementini}, {Damerdji}, {Davidson}, {de Teodoro}, {Nu{\~n}ez
  Campos}, {Delchambre}, {Dell'Oro}, {Esquej}, {Fern{\'a}ndez-Hern{\'a}ndez},
  {Fraile}, {Garabato}, {Garc{\'\i}a-Lario}, {Gosset}, {Haigron}, {Halbwachs},
  {Hambly}, {Harrison}, {Hern{\'a}ndez}, {Hestroffer}, {Hodgkin}, {Holl},
  {Jan{\ss}en}, {Jevardat de Fombelle}, {Jordan}, {Krone-Martins}, {Lanzafame},
  {L{\"o}ffler}, {Marchal}, {Marrese}, {Moitinho}, {Muinonen}, {Osborne},
  {Pancino}, {Pauwels}, {Recio-Blanco}, {Reyl{\'e}}, {Riello}, {Rimoldini},
  {Roegiers}, {Rybizki}, {Sarro}, {Siopis}, {Smith}, {Sozzetti}, {Utrilla},
  {van Leeuwen}, {Abbas}, {{\'A}brah{\'a}m}, {Abreu Aramburu}, {Aerts},
  {Aguado}, {Ajaj}, {Aldea-Montero}, {Altavilla}, {{\'A}lvarez}, {Alves},
  {Anders}, {Anderson}, {Anglada Varela}, {Antoja}, {Baines}, {Baker},
  {Balaguer-N{\'u}{\~n}ez}, {Balbinot}, {Balog}, {Barache}, {Barbato},
  {Barros}, {Barstow}, {Bartolom{\'e}}, {Bassilana}, {Bauchet}, {Becciani},
  {Bellazzini}, {Berihuete}, {Bernet}, {Bertone}, {Bianchi}, {Binnenfeld},
  {Blanco-Cuaresma}, {Blazere}, {Boch}, {Bombrun}, {Bossini}, {Bouquillon},
  {Bragaglia}, {Bramante}, {Breedt}, {Bressan}, {Brouillet}, {Brugaletta},
  {Bucciarelli}, {Burlacu}, {Butkevich}, {Buzzi}, {Caffau}, {Cancelliere},
  {Cantat-Gaudin}, {Carballo}, {Carlucci}, {Carnerero}, {Carrasco},
  {Casamiquela}, {Castellani}, {Castro-Ginard}, {Chaoul}, {Charlot}, {Chemin},
  {Chiaramida}, {Chiavassa}, {Chornay}, {Comoretto}, {Contursi}, {Cooper},
  {Cornez}, {Cowell}, {Crifo}, {Cropper}, {Crosta}, {Crowley}, {Dafonte},
  {Dapergolas}, {David}, {David}, {de Laverny}, {De Luise}, {De March}, {De
  Ridder}, {de Souza}, {de Torres}, {del Peloso}, {del Pozo}, {Delbo},
  {Delgado}, {Delisle}, {Demouchy}, {Dharmawardena}, {Di Matteo}, {Diakite},
  {Diener}, {Distefano}, {Dolding}, {Edvardsson}, {Enke}, {Fabre}, {Fabrizio},
  {Faigler}, {Fedorets}, {Fernique}, {Fienga}, {Figueras}, {Fournier},
  {Fouron}, {Fragkoudi}, {Gai}, {Garcia-Gutierrez}, {Garcia-Reinaldos},
  {Garc{\'\i}a-Torres}, {Garofalo}, {Gavel}, {Gavras}, {Gerlach}, {Geyer},
  {Giacobbe}, {Gilmore}, {Girona}, {Giuffrida}, {Gomel}, {Gomez},
  {Gonz{\'a}lez-N{\'u}{\~n}ez}, {Gonz{\'a}lez-Santamar{\'\i}a},
  {Gonz{\'a}lez-Vidal}, {Granvik}, {Guillout}, {Guiraud},
  {Guti{\'e}rrez-S{\'a}nchez}, {Guy}, {Hatzidimitriou}, {Hauser}, {Haywood},
  {Helmer}, {Helmi}, {Sarmiento}, {Hidalgo}, {Hilger}, {H{\l}adczuk}, {Hobbs},
  {Holland}, {Huckle}, {Jardine}, {Jasniewicz}, {Jean-Antoine Piccolo},
  {Jim{\'e}nez-Arranz}, {Jorissen}, {Juaristi Campillo}, {Julbe}, {Karbevska},
  {Kervella}, {Khanna}, {Kontizas}, {Kordopatis}, {Korn}, {K{\'o}sp{\'a}l},
  {Kostrzewa-Rutkowska}, {Kruszy{\'n}ska}, {Kun}, {Laizeau}, {Lambert},
  {Lanza}, {Lasne}, {Le Campion}, {Lebreton}, {Lebzelter}, {Leccia}, {Leclerc},
  {Lecoeur-Taibi}, {Liao}, {Licata}, {Lindstr{\o}m}, {Lister}, {Livanou},
  {Lobel}, {Lorca}, {Loup}, {Madrero Pardo}, {Magdaleno Romeo}, {Managau},
  {Mann}, {Manteiga}, {Marchant}, {Marconi}, {Marcos}, {Marcos Santos},
  {Mar{\'\i}n Pina}, {Marinoni}, {Marocco}, {Marshall}, {Martin Polo},
  {Mart{\'\i}n-Fleitas}, {Marton}, {Mary}, {Masip}, {Massari},
  {Mastrobuono-Battisti}, {Mazeh}, {McMillan}, {Messina}, {Michalik}, {Millar},
  {Mints}, {Molina}, {Molinaro}, {Moln{\'a}r}, {Monari}, {Mongui{\'o}},
  {Montegriffo}, {Montero}, {Mor}, {Mora}, {Morbidelli}, {Morel}, {Morris},
  {Muraveva}, {Murphy}, {Musella}, {Nagy}, {Noval}, {Oca{\~n}a}, {Ogden},
  {Ordenovic}, {Osinde}, {Pagani}, {Pagano}, {Palaversa}, {Palicio},
  {Pallas-Quintela}, {Panahi}, {Payne-Wardenaar}, {Pe{\~n}alosa Esteller},
  {Penttil{\"a}}, {Pichon}, {Piersimoni}, {Pineau}, {Plachy}, {Plum}, {Poggio},
  {Pr{\v{s}}a}, {Pulone}, {Racero}, {Ragaini}, {Rainer}, {Raiteri}, {Rambaux},
  {Ramos}, {Ramos-Lerate}, {Re Fiorentin}, {Regibo}, {Richards}, {Rios Diaz},
  {Ripepi}, {Riva}, {Rix}, {Rixon}, {Robichon}, {Robin}, {Robin}, {Roelens},
  {Rogues}, {Rohrbasser}, {Romero-G{\'o}mez}, {Rowell}, {Royer}, {Ruz Mieres},
  {Rybicki}, {Sadowski}, {S{\'a}ez N{\'u}{\~n}ez}, {Sagrist{\`a} Sell{\'e}s},
  {Sahlmann}, {Salguero}, {Samaras}, {Sanchez Gimenez}, {Sanna},
  {Santove{\~n}a}, {Sarasso}, {Schultheis}, {Sciacca}, {Segol}, {Segovia},
  {S{\'e}gransan}, {Semeux}, {Shahaf}, {Siddiqui}, {Siebert}, {Siltala},
  {Silvelo}, {Slezak}, {Slezak}, {Smart}, {Snaith}, {Solano}, {Solitro},
  {Souami}, {Souchay}, {Spagna}, {Spina}, {Spoto}, {Steele},
  {Steidelm{\"u}ller}, {Stephenson}, {S{\"u}veges}, {Surdej}, {Szabados},
  {Szegedi-Elek}, {Taris}, {Taylor}, {Teixeira}, {Tolomei}, {Tonello}, {Torra},
  {Torra}, {Torralba Elipe}, {Trabucchi}, {Tsounis}, {Turon}, {Ulla}, {Unger},
  {Vaillant}, {van Dillen}, {van Reeven}, {Vanel}, {Vecchiato}, {Viala},
  {Vicente}, {Voutsinas}, {Weiler}, {Wevers}, {Wyrzykowski}, {Yoldas}, {Yvard},
  {Zhao}, {Zorec}, {Zucker}, \& {Zwitter}}]{gaiadr3}
{Gaia Collaboration}, {Vallenari}, A., {Brown}, A.~G.~A., {et~al.} 2023, \aap,
  674, A1

\bibitem[{{Garc{\'\i}a-Berro} \& {Oswalt}(2016)}]{garcia-berro2016}
{Garc{\'\i}a-Berro}, E. \& {Oswalt}, T.~D. 2016, \nar, 72, 1

\bibitem[{{Gentile Fusillo} {et~al.}(2021){Gentile Fusillo}, {Tremblay},
  {Cukanovaite}, {Vorontseva}, {Lallement}, {Hollands}, {G{\"a}nsicke},
  {Burdge}, {McCleery}, \& {Jordan}}]{gentilefusillo2021}
{Gentile Fusillo}, N.~P., {Tremblay}, P.~E., {Cukanovaite}, E., {et~al.} 2021,
  \mnras, 508, 3877

\bibitem[{{Gentile Fusillo} {et~al.}(2019){Gentile Fusillo}, {Tremblay},
  {G{\"a}nsicke}, {Manser}, {Cunningham}, {Cukanovaite}, {Hollands}, {Marsh},
  {Raddi}, {Jordan}, {Toonen}, {Geier}, {Barstow}, \&
  {Cummings}}]{gentilefusillo2019}
{Gentile Fusillo}, N.~P., {Tremblay}, P.-E., {G{\"a}nsicke}, B.~T., {et~al.}
  2019, \mnras, 482, 4570

\bibitem[{{Gentile Fusillo} {et~al.}(2018){Gentile Fusillo}, {Tremblay},
  {Jordan}, {G{\"a}nsicke}, {Kalirai}, \& {Cummings}}]{gentilefusillo2018}
{Gentile Fusillo}, N.~P., {Tremblay}, P.~E., {Jordan}, S., {et~al.} 2018,
  \mnras, 473, 3693

\bibitem[{{Gianninas} {et~al.}(2011){Gianninas}, {Bergeron}, \&
  {Ruiz}}]{gianninas20122}
{Gianninas}, A., {Bergeron}, P., \& {Ruiz}, M.~T. 2011, \apj, 743, 138

\bibitem[{{Grabowski} {et~al.}(1987){Grabowski}, {Madej}, \&
  {Halenka}}]{grabowski1987}
{Grabowski}, B., {Madej}, J., \& {Halenka}, J. 1987, \apj, 313, 750

\bibitem[{{Greenstein} \& {Trimble}(1972)}]{greenstein1972}
{Greenstein}, J.~L. \& {Trimble}, V. 1972, \apjl, 175, L1

\bibitem[{{Greenstein} \& {Trimble}(1967)}]{greenstein1967}
{Greenstein}, J.~L. \& {Trimble}, V.~L. 1967, \apj, 149, 283

\bibitem[{{Halenka} {et~al.}(2015){Halenka}, {Olchawa}, {Madej}, \&
  {Grabowski}}]{halenka2015}
{Halenka}, J., {Olchawa}, W., {Madej}, J., \& {Grabowski}, B. 2015, \apj, 808,
  131

\bibitem[{Harris {et~al.}(2020)Harris, Millman, van~der Walt, Gommers,
  Virtanen, Cournapeau, Wieser, Taylor, Berg, Smith, Kern, Picus, Hoyer, van
  Kerkwijk, Brett, Haldane, del R{\'{i}}o, Wiebe, Peterson,
  G{\'{e}}rard-Marchant, Sheppard, Reddy, Weckesser, Abbasi, Gohlke, \&
  Oliphant}]{numpy}
Harris, C.~R., Millman, K.~J., van~der Walt, S.~J., {et~al.} 2020, Nature, 585,
  357

\bibitem[{{Heber} {et~al.}(1997){Heber}, {Napiwotzki}, \& {Reid}}]{heber1997}
{Heber}, U., {Napiwotzki}, R., \& {Reid}, I.~N. 1997, \aap, 323, 819

\bibitem[{{Hidalgo} {et~al.}(2018){Hidalgo}, {Pietrinferni}, {Cassisi},
  {Salaris}, {Mucciarelli}, {Savino}, {Aparicio}, {Silva Aguirre}, \&
  {Verma}}]{hidalgo2018}
{Hidalgo}, S.~L., {Pietrinferni}, A., {Cassisi}, S., {et~al.} 2018, \apj, 856,
  125

\bibitem[{{Holberg} {et~al.}(2012){Holberg}, {Oswalt}, \&
  {Barstow}}]{holberg2012}
{Holberg}, J.~B., {Oswalt}, T.~D., \& {Barstow}, M.~A. 2012, \aj, 143, 68

\bibitem[{Hunter(2007)}]{matplotlib}
Hunter, J.~D. 2007, Computing in Science \& Engineering, 9, 90

\bibitem[{{Hurley} {et~al.}(2000){Hurley}, {Pols}, \& {Tout}}]{hurley2000}
{Hurley}, J.~R., {Pols}, O.~R., \& {Tout}, C.~A. 2000, \mnras, 315, 543

\bibitem[{{Isern} {et~al.}(1991){Isern}, {Canal}, \& {Labay}}]{isern1991}
{Isern}, J., {Canal}, R., \& {Labay}, J. 1991, \apjl, 372, L83

\bibitem[{{Jim{\'e}nez-Esteban} {et~al.}(2023){Jim{\'e}nez-Esteban}, {Torres},
  {Rebassa-Mansergas}, {Cruz}, {Murillo-Ojeda}, {Solano}, {Rodrigo}, \&
  {Camisassa}}]{jimenez-esteban2023}
{Jim{\'e}nez-Esteban}, F.~M., {Torres}, S., {Rebassa-Mansergas}, A., {et~al.}
  2023, \mnras, 518, 5106

\bibitem[{{Jim{\'e}nez-Esteban} {et~al.}(2018){Jim{\'e}nez-Esteban}, {Torres},
  {Rebassa-Mansergas}, {Skorobogatov}, {Solano}, {Cantero}, \&
  {Rodrigo}}]{jimenez-esteban2018}
{Jim{\'e}nez-Esteban}, F.~M., {Torres}, S., {Rebassa-Mansergas}, A., {et~al.}
  2018, \mnras, 480, 4505

\bibitem[{{Jin} {et~al.}(2024){Jin}, {Trager}, {Dalton}, {Aguerri}, {Drew},
  {Falc{\'o}n-Barroso}, {G{\"a}nsicke}, {Hill}, {Iovino}, {Pieri}, {Poggianti},
  {Smith}, {Vallenari}, {Abrams}, {Aguado}, {Antoja}, {Arag{\'o}n-Salamanca},
  {Ascasibar}, {Babusiaux}, {Balcells}, {Barrena}, {Battaglia}, {Belokurov},
  {Bensby}, {Bonifacio}, {Bragaglia}, {Carrasco}, {Carrera}, {Cornwell},
  {Dom{\'\i}nguez-Palmero}, {Duncan}, {Famaey}, {Fari{\~n}a}, {Gonzalez},
  {Guest}, {Hatch}, {Hess}, {Hoskin}, {Irwin}, {Knapen}, {Koposov}, {Kuchner},
  {Laigle}, {Lewis}, {Longhetti}, {Lucatello}, {M{\'e}ndez-Abreu}, {Mercurio},
  {Molaeinezhad}, {Mongui{\'o}}, {Morrison}, {Murphy}, {Peralta de Arriba},
  {P{\'e}rez}, {P{\'e}rez-R{\`a}fols}, {Pic{\'o}}, {Raddi}, {Romero-G{\'o}mez},
  {Royer}, {Siebert}, {Seabroke}, {Som}, {Terrett}, {Thomas}, {Wesson},
  {Worley}, {Alfaro}, {Allende Prieto}, {Alonso-Santiago}, {Amos}, {Ashley},
  {Balaguer-N{\'u}{\~n}ez}, {Balbinot}, {Bellazzini}, {Benn}, {Berlanas},
  {Bernard}, {Best}, {Bettoni}, {Bianco}, {Bishop}, {Blomqvist}, {Boeche},
  {Bolzonella}, {Bonoli}, {Bosma}, {Britavskiy}, {Busarello}, {Caffau},
  {Cantat-Gaudin}, {Castro-Ginard}, {Couto}, {Carbajo-Hijarrubia}, {Carter},
  {Casamiquela}, {Conrado}, {Corcho-Caballero}, {Costantin}, {Deason}, {de
  Burgos}, {De Grandi}, {Di Matteo}, {Dom{\'\i}nguez-G{\'o}mez}, {Dorda},
  {Drake}, {Dutta}, {Erkal}, {Feltzing}, {Ferr{\'e}-Mateu}, {Feuillet},
  {Figueras}, {Fossati}, {Franciosini}, {Frasca}, {Fumagalli}, {Gallazzi},
  {Garc{\'\i}a-Benito}, {Gentile Fusillo}, {Gebran}, {Gilbert}, {Gledhill},
  {Gonz{\'a}lez Delgado}, {Greimel}, {Guarcello}, {Guerra}, {Gullieuszik},
  {Haines}, {Hardcastle}, {Harris}, {Haywood}, {Helmi}, {Hernandez}, {Herrero},
  {Hughes}, {Ir{\v{s}}i{\v{c}}}, {Jablonka}, {Jarvis}, {Jordi}, {Kondapally},
  {Kordopatis}, {Krogager}, {La Barbera}, {Lam}, {Larsen}, {Lemasle}, {Lewis},
  {Lhom{\'e}}, {Lind}, {Lodi}, {Longobardi}, {Lonoce}, {Magrini}, {Ma{\'\i}z
  Apell{\'a}niz}, {Marchal}, {Marco}, {Martin}, {Matsuno}, {Maurogordato},
  {Merluzzi}, {Miralda-Escud{\'e}}, {Molinari}, {Monari}, {Morelli}, {Mottram},
  {Naylor}, {Negueruela}, {O{\~n}orbe}, {Pancino}, {Peirani}, {Peletier},
  {Pozzetti}, {Rainer}, {Ramos}, {Read}, {Rossi}, {R{\"o}ttgering},
  {Rubi{\~n}o-Mart{\'\i}n}, {Sabater}, {San Juan}, {Sanna}, {Schallig},
  {Schiavon}, {Schultheis}, {Serra}, {Shimwell}, {Sim{\'o}n-D{\'\i}az},
  {Smith}, {Sordo}, {Sorini}, {Soubiran}, {Starkenburg}, {Steele}, {Stott},
  {Stuik}, {Tolstoy}, {Tortora}, {Tsantaki}, {Van der Swaelmen}, {van Weeren},
  {Vergani}, {Verheijen}, {Verro}, {Vink}, {Vioque}, {Walcher}, {Walton},
  {Wegg}, {Weijmans}, {Williams}, {Wilson}, {Wright}, {Xylakis-Dornbusch},
  {Youakim}, {Zibetti}, \& {Zurita}}]{weave}
{Jin}, S., {Trager}, S.~C., {Dalton}, G.~B., {et~al.} 2024, \mnras, 530, 2688

\bibitem[{{Jones} {et~al.}(2019){Jones}, {R{\"o}pke}, {Fryer}, {Ruiter},
  {Seitenzahl}, {Nittler}, {Ohlmann}, {Reifarth}, {Pignatari}, \&
  {Belczynski}}]{jones2019}
{Jones}, S., {R{\"o}pke}, F.~K., {Fryer}, C., {et~al.} 2019, \aap, 622, A74

\bibitem[{{J{\"o}nsson} {et~al.}(2020){J{\"o}nsson}, {Holtzman}, {Allende
  Prieto}, {Cunha}, {Garc{\'\i}a-Hern{\'a}ndez}, {Hasselquist}, {Masseron},
  {Osorio}, {Shetrone}, {Smith}, {Stringfellow}, {Bizyaev}, {Edvardsson},
  {Majewski}, {M{\'e}sz{\'a}ros}, {Souto}, {Zamora}, {Beaton}, {Bovy}, {Donor},
  {Pinsonneault}, {Poovelil}, \& {Sobeck}}]{jonsson2020}
{J{\"o}nsson}, H., {Holtzman}, J.~A., {Allende Prieto}, C., {et~al.} 2020, \aj,
  160, 120

\bibitem[{{Joyce} {et~al.}(2018){Joyce}, {Barstow}, {Holberg}, {Bond},
  {Casewell}, \& {Burleigh}}]{joyce2018}
{Joyce}, S.~R.~G., {Barstow}, M.~A., {Holberg}, J.~B., {et~al.} 2018, \mnras,
  481, 2361

\bibitem[{{Karl} {et~al.}(2005){Karl}, {Napiwotzki}, {Heber}, {Dreizler},
  {Koester}, \& {Reid}}]{karl2005}
{Karl}, C.~A., {Napiwotzki}, R., {Heber}, U., {et~al.} 2005, \aap, 434, 637

\bibitem[{{Katz} {et~al.}(2019){Katz}, {Sartoretti}, {Cropper}, {Panuzzo},
  {Seabroke}, {Viala}, {Benson}, {Blomme}, {Jasniewicz}, {Jean-Antoine},
  {Huckle}, {Smith}, {Baker}, {Crifo}, {Damerdji}, {David}, {Dolding},
  {Fr{\'e}mat}, {Gosset}, {Guerrier}, {Guy}, {Haigron}, {Jan{\ss}en},
  {Marchal}, {Plum}, {Soubiran}, {Th{\'e}venin}, {Ajaj}, {Allende Prieto},
  {Babusiaux}, {Boudreault}, {Chemin}, {Delle Luche}, {Fabre}, {Gueguen},
  {Hambly}, {Lasne}, {Meynadier}, {Pailler}, {Panem}, {Royer}, {Tauran},
  {Zurbach}, {Zwitter}, {Arenou}, {Bossini}, {Gerssen}, {G{\'o}mez},
  {Lemaitre}, {Leclerc}, {Morel}, {Munari}, {Turon}, {Vallenari}, \&
  {{\v{Z}}erjal}}]{katz2019}
{Katz}, D., {Sartoretti}, P., {Cropper}, M., {et~al.} 2019, \aap, 622, A205

\bibitem[{{Katz} {et~al.}(2023){Katz}, {Sartoretti}, {Guerrier}, {Panuzzo},
  {Seabroke}, {Th{\'e}venin}, {Cropper}, {Benson}, {Blomme}, {Haigron},
  {Marchal}, {Smith}, {Baker}, {Chemin}, {Damerdji}, {David}, {Dolding},
  {Fr{\'e}mat}, {Gosset}, {Jan{\ss}en}, {Jasniewicz}, {Lobel}, {Plum},
  {Samaras}, {Snaith}, {Soubiran}, {Vanel}, {Zwitter}, {Antoja}, {Arenou},
  {Babusiaux}, {Brouillet}, {Caffau}, {Di Matteo}, {Fabre}, {Fabricius},
  {Fragkoudi}, {Haywood}, {Huckle}, {Hottier}, {Lasne}, {Leclerc},
  {Mastrobuono-Battisti}, {Royer}, {Teyssier}, {Zorec}, {Crifo}, {Jean-Antoine
  Piccolo}, {Turon}, \& {Viala}}]{katz2023}
{Katz}, D., {Sartoretti}, P., {Guerrier}, A., {et~al.} 2023, \aap, 674, A5

\bibitem[{{Kawka} \& {Vennes}(2012)}]{kawka2012}
{Kawka}, A. \& {Vennes}, S. 2012, \mnras, 425, 1394

\bibitem[{{Koester}(1987)}]{koester1987}
{Koester}, D. 1987, \apj, 322, 852

\bibitem[{{Koester}(2010)}]{koester2010}
{Koester}, D. 2010, \memsai, 81, 921

\bibitem[{{Koester} {et~al.}(1998){Koester}, {Dreizler}, {Weidemann}, \&
  {Allard}}]{koester1998}
{Koester}, D., {Dreizler}, S., {Weidemann}, V., \& {Allard}, N.~F. 1998, \aap,
  338, 612

\bibitem[{{Koester} \& {Herrero}(1988)}]{koester1988}
{Koester}, D. \& {Herrero}, A. 1988, \apj, 332, 910

\bibitem[{{Koester} {et~al.}(2020){Koester}, {Kepler}, \&
  {Irwin}}]{koester2020}
{Koester}, D., {Kepler}, S.~O., \& {Irwin}, A.~W. 2020, \aap, 635, A103

\bibitem[{{Koester} {et~al.}(2009){Koester}, {Voss}, {Napiwotzki},
  {Christlieb}, {Homeier}, {Lisker}, {Reimers}, \& {Heber}}]{koester2009}
{Koester}, D., {Voss}, B., {Napiwotzki}, R., {et~al.} 2009, \aap, 505, 441

\bibitem[{{Kollmeier} {et~al.}(2017){Kollmeier}, {Zasowski}, {Rix}, {Johns},
  {Anderson}, {Drory}, {Johnson}, {Pogge}, {Bird}, {Blanc}, {Brownstein},
  {Crane}, {De Lee}, {Klaene}, {Kreckel}, {MacDonald}, {Merloni}, {Ness},
  {O'Brien}, {Sanchez-Gallego}, {Sayres}, {Shen}, {Thakar}, {Tkachenko},
  {Aerts}, {Blanton}, {Eisenstein}, {Holtzman}, {Maoz}, {Nandra}, {Rockosi},
  {Weinberg}, {Bovy}, {Casey}, {Chaname}, {Clerc}, {Conroy}, {Eracleous},
  {G{\"a}nsicke}, {Hekker}, {Horne}, {Kauffmann}, {McQuinn}, {Pellegrini},
  {Schinnerer}, {Schlafly}, {Schwope}, {Seibert}, {Teske}, \& {van
  Saders}}]{kollmeier2017}
{Kollmeier}, J.~A., {Zasowski}, G., {Rix}, H.-W., {et~al.} 2017, arXiv
  e-prints, arXiv:1711.03234

\bibitem[{{Kunder} {et~al.}(2017){Kunder}, {Kordopatis}, {Steinmetz},
  {Zwitter}, {McMillan}, {Casagrande}, {Enke}, {Wojno}, {Valentini},
  {Chiappini}, {Matijevi{\v{c}}}, {Siviero}, {de Laverny}, {Recio-Blanco},
  {Bijaoui}, {Wyse}, {Binney}, {Grebel}, {Helmi}, {Jofre}, {Antoja}, {Gilmore},
  {Siebert}, {Famaey}, {Bienaym{\'e}}, {Gibson}, {Freeman}, {Navarro},
  {Munari}, {Seabroke}, {Anguiano}, {{\v{Z}}erjal}, {Minchev}, {Reid},
  {Bland-Hawthorn}, {Kos}, {Sharma}, {Watson}, {Parker}, {Scholz}, {Burton},
  {Cass}, {Hartley}, {Fiegert}, {Stupar}, {Ritter}, {Hawkins}, {Gerhard},
  {Chaplin}, {Davies}, {Elsworth}, {Lund}, {Miglio}, \& {Mosser}}]{kunder2017}
{Kunder}, A., {Kordopatis}, G., {Steinmetz}, M., {et~al.} 2017, \aj, 153, 75

\bibitem[{{Lallement} {et~al.}(2022){Lallement}, {Vergely}, {Babusiaux}, \&
  {Cox}}]{lallement2022}
{Lallement}, R., {Vergely}, J.~L., {Babusiaux}, C., \& {Cox}, N.~L.~J. 2022,
  \aap, 661, A147

\bibitem[{{Landstreet} {et~al.}(2016){Landstreet}, {Bagnulo}, {Martin}, \&
  {Valyavin}}]{landstreet2016}
{Landstreet}, J.~D., {Bagnulo}, S., {Martin}, A., \& {Valyavin}, G. 2016, \aap,
  591, A80

\bibitem[{{Lindegren} {et~al.}(2021){Lindegren}, {Bastian}, {Biermann},
  {Bombrun}, {de Torres}, {Gerlach}, {Geyer}, {Hern{\'a}ndez}, {Hilger},
  {Hobbs}, {Klioner}, {Lammers}, {McMillan}, {Ramos-Lerate},
  {Steidelm{\"u}ller}, {Stephenson}, \& {van Leeuwen}}]{lindegren2021}
{Lindegren}, L., {Bastian}, U., {Biermann}, M., {et~al.} 2021, \aap, 649, A4

\bibitem[{{Luo} {et~al.}(2019){Luo}, {Zhao}, {Zhao}, \& {et al.}}]{luo2019}
{Luo}, A.~L., {Zhao}, Y.~H., {Zhao}, G., \& {et al.} 2019, VizieR Online Data
  Catalog, V/164

\bibitem[{{Mainieri} {et~al.}(2024){Mainieri}, {Anderson}, {Brinchmann},
  {Cimatti}, {Ellis}, {Hill}, {Kneib}, {McLeod}, {Opitom}, {Roth},
  {Sanchez-Saez}, {Smiljanic}, {Tolstoy}, {Bacon}, {Randich}, {Adamo},
  {Annibali}, {Arevalo}, {Audard}, {Barsanti}, {Battaglia}, {Bayo Aran},
  {Belfiore}, {Bellazzini}, {Bellini}, {Beltran}, {Berni}, {Bianchi}, {Biazzo},
  {Bisero}, {Bisogni}, {Bland-Hawthorn}, {Blondin}, {Bodensteiner}, {Boffin},
  {Bonito}, {Bono}, {Bouche}, {Bowman}, {Braga}, {Bragaglia}, {Branchesi},
  {Brucalassi}, {Bryant}, {Bryson}, {Busa}, {Camera}, {Carbone}, {Casali},
  {Casali}, {Casasola}, {Castro}, {Catelan}, {Cavallo}, {Chiappini}, {Cioni},
  {Colless}, {Colzi}, {Contarini}, {Couch}, {D'Ammando}, {d'Assignies D.},
  {D'Orazi}, {da Silva}, {Dainotti}, {Damiani}, {Danielski}, {De Cia}, {de
  Jong}, {Dhawan}, {Dierickx}, {Driver}, {Dupletsa}, {Escoffier}, {Escorza},
  {Fabrizio}, {Fiorentino}, {Fontana}, {Fontani}, {Forero Sanchez}, {Franois},
  {Galindo-Guil}, {Gallazzi}, {Galli}, {Garcia}, {Garcia-Rojas}, {Garilli},
  {Grand}, {Guarcello}, {Hazra}, {Helmi}, {Herrero}, {Iglesias}, {Ilic},
  {Irsic}, {Ivanov}, {Izzo}, {Jablonka}, {Joachimi}, {Kakkad}, {Kamann},
  {Koposov}, {Kordopatis}, {Kovacevic}, {Kraljic}, {Kuncarayakti}, {Kwon}, {La
  Forgia}, {Lahav}, {Laigle}, {Lazzarin}, {Leaman}, {Leclercq}, {Lee}, {Lee},
  {Lehnert}, {Lira}, {Loffredo}, {Lucatello}, {Magrini}, {Maguire}, {Mahler},
  {Zahra Majidi}, {Malavasi}, {Mannucci}, {Marconi}, {Martin}, {Marulli},
  {Massari}, {Matsuno}, {Mattheee}, {McGee}, {Merc}, {Merle}, {Miglio},
  {Migliorini}, {Minchev}, {Minniti}, {Miret-Roig}, {Monreal Ibero}, {Montano},
  {Montet}, {Moresco}, {Moretti}, {Moscardini}, {Moya}, {Mueller},
  {Nanayakkara}, {Nicholl}, {Nordlander}, {Onori}, {Padovani}, {Pala}, {Panda},
  {Pandey-Pommier}, {Pasquini}, {Pawlak}, {Pessi}, {Pisani}, {Popovic},
  {Prisinzano}, {Raddi}, {Rainer}, {Rebassa-Mansergas}, {Richard}, {Rigault},
  {Rocher}, {Romano}, {Rosati}, {Sacco}, {Sanchez-Janssen}, {Sander},
  {Sanders}, {Sargent}, {Sarpa}, {Schimd}, {Schipani}, {Sefusatti}, {Smith},
  {Spina}, {Steinmetz}, {Tacchella}, {Tautvaisiene}, {Theissen}, {Thomas},
  {Ting}, {Travouillon}, {Tresse}, {Trivedi}, {Tsantaki}, {Tsedrik}, {Urrutia},
  {Valenti}, {Van der Swaelmen}, {Van Eck}, {Verdiani}, {Verdier}, {Vergani},
  {Verhamme}, {Vernet}, {Verza}, {Viel}, {Vielzeuf}, {Vietri}, {Vink},
  {Viscasillas Vazquez}, {Wang}, {Weilbacher}, {Wendt}, {Wright}, {Ye},
  {Yeche}, {Yu}, {Zafar}, {Zibetti}, {Ziegler}, \& {Zinchenko}}]{wst}
{Mainieri}, V., {Anderson}, R.~I., {Brinchmann}, J., {et~al.} 2024, arXiv
  e-prints, arXiv:2403.05398

\bibitem[{{Manser} {et~al.}(2024){Manser}, {Izquierdo}, {G{\"a}nsicke}, {Swan},
  {Koester}, {Robert}, {Xu}, {Inight}, {Amroota}, {Fusillo}, {Koposov}, {Kim},
  {Dey}, {Prieto}, {Aguilar}, {Ahlen}, {Blum}, {Brooks}, {Claybaugh}, {Cooper},
  {Dawson}, {de la Macorra}, {Doel}, {Forero-Romero}, {Gazta{\~n}aga},
  {Gontcho}, {Honscheid}, {Kisner}, {Kremin}, {Lambert}, {Landriau}, {Le
  Guillou}, {Levi}, {Li}, {Meisner}, {Miquel}, {Moustakas}, {Nie},
  {Palanque-Delabrouille}, {Percival}, {Poppett}, {Prada}, {Rezaie}, {Rossi},
  {Sanchez}, {Schlafly}, {Schlegel}, {Schubnell}, {Seo}, {Silber}, {Tarl{\'e}},
  {Weaver}, {Zhou}, \& {Zou}}]{manser2024}
{Manser}, C.~J., {Izquierdo}, P., {G{\"a}nsicke}, B.~T., {et~al.} 2024, \mnras,
  535, 254

\bibitem[{{Maoz} \& {Hallakoun}(2017)}]{maoz2017}
{Maoz}, D. \& {Hallakoun}, N. 2017, \mnras, 467, 1414

\bibitem[{{Marigo}(2022)}]{marigo2022}
{Marigo}, P. 2022, Universe, 8, 243

\bibitem[{{Montegriffo} {et~al.}(2023){Montegriffo}, {De Angeli}, {Andrae},
  {Riello}, {Pancino}, {Sanna}, {Bellazzini}, {Evans}, {Carrasco}, {Sordo},
  {Busso}, {Cacciari}, {Jordi}, {van Leeuwen}, {Vallenari}, {Altavilla},
  {Barstow}, {Brown}, {Burgess}, {Castellani}, {Cowell}, {Davidson}, {De
  Luise}, {Delchambre}, {Diener}, {Fabricius}, {Fr{\'e}mat}, {Fouesneau},
  {Gilmore}, {Giuffrida}, {Hambly}, {Harrison}, {Hidalgo}, {Hodgkin},
  {Holland}, {Marinoni}, {Osborne}, {Pagani}, {Palaversa}, {Piersimoni},
  {Pulone}, {Ragaini}, {Rainer}, {Richards}, {Rowell}, {Ruz-Mieres}, {Sarro},
  {Walton}, \& {Yoldas}}]{montegrifo2023}
{Montegriffo}, P., {De Angeli}, F., {Andrae}, R., {et~al.} 2023, \aap, 674, A3

\bibitem[{{Napiwotzki} {et~al.}(2001){Napiwotzki}, {Christlieb}, {Drechsel},
  {Hagen}, {Heber}, {Homeier}, {Karl}, {Koester}, {Leibundgut}, {Marsh},
  {Moehler}, {Nelemans}, {Pauli}, {Reimers}, {Renzini}, \&
  {Yungelson}}]{napiwotzki2001}
{Napiwotzki}, R., {Christlieb}, N., {Drechsel}, H., {et~al.} 2001,
  Astronomische Nachrichten, 322, 411

\bibitem[{{Napiwotzki} {et~al.}(1999){Napiwotzki}, {Green}, \&
  {Saffer}}]{napiwotzki1999}
{Napiwotzki}, R., {Green}, P.~J., \& {Saffer}, R.~A. 1999, \apj, 517, 399

\bibitem[{{Napiwotzki} {et~al.}(2020){Napiwotzki}, {Karl}, {Lisker},
  {Catal{\'a}n}, {Drechsel}, {Heber}, {Homeier}, {Koester}, {Leibundgut},
  {Marsh}, {Moehler}, {Nelemans}, {Reimers}, {Renzini}, {Str{\"o}er}, \&
  {Yungelson}}]{napiwotzki2020}
{Napiwotzki}, R., {Karl}, C.~A., {Lisker}, T., {et~al.} 2020, \aap, 638, A131

\bibitem[{Newville {et~al.}(2015)Newville, Stensitzki, Allen, \&
  Ingargiola}]{lmfit}
Newville, M., Stensitzki, T., Allen, D.~B., \& Ingargiola, A. 2015, LMFIT:
  Non-Linear Least-Square Minimization and Curve-Fitting for Python

\bibitem[{{Ochsenbein} {et~al.}(2000){Ochsenbein}, {Bauer}, \&
  {Marcout}}]{ochsenbein20000}
{Ochsenbein}, F., {Bauer}, P., \& {Marcout}, J. 2000, \aaps, 143, 23

\bibitem[{{Panei} {et~al.}(2000){Panei}, {Althaus}, \& {Benvenuto}}]{panei2000}
{Panei}, J.~A., {Althaus}, L.~G., \& {Benvenuto}, O.~G. 2000, \aap, 353, 970

\bibitem[{{Parsons} {et~al.}(2017){Parsons}, {G{\"a}nsicke}, {Marsh}, {Ashley},
  {Bours}, {Breedt}, {Burleigh}, {Copperwheat}, {Dhillon}, {Green}, {Hardy},
  {Hermes}, {Irawati}, {Kerry}, {Littlefair}, {McAllister}, {Rattanasoon},
  {Rebassa-Mansergas}, {Sahman}, \& {Schreiber}}]{parsons2017}
{Parsons}, S.~G., {G{\"a}nsicke}, B.~T., {Marsh}, T.~R., {et~al.} 2017, \mnras,
  470, 4473

\bibitem[{{Pasquini} {et~al.}(2019){Pasquini}, {Pala}, {Ludwig}, {Lẽao}, {de
  Medeiros}, \& {Weiss}}]{pasquini2019}
{Pasquini}, L., {Pala}, A.~F., {Ludwig}, H.~G., {et~al.} 2019, \aap, 627, L8

\bibitem[{{Pasquini} {et~al.}(2023){Pasquini}, {Pala}, {Salaris}, {Ludwig},
  {Le{\~a}o}, {Weiss}, \& {de Medeiros}}]{pasquini2023}
{Pasquini}, L., {Pala}, A.~F., {Salaris}, M., {et~al.} 2023, \mnras, 522, 3710

\bibitem[{{Pauli} {et~al.}(2006){Pauli}, {Napiwotzki}, {Heber}, {Altmann}, \&
  {Odenkirchen}}]{pauli2006}
{Pauli}, E.~M., {Napiwotzki}, R., {Heber}, U., {Altmann}, M., \& {Odenkirchen},
  M. 2006, \aap, 447, 173

\bibitem[{{Popper}(1954)}]{popper1954}
{Popper}, D.~M. 1954, \apj, 120, 316

\bibitem[{{Popper}(1980)}]{popper1980}
{Popper}, D.~M. 1980, \araa, 18, 115

\bibitem[{{Prada Moroni} \& {Straniero}(2009)}]{prada-moroni2009}
{Prada Moroni}, P.~G. \& {Straniero}, O. 2009, \aap, 507, 1575

\bibitem[{{Press} {et~al.}(1992){Press}, {Teukolsky}, {Vetterling}, \&
  {Flannery}}]{press1992}
{Press}, W.~H., {Teukolsky}, S.~A., {Vetterling}, W.~T., \& {Flannery}, B.~P.
  1992, {Numerical recipes in FORTRAN. The art of scientific computing}

\bibitem[{{Provencal} {et~al.}(1998){Provencal}, {Shipman}, {H{\o}g}, \&
  {Thejll}}]{provencal1998}
{Provencal}, J.~L., {Shipman}, H.~L., {H{\o}g}, E., \& {Thejll}, P. 1998, \apj,
  494, 759

\bibitem[{{Provencal} {et~al.}(2002){Provencal}, {Shipman}, {Koester},
  {Wesemael}, \& {Bergeron}}]{provencal2002}
{Provencal}, J.~L., {Shipman}, H.~L., {Koester}, D., {Wesemael}, F., \&
  {Bergeron}, P. 2002, \apj, 568, 324

\bibitem[{{Raddi} {et~al.}(2022){Raddi}, {Torres}, {Rebassa-Mansergas},
  {Maldonado}, {Camisassa}, {Koester}, {Gentile Fusillo}, {Tremblay}, {Dimpel},
  {Heber}, {Cunningham}, \& {Ren}}]{raddi2022}
{Raddi}, R., {Torres}, S., {Rebassa-Mansergas}, A., {et~al.} 2022, \aap, 658,
  A22

\bibitem[{{Rebassa-Mansergas} {et~al.}(2021){Rebassa-Mansergas}, {Maldonado},
  {Raddi}, {Knowles}, {Torres}, {Hoskin}, {Cunningham}, {Hollands}, {Ren},
  {G{\"a}nsicke}, {Tremblay}, {Castro-Rodr{\'\i}guez}, {Camisassa}, \&
  {Koester}}]{rebassa2021}
{Rebassa-Mansergas}, A., {Maldonado}, J., {Raddi}, R., {et~al.} 2021, \mnras,
  505, 3165

\bibitem[{{Recio-Blanco} {et~al.}(2023){Recio-Blanco}, {de Laverny}, {Palicio},
  {Kordopatis}, {{\'A}lvarez}, {Schultheis}, {Contursi}, {Zhao}, {Torralba
  Elipe}, {Ordenovic}, {Manteiga}, {Dafonte}, {Oreshina-Slezak}, {Bijaoui},
  {Fr{\'e}mat}, {Seabroke}, {Pailler}, {Spitoni}, {Poggio}, {Creevey}, {Abreu
  Aramburu}, {Accart}, {Andrae}, {Bailer-Jones}, {Bellas-Velidis}, {Brouillet},
  {Brugaletta}, {Burlacu}, {Carballo}, {Casamiquela}, {Chiavassa}, {Cooper},
  {Dapergolas}, {Delchambre}, {Dharmawardena}, {Drimmel}, {Edvardsson},
  {Fouesneau}, {Garabato}, {Garc{\'\i}a-Lario}, {Garc{\'\i}a-Torres}, {Gavel},
  {Gomez}, {Gonz{\'a}lez-Santamar{\'\i}a}, {Hatzidimitriou}, {Heiter},
  {Jean-Antoine Piccolo}, {Kontizas}, {Korn}, {Lanzafame}, {Lebreton}, {Le
  Fustec}, {Licata}, {Lindstr{\o}m}, {Livanou}, {Lobel}, {Lorca}, {Magdaleno
  Romeo}, {Marocco}, {Marshall}, {Mary}, {Nicolas}, {Pallas-Quintela}, {Panem},
  {Pichon}, {Riclet}, {Robin}, {Rybizki}, {Santove{\~n}a}, {Silvelo}, {Smart},
  {Sarro}, {Sordo}, {Soubiran}, {S{\"u}veges}, {Ulla}, {Vallenari}, {Zorec},
  {Utrilla}, \& {Bakker}}]{recio-blanco2023}
{Recio-Blanco}, A., {de Laverny}, P., {Palicio}, P.~A., {et~al.} 2023, \aap,
  674, A29

\bibitem[{{Reid}(1996)}]{reid1996}
{Reid}, I.~N. 1996, \aj, 111, 2000

\bibitem[{{Serenelli} {et~al.}(2021){Serenelli}, {Weiss}, {Aerts}, {Angelou},
  {Baroch}, {Bastian}, {Beck}, {Bergemann}, {Bestenlehner}, {Czekala},
  {Elias-Rosa}, {Escorza}, {Van Eylen}, {Feuillet}, {Gandolfi}, {Gieles},
  {Girardi}, {Lebreton}, {Lodieu}, {Martig}, {Miller Bertolami}, {Mombarg},
  {Morales}, {Moya}, {Nsamba}, {Pavlovski}, {Pedersen}, {Ribas}, {Schneider},
  {Silva Aguirre}, {Stassun}, {Tolstoy}, {Tremblay}, \&
  {Zwintz}}]{serenelli2021}
{Serenelli}, A., {Weiss}, A., {Aerts}, C., {et~al.} 2021, \aapr, 29, 4

\bibitem[{{Serenelli} {et~al.}(2001){Serenelli}, {Althaus}, {Rohrmann}, \&
  {Benvenuto}}]{serenelli2001}
{Serenelli}, A.~M., {Althaus}, L.~G., {Rohrmann}, R.~D., \& {Benvenuto}, O.~G.
  2001, \mnras, 325, 607

\bibitem[{{Shipman} {et~al.}(1997){Shipman}, {Provencal}, {H{\o}g}, \&
  {Thejll}}]{shipman1997}
{Shipman}, H.~L., {Provencal}, J.~L., {H{\o}g}, E., \& {Thejll}, P. 1997,
  \apjl, 488, L43

\bibitem[{{Silvestri} {et~al.}(2001){Silvestri}, {Oswalt}, {Wood}, {Smith},
  {Reid}, \& {Sion}}]{silvestri2001}
{Silvestri}, N.~M., {Oswalt}, T.~D., {Wood}, M.~A., {et~al.} 2001, \aj, 121,
  503

\bibitem[{{Toloza} {et~al.}(2023){Toloza}, {Rebassa-Mansergas}, {Raddi},
  {Reindl}, {Gaensicke}, {Fusillo}, {Scaringi}, {Belloni}, {Breedt},
  {Camisassa}, {Cunningham}, {de Martino}, {Ederoclite}, {Geier}, {Green},
  {Inight}, {Kupfer}, {Maldonado}, {Marsh}, {Pala}, {Parsons}, {Pelisoli},
  {Ren}, {Rodriguez-Gil}, {Sahu}, {Schmidtobreick}, {Schreiber}, {Schwope},
  {Steeghs}, {Szkody}, {Toonen}, {Tremblay}, \& {Zorotovic}}]{s11}
{Toloza}, O., {Rebassa-Mansergas}, A., {Raddi}, R., {et~al.} 2023, The
  Messenger, 190, 4

\bibitem[{{Torres} {et~al.}(2021){Torres}, {Rebassa-Mansergas}, {Camisassa}, \&
  {Raddi}}]{torres2021}
{Torres}, S., {Rebassa-Mansergas}, A., {Camisassa}, M.~E., \& {Raddi}, R. 2021,
  \mnras, 502, 1753

\bibitem[{{Tremblay} \& {Bergeron}(2009)}]{tremblay2009}
{Tremblay}, P.~E. \& {Bergeron}, P. 2009, \apj, 696, 1755

\bibitem[{{Tremblay} {et~al.}(2011){Tremblay}, {Bergeron}, \&
  {Gianninas}}]{tremblay2011}
{Tremblay}, P.~E., {Bergeron}, P., \& {Gianninas}, A. 2011, \apj, 730, 128

\bibitem[{{Tremblay} {et~al.}(2019){Tremblay}, {Cukanovaite}, {Gentile
  Fusillo}, {Cunningham}, \& {Hollands}}]{tremblay2019}
{Tremblay}, P.~E., {Cukanovaite}, E., {Gentile Fusillo}, N.~P., {Cunningham},
  T., \& {Hollands}, M.~A. 2019, \mnras, 482, 5222

\bibitem[{{Tremblay} {et~al.}(2016){Tremblay}, {Cummings}, {Kalirai},
  {G{\"a}nsicke}, {Gentile-Fusillo}, \& {Raddi}}]{tremblay2016}
{Tremblay}, P.~E., {Cummings}, J., {Kalirai}, J.~S., {et~al.} 2016, \mnras,
  461, 2100

\bibitem[{{Tremblay} {et~al.}(2015{\natexlab{a}}){Tremblay}, {Fontaine},
  {Freytag}, {Steiner}, {Ludwig}, {Steffen}, {Wedemeyer}, \&
  {Brassard}}]{tremblay2015a}
{Tremblay}, P.~E., {Fontaine}, G., {Freytag}, B., {et~al.} 2015{\natexlab{a}},
  \apj, 812, 19

\bibitem[{{Tremblay} {et~al.}(2017){Tremblay}, {Gentile-Fusillo}, {Raddi},
  {Jordan}, {Besson}, {G{\"a}nsicke}, {Parsons}, {Koester}, {Marsh}, {Bohlin},
  {Kalirai}, \& {Deustua}}]{tremblay2017}
{Tremblay}, P.~E., {Gentile-Fusillo}, N., {Raddi}, R., {et~al.} 2017, \mnras,
  465, 2849

\bibitem[{{Tremblay} {et~al.}(2015{\natexlab{b}}){Tremblay}, {Gianninas},
  {Kilic}, {Ludwig}, {Steffen}, {Freytag}, \& {Hermes}}]{tremblay2015}
{Tremblay}, P.~E., {Gianninas}, A., {Kilic}, M., {et~al.} 2015{\natexlab{b}},
  \apj, 809, 148

\bibitem[{{Tremblay} {et~al.}(2014){Tremblay}, {Kalirai}, {Soderblom},
  {Cignoni}, \& {Cummings}}]{tremblay2014}
{Tremblay}, P.~E., {Kalirai}, J.~S., {Soderblom}, D.~R., {Cignoni}, M., \&
  {Cummings}, J. 2014, \apj, 791, 92

\bibitem[{{Tremblay} {et~al.}(2013){Tremblay}, {Ludwig}, {Steffen}, \&
  {Freytag}}]{tremblay2013}
{Tremblay}, P.~E., {Ludwig}, H.~G., {Steffen}, M., \& {Freytag}, B. 2013, \aap,
  552, A13

\bibitem[{{Trimble} \& {Greenstein}(1972)}]{trimble1972}
{Trimble}, V. \& {Greenstein}, J.~L. 1972, \apj, 177, 441

\bibitem[{{Vennes} {et~al.}(2011){Vennes}, {Kawka}, \&
  {N{\'e}meth}}]{vennes2011}
{Vennes}, S., {Kawka}, A., \& {N{\'e}meth}, P. 2011, \mnras, 413, 2545

\bibitem[{{von Hippel}(1996)}]{vonHippel1996}
{von Hippel}, T. 1996, \apjl, 458, L37

\bibitem[{{Wegner}(1974)}]{wegner1974}
{Wegner}, G. 1974, \mnras, 166, 271

\bibitem[{{Wenger} {et~al.}(2000){Wenger}, {Ochsenbein}, {Egret}, {Dubois},
  {Bonnarel}, {Borde}, {Genova}, {Jasniewicz}, {Lalo{\"e}}, {Lesteven}, \&
  {Monier}}]{wenger2000}
{Wenger}, M., {Ochsenbein}, F., {Egret}, D., {et~al.} 2000, \aaps, 143, 9

\bibitem[{{Wiese} \& {Kelleher}(1971)}]{wiese1971}
{Wiese}, W.~L. \& {Kelleher}, D.~E. 1971, \apjl, 166, L59

\bibitem[{{Xiang} {et~al.}(2019){Xiang}, {Ting}, {Rix}, {Sandford}, {Buder},
  {Lind}, {Liu}, {Shi}, \& {Zhang}}]{xiang2019}
{Xiang}, M., {Ting}, Y.-S., {Rix}, H.-W., {et~al.} 2019, \apjs, 245, 34

\end{thebibliography}

\begin{appendix}
\twocolumn
\newpage
\section{SPY white dwarfs in wide binaries}
\label{app:spy}
Table\,\ref{tab:napiwotzki} summarize the relevant physical parameters of the 17 SPY white dwarfs from \citet{napiwotzki2020}, which we have reanalyzed in this work. Those authors measured the atmospheric parameters of white dwarfs with $T_{\rm eff} \geq 20\,000$\,K using their own grid of pure hydrogen NLTE model atmospheres for DA white dwarfs \citep{napiwotzki1999}, while the cooler stars were analyzed with a grid of LTE Koester model atmospheres whose spectral-synthesis was supplemented with NLTE line formation physics.

\begin{figure*}[th!]
    \centering
\includegraphics[width=\linewidth]{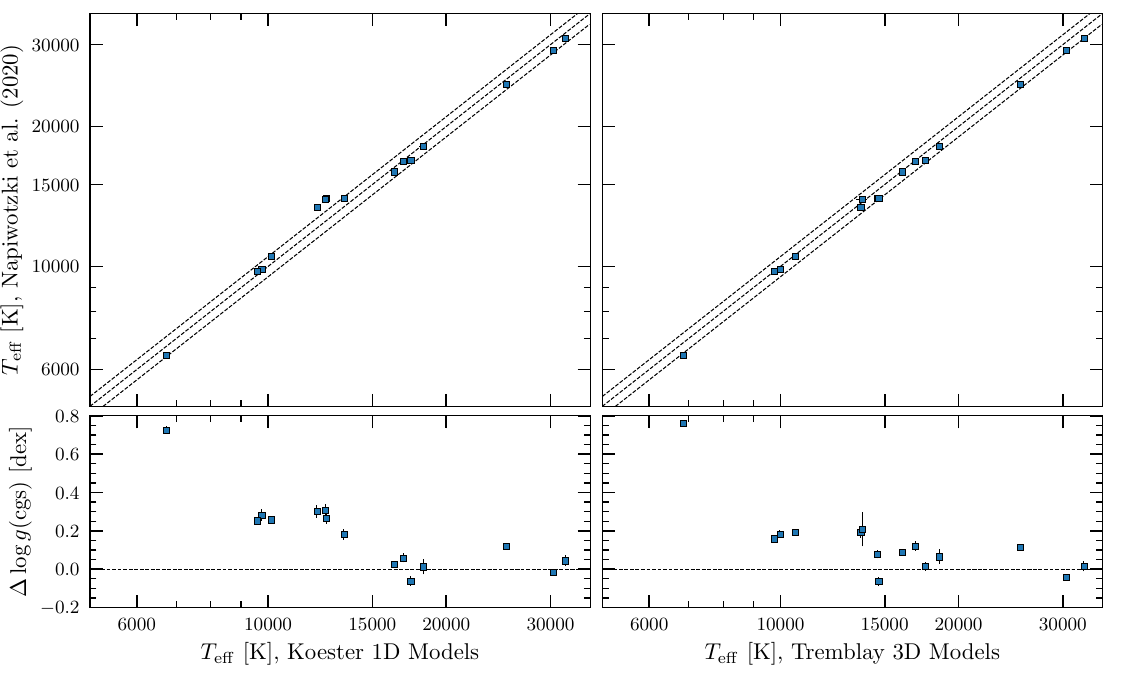}
    \caption{Comparison of our measured spectroscopic parameters against those published in \citet{napiwotzki2020}, which are listed in Table\,\ref{tab:napiwotzki}. The equality lines and $\pm 5$\,\% $T_{\rm eff}$ boundaries are plotted as dashed lines.}
    \label{fig:comparison_napiwotzki}
\end{figure*}

Fig.\,\ref{fig:comparison_napiwotzki} shows the comparison among our measured atmospheric parameters with the results published by \citet{napiwotzki2020}. The comparison among $T_{\rm eff}$ determinations is very good with respect to the Tremblay models; small discrepancies are observed with respect to the Koester models for a small number of objects with $T_{\rm eff} \approx 13\,000$\,K. The comparison among $\log{g}$ determinations is worse for the coolest objects ($T_{\rm eff} < 15\,000$\,K) and, again, it appears better for the 3D models rather than the 1D ones. Nevertheless, we note that the parameter estimates presented in Sect.\,\ref{sec:spectroscopic_analysis} are rather consistent with each other, apart from the known differences due to the 1D vs 3D modeling.

\begin{figure}[th!]
    \centering
\includegraphics[width=\linewidth]{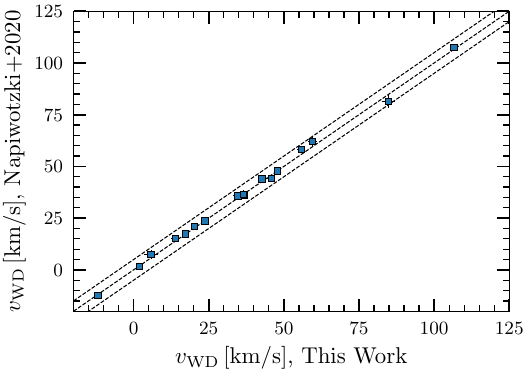}
    \caption{Comparison of our measured spectroscopic parameters against those published in \citet{napiwotzki2020}, which are listed in Table\,\ref{tab:napiwotzki}. The equality lines and $\pm 5$\,\% $T_{\rm eff}$ boundaries are plotted as dashed lines.}
    \label{fig:rv_napiwotzki}
\end{figure}

More important is the comparison among radial velocity measurements for this sample, which confirms our consistent approach as it is shown in Fig.\,\ref{fig:rv_napiwotzki}. Our measurements based on the combined H$\alpha$+H$\beta$ lines are compatible within the small uncertainties with the results presented by \citet{napiwotzki2020}.

Finally, in Table\,\ref{tab:wd_excluded} we list the 11 additional white dwarfs from the SPY project that have wide companions matching their parallaxes and proper motions in {\em Gaia} DR3. They were excluded from our analysis because their companions do not have radial velocity measurements in {\em Gaia} DR3.

\begin{table}[h!]
    \centering
    \caption{Atmospheric parameters of the SPY white dwarfs of DA spectral type \citep{napiwotzki2020}.}
    \label{tab:spy}    
    \begin{tabular}{@{}l s{2.4} c c m{5.3} @{}}
    \hline
    \hline
    \noalign{\smallskip}    
     \noalign{\smallskip}
    Name  & \multicolumn{1}{c}{$T_{\rm eff}$\,(K)} & $\log{g}$\,(cgs)  & $M/$M$_{\odot}$ & \multicolumn{1}{c}{$\varv_{\rm WD}$\,(km/s)} \\
      \noalign{\smallskip}
      \hline
\noalign{\smallskip}
  HE\,0204$-$3821         & 14,038 & 7.79	&  0.512 & 36.1 , 1.3  \\
  HE\,0330$-$4736         & 13,437 & 7.95	&  0.585 & 44.3 , 1.5	\\
  HS\,0145+1737           & 18,125 & 7.89	&  0.566 & 23.7 , 0.8	\\
  WD\,0106$-$358          & 29,198 & 7.86	&  0.576 & 35.8 , 2.0	\\
  WD\,0204$-$306$^{a}$    & 5,640  & 8.00	&	 & 81.5 , 3.4	\\
  WD\,0628$-$020          & 6,443  & 7.20	&	 &107.7 , 1.0	\\
  WD\,1015$-$216          & 30,937 & 7.89	&  0.593 &  7.6 , 2.2	\\
  WD\,1105$-$048          & 15,995 & 7.75	&  0.503 & 47.9 , 0.6	\\
  WD\,1147+255            & 9,863  & 7.78	&  0.497 & 62.0 , 2.4	\\
  WD\,1348$-$273          & 9,787  & 7.78	&  0.495 & 58.2 , 1.5	\\
  WD\,1544$-$377          & 10,525 & 7.83	&  0.517 & 21.1 , 0.6	\\
  WD\,1619+123            & 16,853 & 7.68	&  0.483 & 15.3 , 1.0	\\
  WD\,1620$-$391          & 24,677 & 7.93	&  0.596 & 44.0 , 0.6	\\
  WD\,1911+135            & 14,004 & 7.86	&  0.539 & 17.4 , 1.0	\\
  WD\,1932$-$136          & 16,931 & 7.73	&  0.497 &  1.8 , 1.2	\\
  WD\,2253+054$^{b}$      & 6,244  & 8.64	&	 & 36.4 , 2.0	\\
  WD\,2318+126            & 13,965 & 7.80	&  0.512 &-12.3 , 1.1	\\
      \noalign{\smallskip}    
    \hline 
\multicolumn{5}{p{3.1in}}{$a$: fixed $\log{g}$ from \citet{kawka2012};${b}$: $\log{g}$ from \citet{gianninas20122}}
    \end{tabular} 
    \label{tab:napiwotzki}
\end{table}
\begin{table}[h!]
    \caption{SPY white dwarfs in wide binaries that have been excluded from
    our analysis.}
    \small
    \centering
    \begin{tabular}{@{}lll@{}}
    \hline
    \hline
     \noalign{\smallskip}
Short Name & {\em Gaia} ID white dwarf & {\em Gaia} ID companion \\
 \noalign{\smallskip}
\hline
 \noalign{\smallskip}
HE\,0516$-$1804 & 2981590730954538112 & 2981590730954537984 \\
HE\,0409$-$5154 & 4780544792270137088 & 4780544860989614080 \\
HS\,0949+0935 & 3878937457832171776 & 3878937487896332416 \\
HS\,2244+0305 & 2656542452030717952 & 2656542447738589312 \\
WD\,0032$-$177 & 2364311331022888704 & 2364311331022888576 \\
WD\,0220+222 & 99915890086770176 & 99915890086770560 \\
WD\,0229$-$481 & 4939012317940174464 & 4939012317940174592 \\
WD\,1126$-$222 & 3541237717085787008 & 3541237717085786880 \\
WD\,1257+047 & 3705070756419217408 & 3705070756418520064 \\
WD\,2020$-$425 & 6679362959252072832 & 6679361138186074240 \\
WD\,2353+026 & 2739782629080048000 & 2739782968381225600 \\
 \noalign{\smallskip}
\hline
\multicolumn{3}{p{8cm}}{Note: WD2020$-$425 is a candidate double-degenerate, hence it is a possible hierarchical triple system.}
    \end{tabular}
    \label{tab:wd_excluded}
\end{table}
\onecolumn
\section{{\em Gaia} parameters of the studied sample}

\begin{table}[h!]
    \caption{Relevant {\it Gaia} parameters of the studied white dwarfs.}
    \scriptsize
    \centering
    \begin{tabular}{@{}l l l m{5.4}  m{5.4}  m{5.4} c c c @{}}
\hline
\hline
 \noalign{\smallskip}
Short Name  & {\em Gaia} ID & Type 
& \multicolumn{1}{c}{$\varpi$} & \multicolumn{1}{c}{$\mu_\alpha \cos{\delta}$} & \multicolumn{1}{c}{$\mu_\delta$ } & $G$  & $BP$  & $RP$  \\
  &  & & \multicolumn{1}{c}{(arcsec)} & \multicolumn{1}{c}{ (mas/yr)} & \multicolumn{1}{c}{(mas/yr)} & (mag) & (mag) & (mag) \\
   \noalign{\smallskip}
\hline
 \noalign{\smallskip}
0003+1713 & 2772926204507772544 &  DA &  9.33 ,  0.05  &   -52.67 ,    0.06  &   -62.36 ,    0.06  & $ 16.17 $ & $ 16.12  $ & $ 16.30 $ \\
0030$-$5856 & 4906307207132391040 &  DBZ & 11.95 ,  0.04  &    17.95 ,    0.03  &    -7.10 ,    0.03  & $ 15.90 $ & $ 15.84  $ & $ 16.00 $ \\
0042$-$5820 & 4907169670925418752 &  DA & 16.66 ,  0.03  &   -89.67 ,    0.03  &   -68.07 ,    0.03  & $ 16.07 $ & $ 16.12  $ & $ 16.00 $ \\
0115+1534 & 2591573203298626432 &  DA &  9.49 ,  0.06  &   -26.34 ,    0.08  &    -4.74 ,    0.05  & $ 15.51 $ & $ 15.38  $ & $ 15.72 $ \\
0125$-$7619 & 4637598659932556928 &  DA &  5.69 ,  0.04  &     8.88 ,    0.05  &     0.52 ,    0.05  & $ 16.61 $ & $ 16.50  $ & $ 16.79 $ \\
0141+0711 & 2568552487829583616 &  DZ & 12.80 ,  0.06  &   -61.37 ,    0.08  &   -78.10 ,    0.05  & $ 16.59 $ & $ 16.61  $ & $ 16.57 $ \\
0156$-$4722 & 4942195988218119168 &  DA & 11.92 ,  0.04  &    32.31 ,    0.03  &     3.03 ,    0.04  & $ 16.13 $ & $ 16.13  $ & $ 16.17 $ \\
0207+0355 & 2517771401566875264 &  DA &  5.27 ,  0.22  &     5.49 ,    0.24  &    -0.90 ,    0.18  & $ 18.69 $ & $ 18.65  $ & $ 18.83 $ \\
0209$-$0140 & 2494399362068211328 &  DAZ & 17.98 ,  0.06  &   -33.97 ,    0.06  &   -55.05 ,    0.05  & $ 16.15 $ & $ 16.22  $ & $ 16.02 $ \\
0319$-$7254 & 4642153627368210560 &  DA & 10.36 ,  0.04  &   123.51 ,    0.06  &   -22.44 ,    0.05  & $ 16.39 $ & $ 16.38  $ & $ 16.44 $ \\
0439$-$0405 & 3201594074140023936 &  DA & 19.00 ,  0.05  &    98.49 ,    0.06  &   -38.72 ,    0.05  & $ 16.44 $ & $ 16.57  $ & $ 16.22 $ \\
0608+0204 & 3315373564607614208 &  DC & 22.06 ,  0.05  &    18.06 ,    0.06  &   -77.06 ,    0.05  & $ 15.87 $ & $ 15.91  $ & $ 15.76 $ \\
0608$-$0059 & 3121385658671190784 &  DA & 16.17 ,  0.09  &     2.76 ,    0.10  &   -36.89 ,    0.08  & $ 17.24 $ & $ 17.20  $ & $ 17.38 $ \\
0608$-$5301 & 5548648634588556288 &  DA & 17.35 ,  0.03  &  -144.24 ,    0.05  &   216.45 ,    0.04  & $ 15.96 $ & $ 16.02  $ & $ 15.86 $ \\
0616$-$5912 & 5482551252566796928 &  DB & 27.73 ,  0.02  &   -48.35 ,    0.03  &  -312.18 ,    0.03  & $ 13.95 $ & $ 13.89  $ & $ 14.06 $ \\
0621$-$5456 & 5499525444356625408 &  DA & 10.75 ,  0.04  &    27.37 ,    0.06  &   -91.32 ,    0.04  & $ 15.88 $ & $ 15.80  $ & $ 15.99 $ \\
0653+0239 & 3126316040609035008 &  DA & 19.43 ,  0.07  &    -2.32 ,    0.07  &   207.02 ,    0.06  & $ 17.00 $ & $ 17.23  $ & $ 16.61 $ \\
0715$-$2342 & 5617978855684923136 &  DA &  8.95 ,  0.05  &    -7.10 ,    0.03  &   -20.10 ,    0.05  & $ 16.51 $ & $ 16.48  $ & $ 16.59 $ \\
0746$-$0642 & 3043761409059739776 &  DA &  9.87 ,  0.05  &     8.26 ,    0.05  &   -27.35 ,    0.04  & $ 16.56 $ & $ 16.54  $ & $ 16.64 $ \\
0806$-$0006 & 3083590225639363712 &  DA & 16.32 ,  0.07  &   105.13 ,    0.07  &  -232.60 ,    0.04  & $ 16.28 $ & $ 16.28  $ & $ 16.05 $ \\
0807$-$3622 & 5544588928434374016 &  DA & 20.52 ,  0.03  &   -38.38 ,    0.02  &   102.73 ,    0.03  & $ 14.51 $ & $ 14.45  $ & $ 14.65 $ \\
0818+1211 & 649472245695986816 &  DA &  3.53 ,  0.10  &   -16.99 ,    0.12  &    -2.52 ,    0.07  & $ 17.44 $ & $ 17.31  $ & $ 17.74 $ \\
0847$-$1859 & 5705571754442134400 &  DB & 11.59 ,  0.04  &  -184.88 ,    0.04  &    34.64 ,    0.04  & $ 15.62 $ & $ 15.55  $ & $ 15.75 $ \\
1011+0536 & 3861237313489649280 &  DA &  5.24 ,  0.09  &     0.93 ,    0.09  &   -17.09 ,    0.07  & $ 17.00 $ & $ 16.98  $ & $ 17.07 $ \\
1024$-$0023 & 3830990156631488128 &  DC & 13.54 ,  0.12  &   -39.20 ,    0.12  &   -24.10 ,    0.15  & $ 17.09 $ & $ 17.14  $ & $ 17.01 $ \\
1029$-$3748 & 5442124615196939392 &  DA & 16.00 ,  0.03  &  -119.08 ,    0.03  &   -10.57 ,    0.03  & $ 15.73 $ & $ 15.73  $ & $ 15.74 $ \\
1054$-$4123 & 5390094110062904704 &  DA &  8.05 ,  0.05  &   -33.37 ,    0.03  &     5.64 ,    0.04  & $ 16.67 $ & $ 16.61  $ & $ 16.78 $ \\
1126$-$7631 & 5224570537243383296 &  DA &  6.91 ,  0.04  &   -67.11 ,    0.05  &    19.14 ,    0.05  & $ 16.27 $ & $ 16.13  $ & $ 16.37 $ \\
1140$-$3142 & 3478629978813108864 &  DA &  3.74 ,  0.07  &     4.72 ,    0.07  &    -1.41 ,    0.05  & $ 16.88 $ & $ 16.79  $ & $ 17.07 $ \\
1206$-$3805 & 3459790225027896448 &  DA & 17.01 ,  0.04  &   -82.13 ,    0.03  &   -46.88 ,    0.02  & $ 15.67 $ & $ 15.69  $ & $ 15.69 $ \\
1207$-$0412 & 3597718495533084416 &  DA &  7.85 ,  0.06  &    -8.69 ,    0.08  &     7.56 ,    0.06  & $ 16.66 $ & $ 16.63  $ & $ 16.76 $ \\
1234$-$4440 & 6133033635916500608 &  DC & 35.12 ,  0.04  &   -78.76 ,    0.04  &  -200.46 ,    0.05  & $ 16.18 $ & $ 16.40  $ & $ 15.83 $ \\
1247+1442 & 3930922058356114560 &  DA & 16.39 ,  0.05  &  -354.38 ,    0.05  &   148.67 ,    0.05  & $ 15.89 $ & $ 15.93  $ & $ 15.83 $ \\
1344$-$4154 & 6112283308883211264 &  DB &  6.00 ,  0.08  &   -46.93 ,    0.08  &     4.46 ,    0.06  & $ 16.88 $ & $ 16.80  $ & $ 17.03 $ \\
1350+0522 & 3720064899566194560 &  DA &  5.44 ,  0.08  &     0.52 ,    0.09  &   -56.86 ,    0.05  & $ 16.99 $ & $ 16.90  $ & $ 17.18 $ \\
1426$-$5716 & 5892401599200616064 &  DAH & 14.24 ,  0.05  &   -95.31 ,    0.04  &   -68.68 ,    0.05  & $ 16.38 $ & $ 16.39  $ & $ 16.41 $ \\
1434$-$3256 & 6216266734853792000 &  DA &  8.71 ,  0.08  &   -83.18 ,    0.09  &   -23.55 ,    0.10  & $ 16.82 $ & $ 16.81  $ & $ 16.86 $ \\
1524$-$2318 & 6251221515175765888 &  DA &  6.40 ,  0.08  &   -36.65 ,    0.09  &   -40.17 ,    0.08  & $ 16.45 $ & $ 16.36  $ & $ 16.57 $ \\
1559+1904 & 1203312813999856896 &  DBA &  7.11 ,  0.05  &    -7.39 ,    0.04  &   -45.72 ,    0.05  & $ 16.52 $ & $ 16.45  $ & $ 16.66 $ \\
1601$-$0345 & 4398887092042907904 &  DBA & 18.61 ,  0.04  &   -77.00 ,    0.05  &    31.18 ,    0.03  & $ 15.55 $ & $ 15.51  $ & $ 15.58 $ \\
1743+1434 & 4500646618315862144 &  DC & 29.13 ,  0.03  &  -256.64 ,    0.02  &   292.40 ,    0.03  & $ 14.85 $ & $ 14.87  $ & $ 14.79 $ \\
1805$-$0813 & 4170899954423211776 &  DZ & 14.45 ,  0.06  &   112.51 ,    0.06  &   -68.02 ,    0.05  & $ 16.44 $ & $ 16.46  $ & $ 16.41 $ \\
1841$-$5056 & 6655404287348972928 &  DA &  7.38 ,  0.06  &    -5.36 ,    0.07  &   -54.98 ,    0.07  & $ 16.74 $ & $ 16.70  $ & $ 16.87 $ \\
2041$-$2628 & 6799271058813651968 &  DC & 15.84 ,  0.09  &    38.57 ,    0.08  &  -111.98 ,    0.05  & $ 16.69 $ & $ 16.72  $ & $ 16.58 $ \\
2227$-$3411 & 6600365522695593472 &  DBA & 18.80 ,  0.03  &   208.61 ,    0.04  &   -24.83 ,    0.03  & $ 14.45 $ & $ 14.36  $ & $ 14.58 $ \\
2234+1456 & 2733055335904034432 &  DA & 29.52 ,  0.06  &   133.43 ,    0.08  &   -32.40 ,    0.07  & $ 16.50 $ & $ 16.77  $ & $ 16.09 $ \\
2245$-$1002 & 2608247533357159424 &  DA & 17.66 ,  0.09  &    54.05 ,    0.10  &   -47.43 ,    0.09  & $ 16.94 $ & $ 17.06  $ & $ 16.75 $ \\
2310$-$6850 & 6387649708219253248 &  DC & 47.46 ,  0.02  &    65.12 ,    0.02  &   320.69 ,    0.02  & $ 13.57 $ & $ 13.58  $ & $ 13.54 $ \\
 \noalign{\smallskip}
\hline
 \noalign{\smallskip}
HE\,0204$-$3821 & 4964509614631078272 &  DA & 10.56 ,  0.04  &    59.35 ,    0.03  &   -10.34 ,    0.04  & $ 16.25 $ & $ 16.23  $ & $ 16.25 $ \\
HE\,0330$-$4736 & 4833967725801565696 &  DA & 13.98 ,  0.03  &   -28.57 ,    0.03  &    56.86 ,    0.04  & $ 15.95 $ & $ 15.95  $ & $ 16.01 $ \\
HS\,0145+1737 & 91690164426711040 &  DA & 11.35 ,  0.05  &    60.28 ,    0.09  &   -11.13 ,    0.05  & $ 15.79 $ & $ 15.71  $ & $ 15.91 $ \\
WD\,0106$-$358 & 5014009353235167744 &  DA & 10.69 ,  0.04  &   -38.56 ,    0.02  &   -52.06 ,    0.03  & $ 14.70 $ & $ 14.56  $ & $ 14.98 $ \\
WD\,0204$-$306 & 5020119579868434944 &  DA & 34.19 ,  0.06  &   246.63 ,    0.07  &  -133.03 ,    0.06  & $ 16.70 $ & $ 17.06  $ & $ 16.18 $ \\
WD\,0628$-$020 & 3117320802840630400 &  DA & 46.72 ,  0.03  &  -113.14 ,    0.03  &  -172.04 ,    0.03  & $ 15.21 $ & $ 15.42  $ & $ 14.82 $ \\
WD\,1015$-$216 & 5666295485406677632 &  DA &  6.66 ,  0.05  &   -23.26 ,    0.05  &   -17.83 ,    0.05  & $ 15.65 $ & $ 15.51  $ & $ 15.94 $ \\
W\,D1105$-$048 & 3788194488314248832 &  DA & 40.29 ,  0.03  &   -55.55 ,    0.03  &  -442.63 ,    0.03  & $ 13.09 $ & $ 13.05  $ & $ 13.21 $ \\
WD\,1147+255 & 4005438916307756928 &  DA & 19.85 ,  0.05  &  -311.30 ,    0.05  &  -108.67 ,    0.05  & $ 15.66 $ & $ 15.71  $ & $ 15.56 $ \\
WD\,1348$-$273 & 6177238676273826304 &  DA & 26.59 ,  0.03  &    62.95 ,    0.04  &  -196.81 ,    0.04  & $ 15.12 $ & $ 15.18  $ & $ 15.01 $ \\
WD\,1544$-$377 & 6009537829925128064 &  DA & 65.69 ,  0.03  &  -423.69 ,    0.03  &  -209.11 ,    0.03  & $ 13.00 $ & $ 13.04  $ & $ 12.93 $ \\
WD\,1619+123 & 4460086458999242368 &  DA & 17.91 ,  0.03  &    63.66 ,    0.03  &   -66.72 ,    0.02  & $ 14.66 $ & $ 14.60  $ & $ 14.80 $ \\
WD\,1620$-$391 & 6018034958869558912 &  DA & 77.45 ,  0.04  &    77.40 ,    0.05  &     0.39 ,    0.04  & $ 11.00 $ & $ 10.88  $ & $ 11.23 $ \\
WD\,1911+135 & 4320094439580536320 &  DA & 29.65 ,  0.02  &    15.57 ,    0.02  &  -180.88 ,    0.02  & $ 14.09 $ & $ 14.07  $ & $ 14.17 $ \\
WD\,1932$-$136 & 4183272552601606400 &  DA &  9.54 ,  0.05  &   -36.24 ,    0.06  &  -124.18 ,    0.05  & $ 15.98 $ & $ 15.94  $ & $ 16.12 $ \\
WD\,2253+054 & 2711324446359728384 &  DA & 40.33 ,  0.07  &   350.81 ,    0.05  &  -264.71 ,    0.05  & $ 16.04 $ & $ 16.33  $ & $ 15.56 $ \\
WD\,2318+126 & 2811363550466857344 &  DA & 11.11 ,  0.06  &   198.59 ,    0.07  &    38.02 ,    0.05  & $ 16.29 $ & $ 16.27  $ & $ 16.35 $ \\
 \noalign{\smallskip}
\hline
    \noalign{\smallskip}
\multicolumn{9}{p{13cm}}{{\bf Notes.} The line separates the new targets (top) from the SPY sample (bottom); WD\,1544$-$377 was re-observed for comparison purposes.
The short name of the new targets is based on the {\it Gaia} DR3 coordinates in the {\it hhmm}$\pm${\it ddmm} format.
The white dwarf type is assigned upon visual inspection of the spectra.}
    \end{tabular}
    \label{tab:wd_gaia}
\end{table}
\begin{table}
\scriptsize
    \caption{Relevant {\it Gaia} parameters of the wide binary companions.}
    \centering  
    \begin{tabular}{@{}l  l m{5.4}  m{5.4}  m{5.4} c c c c c c m{5.4}@{}}
\hline
\hline
 \noalign{\smallskip}
Short Name & {\em Gaia} ID & \multicolumn{1}{c}{$\varpi$} & \multicolumn{1}{c}{$\mu_\alpha \cos{\delta}$} & \multicolumn{1}{c}{$\mu_\delta$} & $G$ & $BP$  & $RP$ 
 & Separation  & $a$ & $\Delta \varv$  & \multicolumn{1}{c}{$\varv_{\rm rad}$}\\
 &  & \multicolumn{1}{c}{(arcsec)} & \multicolumn{1}{c}{(mas/yr)} & \multicolumn{1}{c}{(mas/yr)} & (mag) & (mag) & (mag) & (arcsec) & (au) &  (km/s) & \multicolumn{1}{c}{(km/s)}\\
  \noalign{\smallskip}
\hline
 \noalign{\smallskip}
0003+1713$^{{n}}$ & 2772926200211837056 &   9.42 ,  0.02  &   -52.97 ,    0.02  &   -61.94 ,    0.02  & $ 12.59 $ & $ 13.32  $ & $ 11.76 $ & 12.7 & 1359 &  0.26 &  -18.76 ,  0.75 \\
0030$-$5856 & 4906307172772653056 &  12.04 ,  0.01  &    16.92 ,    0.01  &    -7.13 ,    0.01  & $ 11.13 $ & $ 11.70  $ & $ 10.42 $ & 32.4 & 2714 &  0.41 &  -19.50 ,  0.25 \\
0042$-$5820 & 4907169670925418880 &  16.70 ,  0.01  &   -91.86 ,    0.01  &   -68.41 ,    0.01  & $ 12.04 $ & $ 12.99  $ & $ 11.08 $ &  9.8 & 588 &  0.63 &  18.44 ,  0.32 \\
0115+1534 & 2591573199003071616 &   9.55 ,  0.03  &   -26.41 ,    0.03  &    -3.68 ,    0.02  & $ 11.71 $ & $ 12.29  $ & $ 10.99 $ &  6.4 & 679 &  0.53 &   2.22 ,  0.52 \\
0125$-$7619 & 4637598655635818368 &   5.64 ,  0.01  &     9.80 ,    0.01  &     0.81 ,    0.01  & $ 10.31 $ & $ 10.58  $ & $  9.88 $ &  9.8 & 1720 &  0.81 &  -2.14 ,  0.31 \\
0141+0711 & 2568552556549060224 &  12.75 ,  0.02  &   -61.27 ,    0.02  &   -78.95 ,    0.01  & $ 11.01 $ & $ 11.55  $ & $ 10.33 $ & 41.8 & 3267 &  0.31 &  -24.42 ,  0.34 \\
0156$-$4722 & 4942194923066210944 &  11.93 ,  0.01  &    32.54 ,    0.01  &     3.50 ,    0.01  & $ 11.51 $ & $ 12.12  $ & $ 10.77 $ & 174.8 & 14660 &  0.21 &  -10.44 ,  0.31 \\
0207+0355 & 2517771234063151360 &   4.94 ,  0.02  &     5.70 ,    0.02  &    -0.99 ,    0.01  & $ 13.09 $ & $ 13.64  $ & $ 12.39 $ & 52.4 & 9939 &  0.22 &   9.81 ,  1.57 \\
0209$-$0140 & 2494399362068211456 &  17.94 ,  0.02  &   -34.97 ,    0.02  &   -54.00 ,    0.01  & $  9.70 $ & $ 10.14  $ & $  9.10 $ & 20.5 & 1140 &  0.38 &  -17.08 ,  0.38 \\
0319$-$7254 & 4642153696087687168 &  10.43 ,  0.01  &   123.96 ,    0.01  &   -21.77 ,    0.01  & $ 10.53 $ & $ 11.01  $ & $  9.90 $ &  9.5 & 916 &  0.37 &  28.42 ,  0.24 \\
0439$-$0405 & 3201594035484285440 &  18.92 ,  0.01  &   100.44 ,    0.02  &   -38.00 ,    0.01  & $  8.79 $ & $  9.17  $ & $  8.25 $ & 37.4 & 1969 &  0.52 &  38.69 ,  0.14 \\
0608+0204 & 3315373564607612416 &  22.05 ,  0.02  &    18.43 ,    0.02  &   -78.69 ,    0.02  & $ 11.59 $ & $ 12.55  $ & $ 10.63 $ & 15.0 & 681 &  0.36 &  90.14 ,  0.91 \\
0608$-$0059 & 3121391740343635328 &  16.22 ,  0.02  &     3.26 ,    0.02  &   -36.32 ,    0.01  & $ 12.26 $ & $ 13.30  $ & $ 11.25 $ & 43.6 & 2694 &  0.22 &  28.80 ,  0.78 \\
0608$-$5301 & 5548648600228817792 &  17.41 ,  0.01  &  -142.24 ,    0.02  &   215.88 ,    0.02  & $ 12.42 $ & $ 13.50  $ & $ 11.38 $ & 21.3 & 1225 &  0.57 &  33.00 ,  0.42 \\
0616$-$5912$^{{n}}$ & 5482551183847322752 &  27.76 ,  0.02  &   -45.19 ,    0.02  &  -316.39 ,    0.02  & $  6.30 $ & $  6.59  $ & $  5.84 $ & 40.8 & 1471 &  0.90 &  -2.06 ,  0.13 \\
0621$-$5456 & 5499525238198195584 &  10.74 ,  0.02  &    28.29 ,    0.03  &   -91.77 ,    0.02  & $  8.30 $ & $  8.59  $ & $  7.84 $ & 13.4 & 1245 &  0.45 &  25.48 ,  0.18 \\
0653+0239 & 3126315044174917376 &  19.59 ,  0.02  &    -2.91 ,    0.02  &   207.50 ,    0.02  & $ 12.38 $ & $ 13.43  $ & $ 11.36 $ & 140.0 & 7204 &  0.18 &  59.42 ,  0.94 \\
0715$-$2342 & 5617978924404406144 &   9.03 ,  0.02  &    -6.94 ,    0.01  &   -19.16 ,    0.02  & $  8.21 $ & $  8.49  $ & $  7.76 $ & 25.3 & 2829 &  0.50 &  10.56 ,  0.12 \\
0746$-$0642 & 3043761413359412992 &   9.81 ,  0.02  &     8.05 ,    0.02  &   -27.63 ,    0.01  & $ 11.30 $ & $ 11.80  $ & $ 10.65 $ & 20.7 & 2100 &  0.17 &  -11.05 ,  0.31 \\
0806$-$0006 & 3083590225641742080 &  16.32 ,  0.02  &   110.41 ,    0.02  &  -230.10 ,    0.01  & $  9.68 $ & $ 10.13  $ & $  9.07 $ &  6.7 & 413 &  1.69 &  -29.94 ,  0.16 \\
0807$-$3622$^{{n}}$ & 5544588997155925248 &  20.60 ,  0.03  &   -37.48 ,    0.03  &   101.14 ,    0.03  & $  7.18 $ & $  7.44  $ & $  6.76 $ & 20.6 & 1003 &  0.42 &  26.12 ,  0.12 \\
0818+1211 & 649472486214811776 &   3.64 ,  0.02  &   -17.15 ,    0.03  &    -2.72 ,    0.02  & $ 10.48 $ & $ 10.68  $ & $ 10.13 $ & 66.3 & 18757 &  0.34 &  33.17 ,  1.12 \\
0847$-$1859$^{{n}}$ & 5705571861818011008 &  11.59 ,  0.02  &  -185.57 ,    0.02  &    35.53 ,    0.02  & $ 11.22 $ & $ 11.78  $ & $ 10.52 $ & 30.4 & 2620 &  0.46 &  65.18 ,  0.22 \\
1011+0536 & 3861237313489649536 &   5.49 ,  0.02  &     0.27 ,    0.02  &   -17.12 ,    0.02  & $ 10.95 $ & $ 11.27  $ & $ 10.46 $ & 17.8 & 3394 &  0.57 &   3.06 ,  0.57 \\
1024$-$0023 & 3830984079252585600 &  13.60 ,  0.03  &   -38.78 ,    0.03  &   -23.03 ,    0.03  & $  7.93 $ & $  8.19  $ & $  7.51 $ & 64.3 & 4747 &  0.40 &  17.40 ,  0.14 \\
1029$-$3748 & 5442124610898498816 &  16.05 ,  0.01  &  -118.75 ,    0.01  &    -9.63 ,    0.01  & $ 12.02 $ & $ 12.92  $ & $ 11.08 $ &  8.1 & 508 &  0.29 &   2.90 ,  0.52 \\
1054$-$4123 & 5390094110062904960 &   8.04 ,  0.01  &   -33.44 ,    0.01  &     4.31 ,    0.01  & $ 13.16 $ & $ 14.00  $ & $ 12.27 $ &  8.9 & 1105 &  0.78 &  -0.21 ,  1.04 \\
1126$-$7631 & 5224570537243382784 &   6.99 ,  0.01  &   -67.93 ,    0.01  &    19.73 ,    0.01  & $ 11.45 $ & $ 11.89  $ & $ 10.85 $ &  5.7 & 821 &  0.68 &  -0.71 ,  0.42 \\
1140$-$3142 & 3478630047532585216 &   3.81 ,  0.01  &     5.00 ,    0.01  &    -1.04 ,    0.01  & $ 10.02 $ & $ 10.37  $ & $  9.49 $ & 12.9 & 3442 &  0.57 &  15.45 ,  0.16 \\
1206$-$3805 & 3459790018869465984 &  16.98 ,  0.02  &   -82.78 ,    0.01  &   -46.44 ,    0.01  & $  9.15 $ & $  9.53  $ & $  8.59 $ & 56.3 & 3309 &  0.22 &  10.15 ,  0.14 \\
1207$-$0412 & 3597718564252561408 &   7.78 ,  0.02  &    -8.79 ,    0.02  &     8.30 ,    0.02  & $ 12.75 $ & $ 13.47  $ & $ 11.93 $ & 18.5 & 2352 &  0.45 &  -4.04 ,  1.10 \\
1234$-$4440$^{{n}}$ & 6133033601555979648 &  35.14 ,  0.04  &   -76.82 ,    0.03  &  -206.79 ,    0.03  & $  5.61 $ & $  5.94  $ & $  5.10 $ & 38.1 & 1085 &  0.89 &  18.05 ,  0.12 \\
1247+1442 & 3930922058356114432 &  16.50 ,  0.02  &  -354.72 ,    0.02  &   149.31 ,    0.02  & $ 12.78 $ & $ 13.78  $ & $ 11.78 $ & 25.1 & 1530 &  0.21 &  -24.43 ,  1.59 \\
1344$-$4154 & 6112283308883213696 &   6.08 ,  0.01  &   -46.28 ,    0.01  &     5.38 ,    0.01  & $ 13.06 $ & $ 13.92  $ & $ 12.14 $ &  9.6 & 1596 &  0.88 &   3.39 ,  2.01 \\
1350+0522 & 3720062288226076672 &   5.46 ,  0.02  &     0.61 ,    0.02  &   -56.81 ,    0.01  & $ 12.16 $ & $ 12.62  $ & $ 11.54 $ & 219.3 & 40316 &  0.09 &  -11.38 ,  0.64 \\
1426$-$5716$^{{n}}$ & 5892398644263108096 &  14.22 ,  0.02  &   -94.94 ,    0.02  &   -69.22 ,    0.02  & $  7.74 $ & $  7.98  $ & $  7.33 $ & 46.4 & 3257 &  0.22 &  -11.17 ,  0.13 \\
1434$-$3256 & 6216266734857178624 &   8.53 ,  0.02  &   -83.79 ,    0.02  &   -23.11 ,    0.02  & $ 13.12 $ & $ 13.99  $ & $ 12.20 $ &  5.7 & 654 &  0.42 &  -22.95 ,  2.60 \\
1524$-$2318 & 6251221515175765632 &   6.34 ,  0.03  &   -35.99 ,    0.03  &   -40.34 ,    0.03  & $ 11.47 $ & $ 11.88  $ & $ 10.89 $ &  6.5 & 1015 &  0.51 &  -15.78 ,  0.70 \\
1559+1904 & 1203312813999858304 &   7.10 ,  0.02  &    -7.61 ,    0.01  &   -45.90 ,    0.02  & $ 11.50 $ & $ 11.96  $ & $ 10.87 $ & 28.9 & 4063 &  0.19 &   0.12 ,  0.41 \\
1601$-$0345$^{{n}}$ & 4398887092042905856 &  18.65 ,  0.02  &   -74.31 ,    0.03  &    29.49 ,    0.02  & $  6.77 $ & $  6.97  $ & $  6.43 $ & 22.9 & 1229 &  0.81 &  -20.01 ,  0.16 \\
1640$-$7524 & 5781107795253788928 &   2.02 ,  0.01  &    -2.11 ,    0.01  &    -1.95 ,    0.01  & $ 12.20 $ & $ 12.48  $ & $ 11.76 $ &  5.3 & 2657 &  1.05 &  -8.98 ,  1.31 \\
1743+1434 & 4500646618315862400 &  29.08 ,  0.02  &  -262.08 ,    0.02  &   289.48 ,    0.02  & $ 12.65 $ & $ 14.05  $ & $ 11.48 $ & 10.2 & 349 &  1.01 &  -70.20 ,  0.46 \\
1805$-$0813 & 4170525360244678016 &  14.53 ,  0.02  &   113.10 ,    0.02  &   -68.15 ,    0.02  & $ 11.09 $ & $ 11.76  $ & $ 10.31 $ & 666.2 & 46109 &  0.19 &   4.47 ,  0.44 \\
1841$-$5056 & 6655405013202948096 &   7.53 ,  0.02  &    -5.45 ,    0.02  &   -54.87 ,    0.02  & $ 11.11 $ & $ 11.49  $ & $ 10.56 $ & 84.5 & 11448 &  0.09 &  11.88 ,  0.60 \\
2041$-$2628 & 6799271058813652864 &  15.83 ,  0.02  &    37.20 ,    0.02  &  -111.27 ,    0.01  & $ 10.55 $ & $ 11.13  $ & $  9.83 $ & 10.0 & 631 &  0.46 &   6.77 ,  0.18 \\
2227$-$3411 & 6600365518402656256 &  18.88 ,  0.02  &   212.37 ,    0.02  &   -25.52 ,    0.01  & $ 12.15 $ & $ 13.22  $ & $ 11.13 $ &  8.7 & 460 &  0.96 &  -10.71 ,  0.46 \\
2234+1456 & 2733058393920750080 &  29.55 ,  0.02  &   134.06 ,    0.02  &   -30.98 ,    0.02  & $ 12.13 $ & $ 13.30  $ & $ 11.05 $ & 65.3 & 2212 &  0.25 &  -8.53 ,  0.43 \\
2245$-$1002 & 2608249015120878976 &  17.70 ,  0.02  &    55.67 ,    0.03  &   -46.75 ,    0.02  & $  9.99 $ & $ 10.47  $ & $  9.34 $ & 60.5 & 3428 &  0.47 &   1.65 ,  0.25 \\
2310$-$6850 & 6387649914378668160 &  47.48 ,  0.02  &    66.96 ,    0.02  &   319.37 ,    0.02  & $  8.38 $ & $  8.94  $ & $  7.67 $ & 12.9 & 272 &  0.23 &   5.21 ,  0.13 \\
HE\,0204$-$3821 & 4964509614631078400 &  10.60 ,  0.03  &    61.34 ,	0.02  &   -11.13 ,    0.02  & $ 15.26 $ & $ 16.60  $ & $ 14.11 $ &  2.9 & 271 &  0.96 &   6.93 ,  4.90 \\
HE\,0330$-$4736 & 4833967687145469696 &  14.05 ,  0.01  &   -28.91 ,	0.01  &    55.95 ,    0.01  & $ 12.13 $ & $ 13.00  $ & $ 11.22 $ & 20.3 & 1450 &  0.33 &  11.73 ,  0.55 \\
HS\,0145+1737 & 91690164426524288 &  11.28 ,  0.02  &    60.70 ,    0.03  &   -10.82 ,    0.02  & $ 11.63 $ & $ 12.30  $ & $ 10.85 $ &  9.4 & 832 &  0.22 &  -4.97 ,  0.48 \\
WD\,0106$-$358 & 5014008975278023040 &  10.73 ,  0.02  &   -38.26 ,    0.01  &   -52.46 ,    0.02  & $ 14.72 $ & $ 16.04  $ & $ 13.58 $ & 111.1 & 10390 &  0.22 &  11.89 ,  2.14 \\
WD\,0204$-$306 & 5020116556211470848 &  34.30 ,  0.02  &   247.00 ,    0.03  &  -130.07 ,    0.02  & $ 12.89 $ & $ 14.36  $ & $ 11.71 $ & 72.8 & 2129 &  0.41 &  51.56 ,  0.77 \\
WD\,0628$-$020$^{{n}}$ & 3117320802840630784 &  46.80 ,  0.02  &  -116.08 ,    0.02  &  -161.37 ,    0.02  & $ 13.88 $ & $ 15.73  $ & $ 12.60 $ &  4.5 & 95 &  1.12 &  81.42 ,  2.39 \\
WD\,1015$-$216 & 5666295485406677376 &   6.71 ,  0.03  &   -23.55 ,    0.03  &   -18.21 ,    0.03  & $ 15.16 $ & $ 16.32  $ & $ 14.10 $ & 13.4 & 2007 &  0.34 &  -26.15 ,  6.74 \\
WD\,1105$-$048 & 3788190605663811840 &  40.18 ,  0.03  &   -55.03 ,    0.03  &  -439.74 ,    0.02  & $ 11.47 $ & $ 12.80  $ & $ 10.33 $ & 279.0 & 6924 &  0.35 &  22.83 ,  0.23 \\
WD\,1147+255 & 4005438881948018176 &  19.74 ,  0.03  &  -312.29 ,    0.03  &  -109.02 ,    0.03  & $ 14.24 $ & $ 15.74  $ & $ 13.03 $ & 36.8 & 1855 &  0.25 &  -65.46 ,  4.74 \\
WD\,1348$-$273 & 6177238671977087360 &  26.68 ,  0.02  &    71.42 ,    0.02  &  -197.31 ,    0.02  & $ 11.63 $ & $ 12.73  $ & $ 10.59 $ &  9.1 & 342 &  1.51 &  26.51 ,  0.50 \\
WD\,1544$-$377$^{{n}}$ & 6009538585839374336 &  65.59 ,  0.03  &  -415.51 ,    0.04  &  -213.99 ,    0.02  & $  5.83 $ & $  6.19  $ & $  5.31 $ & 14.5 & 220 &  0.69 &  -6.92 ,  0.12 \\
WD\,1619+123$^{{n}}$ & 4460085703084994944 &  17.94 ,  0.02  &    64.90 ,    0.02  &   -68.12 ,    0.02  & $  8.05 $ & $  8.32  $ & $  7.61 $ & 62.5 & 3487 &  0.49 &  -6.70 ,  0.14 \\
WD\,1620$-$391$^{{n}}$ & 6018047019138644480 &  77.57 ,  0.07  &    73.75 ,    0.08  &     3.37 ,    0.06  & $  5.22 $ & $  5.54  $ & $  4.73 $ & 345.0 & 4454 &  0.29 &  12.85 ,  0.13 \\
WD\,1911+135 & 4320094435229636224 &  29.67 ,  0.02  &    14.95 ,    0.02  &  -180.83 ,    0.02  & $ 11.78 $ & $ 12.96  $ & $ 10.70 $ & 17.8 & 599 &  0.10 &  -11.59 ,  0.30 \\
WD\,1932$-$136 & 4183273136713242624 &   9.57 ,  0.02  &   -36.22 ,    0.02  &  -125.22 ,    0.02  & $ 12.91 $ & $ 13.84  $ & $ 11.95 $ & 28.7 & 3006 &  0.52 &  -22.01 ,  0.89 \\
WD\,2253+054 & 2711324381934051200 &  40.35 ,  0.02  &   355.83 ,    0.02  &  -268.96 ,    0.02  & $ 10.42 $ & $ 11.46  $ & $  9.40 $ & 17.2 & 425 &  0.77 &   9.48 ,  0.23 \\
WD\,2318+126 & 2811363584826561408 &  11.13 ,  0.04  &   198.59 ,    0.04  &    37.52 ,    0.03  & $ 15.55 $ & $ 17.13  $ & $ 14.34 $ & 35.4 & 3184 &  0.21 &  -37.17 ,  5.61 \\
 \noalign{\smallskip}
\hline
    \noalign{\smallskip}
\multicolumn{12}{l}{{\bf Notes.} The listed short names refer to the corresponding white dwarf companions; ${n}$: no deblended transits.}
    \end{tabular}
    \label{tab:nwd_gaia}
\end{table}
\twocolumn
\newpage
\onecolumn
\section{Observing logs}
\begin{table}[h!]
\caption{Observing logs}
  \tiny
    \centering
    \begin{tabular}{@{}lcccccccc@{}}
    \hline
    \hline
    \noalign{\smallskip}
     & & & & & \multicolumn{2}{c}{Blue Arm} & \multicolumn{2}{c}{Red Arm} \\
      Short Name & Date & Exp. Time & Airmass & Seeing & Range & SNR &  Range & SNR \\
        &   & (sec) &  & (arcsec) & (nm) &  &  (nm) &  \\

      \noalign{\smallskip}
      \hline
      \noalign{\smallskip}
 0003+1713  & 20210930 & 1800 & 2.02 & 1.62 & 328$-$456  &  7  & 458-668  &  11  \\           
            & 20210930 & 1800 & 1.46 & 1.32 & 328$-$456  &  9  & 458-668  &  13  \\
 0030$-$5856  & 20211001 & 1200 & 1.50 & 1.14 & 328$-$456  &  10  & 458-668  &  12  \\
 0042$-$5820  & 20210921 & 1800 & 1.53 & 0.45 & 328$-$456  &  12  & 472-683  &  15  \\
 0115+1534  & 20210928 & 1200 & 1.83 & 0.93 & 328$-$456  &  14  & 458-668  &  17  \\
 0125$-$7619  & 20211001 & 3000 & 1.90 & 1.12 & 328$-$456  &  10  & 458-668  &  12  \\
 0141+0711  & 20210930 & 3000 & 1.24 & 0.93 & 328$-$456  &  13  & 458-668  &  16  \\
 0156$-$4722  & 20210930 & 1800 & 1.18 & 1.40 & 328$-$456  &  10  & 458-668  &  15  \\
 0207+0355  & 20210928 & 3000 & 1.17 & 1.07 & 328$-$456  &  3  & 458-668  &  4  \\
           & 20210928 & 3000 & 1.14 & 1.26 & 328$-$456  &  3  & 458-668  &  3  \\
           & 20210928 & 3000 & 1.18 & 0.78 & 328$-$456  &  3  & 458-668  &  4  \\
 0209$-$0140  & 20210930 & 1800 & 1.30 & 0.99 & 328$-$456  &  10  & 458-668  &  16  \\
 0319$-$7254  & 20211010 & 3000 & 1.62 & 1.26 & 328$-$456  &  9  & 458-668  &  15  \\
 0439$-$0405  & 20211101 & 2400 & 1.14 & 2.35 & 328$-$456  &  5  & 458-668  &  9  \\
 0608+0204  & 20211203 & 1800 & 1.21 & 1.76 & 328$-$456  &  8  & 458-668  &  12  \\
 0608$-$0059  & 20220112 & 3600 & 1.16 & 0.60 & 328$-$456  &  5  & 458-668  &  4  \\
 0608$-$5301  & 20211212 & 1200 & 1.37 & 1.56 & 328$-$456  &  7  & 458-668  &  10  \\
           & 20211211 & 1200 & 1.31 & 1.84 & 328$-$456  &  6  & 458-668  &  13  \\
 0616$-$5912  & 20211011 & 600 & 1.25 & 1.69 & 328$-$456  &  23  & 458-668  &  27  \\
 0621$-$5456  & 20211021 & 1200 & 1.20 & 2.52 & 328$-$456  &  5  & 458-668  &  8  \\
           & 20211021 & 1200 & 1.18 & 1.67 & 328$-$456  &  7  & 458-668  &  10  \\
 0653+0239  & 20220119 & 3600 & 1.17 & 0.62 & 328$-$456  &  5  & 458-668  &  9  \\
 0715$-$2342  & 20220113 & 3000 & 1.20 & 0.70 & 328$-$456  &  10  & 458-668  &  15  \\
 0746$-$0642  & 20220117 & 3000 & 1.61 & 0.62 & 328$-$456  &  8  & 458-668  &  13  \\
 0806$-$0006  & 20220116 & 1800 & 1.18 & 0.32 & 328$-$456  &  8  & 458-668  &  11  \\
 0807$-$3622  & 20211023 & 600 & 1.16 & 0.64 & 328$-$456  &  14  & 458-668  &  19  \\
 0818+1211  & 20220113 & 3000 & 1.55 & 0.52 & 328$-$456  &  7  & 458-668  &  7  \\
           & 20220113 & 3000 & 1.31 & 0.41 & 328$-$456  &  9  & 458-668  &  9  \\
           & 20220116 & 3000 & 2.09 & 0.55 & 328$-$456  &  4  & 458-668  &  4  \\
 0847$-$1859  & 20220110 & 1200 & 1.10 & 1.33 & 328$-$456  &  11  & 458-668  &  13  \\
 1011+0536  & 20220116 & 3600 & 1.21 & 0.45 & 328$-$456  &  7  & 458-668  &  12  \\
 1024$-$0023  & 20220117 & 3600 & 2.39 & 0.48 & 328$-$456  &  5  & 458-668  &  7  \\
 1029$-$3748  & 20220117 & 1200 & 2.11 & 0.39 & 328$-$456  &  7  & 458-668  &  12  \\
 1054$-$4123  & 20220118 & 3000 & 2.07 & 0.68 & 328$-$456  &  6  & 458-668  &  9  \\
 1126$-$7631  & 20220116 & 1800 & 2.27 & 0.57 & 328$-$456  &  7  & 458-668  &  9  \\
 1140$-$3142  & 20220117 & 3000 & 1.84 & 0.56 & 328$-$456  &  8  & 458-668  &  10  \\
 1206$-$3805  & 20220116 & 1200 & 1.96 & 0.35 & 328$-$456  &  8  & 458-668  &  15  \\
 1207$-$0412  & 20220117 & 3000 & 1.96 & 0.32 & 328$-$456  &  8  & 458-668  &  13  \\
 1234$-$4440  & 20220116 & 1800 & 1.93 & 0.47 & 328$-$456  &  6  & 458-668  &  13  \\
 1247+1442  & 20220116 & 1200 & 1.42 & 0.87 & 328$-$456  &  8  & 458-668  &  13  \\
 1344$-$4154  & 20220118 & 3600 & 1.88 & 0.92 & 328$-$456  &  4  & 458-668  &  6  \\
 1350+0522  & 20220127 & 3600 & 2.15 & 0.52 & 328$-$456  &  9  & 458-668  &  13  \\
 1426$-$5716  & 20220117 & 3000 & 2.23 & 0.39 & 328$-$456  &  5  & 458-668  &  10  \\
 1434$-$3256  & 20220119 & 3000 & 2.37 & 0.52 & 328$-$456  &  4  & 458-668  &  8  \\
 1524$-$2318  & 20220119 & 1800 & 1.86 & 0.57 & 328$-$456  &  6  & 458-668  &  9  \\
 1559+1904  & 20220227 & 3000 & 1.97 & 0.40 & 328$-$456  &  12  & 458-668  &  15  \\
 1601$-$0345  & 20220124 & 1200 & 2.27 & 0.87 & 328$-$456  &  7  & 458-668  &  10  \\
 1743+1434  & 20220227 & 1200 & 2.37 & 0.46 & 328$-$456  &  12  & 458-668  &  18  \\
 1805$-$0813  & 20220228 & 1800 & 1.68 & 0.38 & 328$-$456  &  9  & 458-668  &  13  \\
 1841$-$5056  & 20220317 & 1200 & 2.15 & 1.52 & 328$-$456  &  3  & 458-668  &  7  \\
 2041$-$2628  & 20211011 & 3000 & 1.01 & 0.70 & 328$-$456  &  12  & 458-668  &  13  \\
 2227$-$3411  & 20211009 & 600 & 1.05 & 1.12 & 328$-$456  &  19  & 458-668  &  22  \\
 2234+1456  & 20211011 & 1800 & 1.49 & 0.76 & 328$-$456  &  6  & 458-668  &  10  \\
 2245$-$1002  & 20210930 & 2527 & 1.07 & 1.25 & 328$-$456  &  6  & 458-668  &  11  \\
           & 20211011 & 3000 & 1.04 & 1.01 & 328$-$456  &  8  & 458-668  &  10  \\
 2310$-$6850  & 20220118 & 600 & 2.25 & 0.61 & 304$-$391  &  10  & 472-683  &  27  \\
 WD1544$-$377  & 20220124 & 600 & 1.75 & 1.08 & 328$-$456  &  15  & 458-668  &  13  \\
         \noalign{\smallskip}
                  \hline
\multicolumn{8}{l}{Notes: SNR is the signal-to-noise ratio per pixel (blue arm: 0.045\,\AA/pixel; red arm 0.03\,\AA/pixel).}
    \end{tabular}
    \label{tab:logs}
\end{table}
\twocolumn
\newpage
\onecolumn
\begin{landscape}
\section{Spectroscopic parameters}
\vspace{-0.3cm}
\begin{table}[h!]
    \caption{Spectroscopic parameters.}
    \scriptsize
    \centering
    \begin{tabular}{@{}l m{3.2} m{3.1} s{4.1} s{4.1} m{3.4} m{3.2} m{3.3} | m{4.2} m{3.2} m{4.4} m{3.2} m{3.2} | m{3.3} m{3.3}@{}}
    \hline
    \hline
    \noalign{\smallskip}
      & \multicolumn{7}{c}{Koester 1D Models} & \multicolumn{5}{c}{Tremblay 3D Models} & \multicolumn{2}{c}{Gaussian fit} \\
Short Name &  \multicolumn{1}{c}{$T_{\rm eff}$} & \multicolumn{1}{c}{$\log{g}$}& \multicolumn{1}{c}{$\Delta T_{\rm eff,\,3D}$} & \multicolumn{1}{c}{$\Delta \log{g}_{\rm 3D}$} & \multicolumn{1}{c}{Mass} & \multicolumn{1}{c}{$\varv$\,(H$\alpha$+H$\beta$)$^{a}_{\rm WD}$} &   \multicolumn{1}{c}{$\varv({\rm H}\alpha)^b_{\rm WD}$} &  \multicolumn{1}{c}{$T_{\rm eff}$} & \multicolumn{1}{c}{$\log{g}$} & \multicolumn{1}{c}{Mass} & \multicolumn{1}{c}{$\varv$\,(H$\alpha$+H$\beta$)$^{a}_{\rm WD}$} &  \multicolumn{1}{c}{$\varv({\rm H}\alpha)^b_{\rm WD}$}  & \multicolumn{1}{c}{$\varv$\,(H$\alpha$+H$\beta$)$^{a}_{\rm WD}$}  & \multicolumn{1}{c}{$\varv({\rm H}\alpha)^b_{\rm WD}$}  \\
 &  \multicolumn{1}{c}{(K)} & \multicolumn{1}{c}{(dex)}& \multicolumn{1}{c}{(K)} & \multicolumn{1}{c}{(dex)} & \multicolumn{1}{c}{(M$_\odot$)} & \multicolumn{1}{c}{(km/s)} &   \multicolumn{1}{c}{(km/s)} &  \multicolumn{1}{c}{(K)} & \multicolumn{1}{c}{(dex)} & \multicolumn{1}{c}{(M$_\odot$)} & \multicolumn{1}{c}{(km/s)} &  \multicolumn{1}{c}{(km/s)}  & \multicolumn{1}{c}{(km/s)}  & \multicolumn{1}{c}{(km/s)} \\
          \noalign{\smallskip}
      \hline
      \noalign{\smallskip}
0003+1713 & 16990 , 85 &  7.71 , 0.02  &       &         & 0.491, 0.005 &  9.7 ,  1.2^{g} & 11.0 ,  1.6^{g}  & 17065 , 105 &  7.78 , 0.02  & 0.514 , 0.006 &  8.5 ,  2.0^{g} & 11.2 ,  1.3^{g}  &  7.6,   2.3 & 10.8,  0.6\\
0042$-$5820 & 10145 , 20 &  8.37 , 0.01  &  -160, & -0.26, & 0.656, 0.008 & 53.1 ,  1.8^{g} & 53.3 ,  1.9^{g}  & 11460 , 45 &  8.07 , 0.01  & 0.641 , 0.005 & 53.0 ,  1.6^{} & 54.0 ,  2.5^{}  &  54.7,   2.8 & 52.9,  1.7\\
0115+1534 & 26810 , 90 &  8.01 , 0.01  &       &         & 0.640, 0.007 & 39.5 ,  1.6^{gl} & 41.4 ,  2.5^{gl}  & 26545 , 80 &  8.00 , 0.02  & 0.631 , 0.008 & 38.1 ,  2.1^{g} & 40.8 ,  2.8^{g}  & 37.6,  2.5 & 40.3,  1.4\\
0125$-$7619 & 23200 , 175 &  7.86 , 0.03  &       &         & 0.558, 0.013 & 24.2 ,  2.1^{gl} & 26.2 ,  1.9^{gl}  & 23470 , 165 &  7.88 , 0.03  & 0.566 , 0.014 & 24.4 ,  2.3^{g} & 26.0 ,  2.4^{g}  &  23.1,  2.2 & 25.6,  1.2\\
0156$-$4722 & 11845 , 80 &  8.12 , 0.02  &  -295, & -0.07, & 0.628, 0.012 & 14.6 ,  1.3^{g} & 15.2 ,  1.5^{g}  & 13000 , 40 &  7.88 , 0.01  & 0.542 , 0.005 & 14.2 ,  1.6^{g} & 15.8 ,  2.0^{g}  &  13.7,   1.0 & 14.8,  0.6\\
0207+0355 & 15215 , 215 &  8.96 , 0.05  &       &         & 1.153, 0.021 & 100.2 ,  5.0^{} & 100.2 ,  5.4^{}  & 15860 , 310 &  8.88 , 0.05  & 1.116 , 0.021 & 100.3 ,  4.4^{} & 100.3 ,  4.6^{}  & 99.5,   6.1 & 92.1,  2.6\\
0209$-$0140 & 9300 , 15 &  8.29 , 0.01  &  -85, & -0.27, & 0.606, 0.007 & 13.4 ,  1.4^{gl} & 14.8 ,  1.6^{gl}  & 9300 , 15 &  7.90 , 0.02  & 0.541 , 0.008 & 13.2 ,  0.9^{} & 14.8 ,  1.4^{}  & 13.1, 1.2 & 14.3,  0.6\\
0319$-$7254 & 12230 , 60 &  8.07 , 0.02  &  -255, & -0.04, & 0.618, 0.013 & 53.1 ,  1.2^{g} & 53.9 ,  1.2^{g}  & 13455 , 410 &  7.92 , 0.14  & 0.558 , 0.069 & 52.8 ,  1.5^{g} & 53.9 ,  2.0^{g}  &  53.2,   1.7 & 53.9,  0.8\\
0439$-$0405 & 8000 , 20 &  8.06 , 0.03  &  0, & -0.17, & 0.534, 0.013 & 65.8 ,  1.7^{g} & 67.3 ,  1.9^{g}  & 8175 , 20 &  7.93 , 0.02  & 0.554 , 0.012 & 66.3 ,  1.3^{g} & 67.6 ,  1.9^{g}  & 64.9, 2.2 & 66.8,  1.3\\
0608$-$0059 & 21285 , 610 &  8.89 , 0.06  &       &         & 1.126, 0.028 & 207.8 ,  8.6^{} & 207.8 , 10.3^{}  & 20485 , 725 &  9.01 , 0.07  & 1.175 , 0.025 & 207.9 ,  7.9^{} & 207.9 ,  9.5^{}  & 207.4,   2.5 & 207.4,  1.8\\
0608$-$5301 & 9860 , 15 &  8.25 , 0.01  &  -145, & -0.26, & 0.591, 0.005 & 59.6 ,  1.7^{g} & 60.7 ,  1.9^{g}  & 10030 , 35 &  7.91 , 0.01  & 0.551 , 0.004 & 59.7 ,  0.9^{} & 61.5 ,  1.2^{}  & 58.8,   2.4 & 60.6,  1.2\\
0621$-$5456 & 17820 , 105 &  7.88 , 0.02  &       &         & 0.555, 0.010 & 54.9 ,  1.6^{g} & 54.4 ,  1.6^{g}  & 18055 , 130 &  7.93 , 0.02  & 0.578 , 0.009 & 54.8 ,  1.8^{g} & 54.5 ,  1.8^{g}  &  57.0,   3.0 & 54.2,  0.9\\
0653+0239 & 6480 , 25 &  7.78 , 0.03  &  25, & -0.03, & 0.471, 0.013 & 79.3 ,  1.6^{g} & 81.4 ,  1.8^{g}  & 6670 , 20 &  7.84 , 0.03  & 0.510 , 0.011 & 79.3 ,  1.8^{} & 82.4 ,  2.0^{}  &  77.6,   3.5 & 81.7,  1.3\\
0715$-$2342 & 15765 , 280 &  7.75 , 0.02  &       &         & 0.501, 0.008 & 39.2 ,  1.6^{g} & 39.7 ,  1.4^{g}  & 14590 , 355 &  7.91 , 0.03  & 0.562 , 0.015 & 39.2 ,  1.4^{g} & 40.1 ,  1.5^{g}  &  38.4,   1.8 & 39.5,  0.9\\
0746$-$0642 & 13145 , 85 &  8.12 , 0.04  &  5, & -0.01, & 0.663, 0.024 & 20.6 ,  1.6^{gl} & 21.2 ,  2.2^{gl}  & 13870 , 20 &  8.08 , 0.01  & 0.650 , 0.009 & 20.3 ,  1.7^{g} & 21.4 ,  1.8^{g}  &  20.2,   1.1 & 21.2,  0.9\\
0806$-$0006 & 9420 , 15 &  8.34 , 0.02  &  -100, & -0.27, & 0.633, 0.012 &  2.5 ,  1.7^{gl} &  4.0 ,  1.8^{gl}  & 9405 , 30 &  7.94 , 0.02  & 0.563 , 0.008 &  1.7 ,  1.5^{g} &  4.5 ,  1.7^{g}  &   0.6,   2.2 &  3.7,  0.2\\
0807$-$3622 & 17070 , 60 &  7.84 , 0.01  &       &         & 0.535, 0.005 & 55.7 ,  1.3^{g} & 54.6 ,  1.3^{g}  & 17135 , 70 &  7.89 , 0.01  & 0.555 , 0.005 & 55.4 ,  1.1^{g} & 54.8 ,  1.3^{g}  &  57.1,   2.4 & 54.3,  0.6\\
0818+1211 & 31850 , 145 &  8.26 , 0.03  &       &         & 0.785, 0.019 & 72.9 ,  7.2^{} & 66.4 , 10.7^{}  & 32775 , 205 &  8.21 , 0.04  & 0.760 , 0.020 & 71.3 ,  7.2^{} & 66.4 , 10.7^{}  &  79.8,   9.8 & 65.8,  2.5\\
1011+0536 & 17745 , 80 &  7.79 , 0.02  &       &         & 0.517, 0.007 & 14.3 ,  1.8^{g} & 15.0 ,  2.0^{g}  & 18040 , 110 &  7.84 , 0.02  & 0.539 , 0.007 & 13.7 ,  2.1^{g} & 15.4 ,  2.5^{g}  &  16.0,   2.2 & 18.7,  1.1\\
1029$-$3748 & 12015 , 100 &  8.28 , 0.03  &  -310, & -0.08, & 0.720, 0.016 & 38.8 ,  1.7^{g} & 39.8 ,  2.1^{g}  & 13625 , 20 &  8.01 , 0.02  & 0.612 , 0.010 & 37.8 ,  2.0^{g} & 39.5 ,  2.3^{g}  &  36.0,   3.0 & 39.5,  1.2\\
1054$-$4123 & 18485 , 235 &  7.86 , 0.03  &       &         & 0.549, 0.015 & 41.1 ,  2.8^{g} & 41.7 ,  3.2^{g}  & 18725 , 200 &  7.91 , 0.03  & 0.569 , 0.014 & 40.5 ,  2.7^{g} & 41.9 ,  2.9^{g}  &  39.9,   1.8 & 40.3,  1.5\\
1126$-$7631 & 24635 , 295 &  7.86 , 0.04  &       &         & 0.561, 0.018 & 33.7 ,  2.4^{g} & 34.3 ,  2.7^{g}  & 24795 , 300 &  7.86 , 0.04  & 0.563 , 0.015 & 33.4 ,  3.2^{g} & 34.3 ,  3.2^{g}  &  33.8,   1.9 & 34.2,  1.9\\
1140$-$3142 & 25830 , 155 &  7.51 , 0.02  &       &         & 0.452, 0.004 & 33.4 ,  2.3^{g} & 34.0 ,  2.6^{g}  & 25805 , 120 &  7.50 , 0.02  & 0.449 , 0.004 & 32.8 ,  4.4^{g} & 34.6 ,  5.7^{g}  &  34.4,   1.7 & 34.1,  1.4\\
1206$-$3805 & 11830 , 45 &  8.27 , 0.02  &  -290, & -0.09, & 0.706, 0.010 & 43.6 ,  1.5^{g} & 45.7 ,  2.2^{g}  & 13590 , 25 &  7.97 , 0.01  & 0.590 , 0.007 & 43.3 ,  2.2^{g} & 46.8 ,  2.7^{g}  &  42.3,   2.4 & 45.0,  1.1\\
1207$-$0412 & 15935 , 165 &  7.64 , 0.03  &       &         & 0.462, 0.010 & 19.5 ,  1.4^{g} & 20.4 ,  1.5^{g}  & 15540 , 390 &  7.74 , 0.05  & 0.496 , 0.019 & 19.4 ,  1.2^{g} & 20.6 ,  1.6^{g}  &  19.4,   1.0 & 19.8,  0.8\\
1247+1442 & 10255 , 30 &  8.27 , 0.02  &  -180, & -0.25, & 0.604, 0.009 &  2.5 ,  2.1^{g} &  5.8 ,  1.8^{g}  & 11420 , 30 &  8.03 , 0.01  & 0.615 , 0.006 &  1.7 ,  1.8^{g} &  9.0 ,  2.6^{g}  &   1.2,   4.0 &  6.9,  0.4\\
1350+0522 & 22075 , 135 &  7.91 , 0.02  &       &         & 0.577, 0.008 & 16.9 ,  1.6^{g} & 18.3 ,  1.5^{g}  & 22205 , 160 &  7.95 , 0.01  & 0.596 , 0.007 & 15.7 ,  2.4^{g} & 19.4 ,  3.1^{g}  &  13.9,   3.2 & 18.3,  1.0\\
1426$-$5716$^{c}$ &  &  & & & & & &  &  &  & & &  & 30.3,  1.7\\
1434$-$3256 & 13250 , 100 &  8.29 , 0.03  &  -60, & -0.01, & 0.770, 0.019 &  3.4 ,  1.9^{g} &  3.2 ,  1.9^{g}  & 14100 , 50 &  8.17 , 0.03  & 0.707 , 0.020 &  6.3 ,  3.0^{} &  6.7 ,  3.2^{}  &   3.8,   1.6 &  6.0,  0.4\\
1524$-$2318 & 25570 , 285 &  7.96 , 0.03  &       &         & 0.609, 0.013 & 15.4 ,  4.8^{} & 15.8 ,  6.6^{}  & 25605 , 315 &  7.95 , 0.04  & 0.605 , 0.019 & 14.2 ,  5.1^{} & 15.8 ,  6.4^{}  &  10.4,   1.5 & 12.3,  0.8\\
1841$-$5056 & 16915 , 570 &  7.97 , 0.09  &       &         & 0.596, 0.048 & 38.3 ,  2.6^{g} & 38.8 ,  2.8^{g}  & 14970 , 495 &  8.32 , 0.10  & 0.801 , 0.059 & 38.7 ,  2.7^{g} & 39.1 ,  2.8^{g}  &  39.4,   4.0 & 39.4,  1.0\\
2234+1456 & 6320 , 15 &  8.02 , 0.04  &  20, & -0.01, & 0.591, 0.023 & 22.3 ,  1.5^{g} & 24.4 ,  2.0^{g}  & 6440 , 15 &  8.06 , 0.04  & 0.620 , 0.022 & 21.5 ,  1.6^{g} & 24.6 ,  2.0^{g}  &  20.3,   3.6 & 25.3,  1.3\\
2245$-$1002 & 8500 , 20 &  8.55 , 0.01  &  -5, & -0.22, & 0.792, 0.007 & 52.2 ,  1.1^{g} & 55.3 ,  1.5^{g}  & 8695 , 25 &  8.28 , 0.01  & 0.763 , 0.008 & 52.8 ,  1.1^{g} & 55.7 ,  1.3^{g}  &  52.6,   1.8 & 54.6,  1.0\\
WD1544$-$377 & 10260 , 150 &  8.37 , 0.03  &  -170, & -0.26, & 0.656, 0.019 & 21.7 ,  1.3^{g} & 22.3 ,  2.0^{g}  & 10625 , 155 &  8.02 , 0.03  & 0.609 , 0.017 & 21.4 ,  1.1^{g} & 22.7 ,  1.1^{g}  &  21.9,   1.8 & 21.9,  0.8\\
HE0204$-$3821 & 13310 , 70 &  7.96 , 0.03  &  110, & 0.01, & 0.590, 0.015 & 37.1 ,  1.6^{g} & 37.8 ,  1.8^{g}  & 14565 , 200 &  7.87 , 0.02  & 0.542 , 0.009 & 36.8 ,  1.6^{g} & 37.9 ,  1.9^{g}  &  35.7,   0.7 & 37.6,  0.7\\
HE0330$-$4736 & 12350 , 105 &  8.31 , 0.03  &  -265, & -0.06, & 0.751, 0.021 & 45.9 ,  1.8^{g} & 45.5 ,  1.9^{g}  & 13655 , 30 &  8.14 , 0.03  & 0.687 , 0.018 & 45.5 ,  2.0^{g} & 45.7 ,  2.2^{g}  &  46.1,   0.9 & 45.0,  0.9\\
HS0145+1737 & 18280 , 265 &  7.90 , 0.04  &       &         & 0.565, 0.019 & 23.8 ,  1.7^{g} & 23.3 ,  2.4^{g}  & 18560 , 260 &  7.96 , 0.04  & 0.592 , 0.020 & 23.7 ,  1.7^{g} & 23.4 ,  2.2^{g}  &  25.4,   0.9 & 23.0,  0.9\\
WD0106$-$358 & 30270 , 165 &  7.84 , 0.02  &       &         & 0.567, 0.007 & 34.8 ,  2.4^{g} & 36.9 ,  2.7^{g}  & 30350 , 230 &  7.82 , 0.02  & 0.557 , 0.007 & 34.6 ,  2.1^{g} & 37.0 ,  2.6^{g}  &  32.4,   1.5 & 36.3,  1.5\\
WD0204$-$306 & 5845 , 45 &  8.38 , 0.08  &  -5, & 0.03, & 0.837, 0.053 & 85.1 ,  4.4^{} & 85.5 ,  5.0^{}  & 5865 , 35 &  8.37 , 0.06  & 0.816 , 0.041 & 84.3 ,  5.3^{} & 84.4 ,  6.7^{}  &  84.2,   2.2 & 82.6,  2.2\\
WD0628$-$020 & 6710 , 10 &  7.98 , 0.02  &  20, & -0.05, & 0.549, 0.010 & 107.4 ,  1.4^{gl} & 110.5 ,  1.7^{gl}  & 6845 , 10 &  7.96 , 0.01  & 0.569 , 0.007 & 105.7 ,  1.4^{g} & 109.8 ,  2.8^{g}  &  109.0,   1.3 & 111.7,  1.3\\
WD1015$-$216 & 31780 , 135 &  7.93 , 0.03  &       &         & 0.612, 0.014 &  6.6 ,  3.6^{g} &  8.6 ,  3.0^{g}  & 32545 , 150 &  7.91 , 0.03  & 0.600 , 0.013 &  5.0 ,  6.7^{} &  9.5 ,  8.4^{}  &  -2.8,   0.5 &  8.5,  0.5\\
WD1105$-$048 & 16325 , 55 &  7.78 , 0.00  &       &         & 0.509, 0.002 & 48.0 ,  0.9^{g} & 49.0 ,  1.0^{g}  & 16060 , 40 &  7.84 , 0.01  & 0.533 , 0.003 & 47.5 ,  0.8^{g} & 49.2 ,  1.1^{g}  &  47.5,   0.4 & 48.8,  0.4\\
WD1147+255 & 9895 , 25 &  8.32 , 0.03  &  -140, & -0.26, & 0.631, 0.017 & 59.8 ,  1.9^{g} & 59.0 ,  3.0^{g}  & 9970 , 55 &  7.96 , 0.02  & 0.577 , 0.011 & 59.3 ,  1.7^{l} & 59.4 ,  1.9^{l}  &  59.7,   1.6 & 57.3,  1.6\\
WD1348$-$273 & 9715 , 25 &  8.29 , 0.02  &  -130, & -0.26, & 0.614, 0.010 & 56.1 ,  2.0^{g} & 56.9 ,  2.0^{g}  & 9750 , 20 &  7.94 , 0.01  & 0.564 , 0.007 & 55.3 ,  1.1^{} & 55.8 ,  1.4^{}  &  55.2,   1.5 & 56.6,  1.5\\
WD1544$-$377$^{d}$ & 10300 , 15 &  8.35 , 0.01  &  -175, & -0.26, & 0.647, 0.004 & 20.7 ,  0.6^{g} & 20.7 ,  1.1^{g}  & 10570 , 15 &  8.02 , 0.01  & 0.610 , 0.005 & 19.6 ,  0.6^{g} & 20.7 ,  0.7^{g}  &  20.3,  0.5 & 20.5,  0.5\\
WD1619+123 & 16885 , 110 &  7.74 , 0.02  &       &         & 0.499, 0.008 & 14.3 ,  1.8^{gl} & 14.9 ,  1.7^{gl}  & 16900 , 110 &  7.80 , 0.02  & 0.519 , 0.009 & 13.5 ,  2.3^{} & 14.9 ,  3.3^{}  &  13.8,   0.7 & 15.3,  0.7\\
WD1620$-$391 & 25265 , 40 &  8.05 , 0.01  &       &         & 0.658, 0.003 & 43.1 ,  0.7^{g} & 44.1 ,  0.9^{g}  & 25395 , 45 &  8.05 , 0.01  & 0.655 , 0.005 & 42.3 ,  0.6^{g} & 44.2 ,  0.8^{g}  &  43.2,   0.4 & 44.3,  0.4\\
WD1911+135 & 12690 , 60 &  8.16 , 0.03  &  -150, & -0.03, & 0.675, 0.018 & 17.4 ,  1.7^{g} & 18.2 ,  1.9^{g}  & 14645 , 200 &  7.80 , 0.03  & 0.513 , 0.009 & 17.1 ,  1.6^{g} & 18.4 ,  1.9^{g}  &  16.5,   1.0 & 17.6,  1.0\\
WD1932$-$136 & 17415 , 145 &  7.67 , 0.03  &       &         & 0.477, 0.009 &  2.0 ,  1.6^{g} &  2.3 ,  1.7^{g}  & 17555 , 165 &  7.75 , 0.02  & 0.503 , 0.008 &  1.9 ,  1.8^{g} &  2.5 ,  1.9^{g}  &   1.6,   0.1 &  2.1,  0.1\\
WD2253+054 & 5985 , 10 &  8.17 , 0.03  &  5, & 0.02, & 0.699, 0.022 & 36.6 ,  1.5^{g} & 38.8 ,  1.6^{g}  & 6055 , 15 &  8.15 , 0.03  & 0.676 , 0.019 & 36.3 ,  1.8^{} & 39.8 ,  2.5^{}  &  34.2,   1.8 & 39.8,  1.8\\
WD2318+126 & 12635 , 40 &  8.14 , 0.03  &  -160, & -0.03, & 0.662, 0.019 & -11.6 ,  1.5^{gl} & -10.5 ,  1.9^{gl}  & 13720 , 305 &  8.01 , 0.09  & 0.609 , 0.049 & -12.2 ,  1.6^{g} & -10.8 ,  1.9^{g}  &  -13.2,   0.1 & -11.0,  0.1\\
         \noalign{\smallskip}
         \hline
             \noalign{\smallskip}
        \multicolumn{15}{p{21cm}}{{\bf Notes.} $a$: $\varv_{\rm WD}$ measured via model fit of both the H$\alpha$ and H$\beta$ lines; $b$: $\varv_{\rm WD}$ measured via model fit of the H$\alpha$ line only; ${c}$: $\varv_{\rm WD}$ measured from the Zeeman $\pi$ component; ${d}$: using the spectra from the SPY project; the $g$ and $l$ superscripts indicate whether the radial-velocity fit using the model spectra also required additional Gaussian and Lorentzian functions to reproduce the line cores. The spectroscopic masses are estimated via interpolation of evolutionary tracks  \citep{althaus2013,camisassa2016,camisassa2019}.}
    \end{tabular}
    \label{tab:spectra_params}
\end{table}
\end{landscape}

\twocolumn
\newpage
\onecolumn
\section{Gallery of Balmer-line best fits}

\begin{figure}[h!]
    \centering
    \includegraphics[height=22cm]{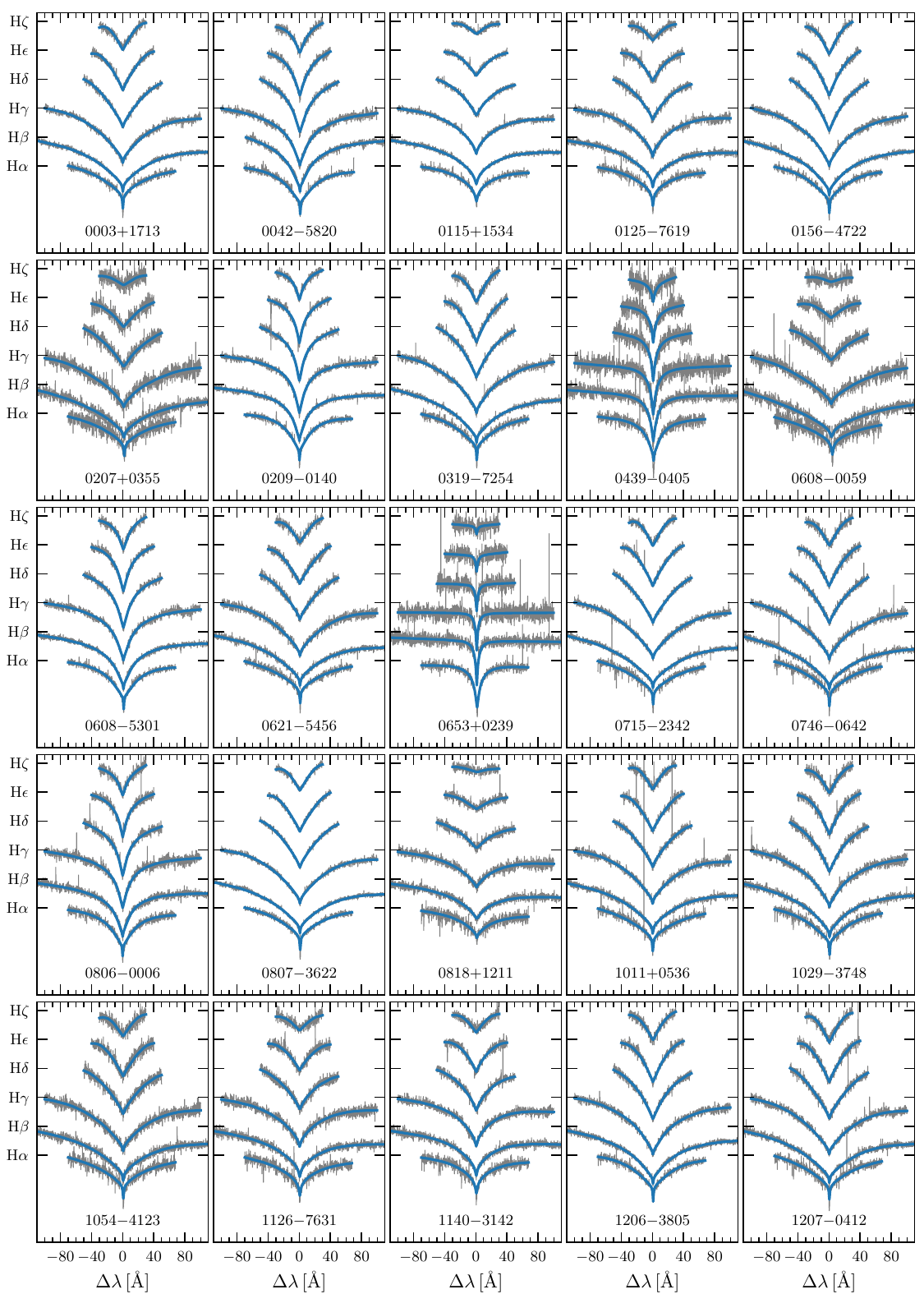}
    \caption{Gallery showing the average, observed spectra in gray and the best-fit models from the Koester grid of synthetic spectra.}
    \label{fig:fig01}
\end{figure}
\begin{figure}
    \centering
    \includegraphics[height=22cm]{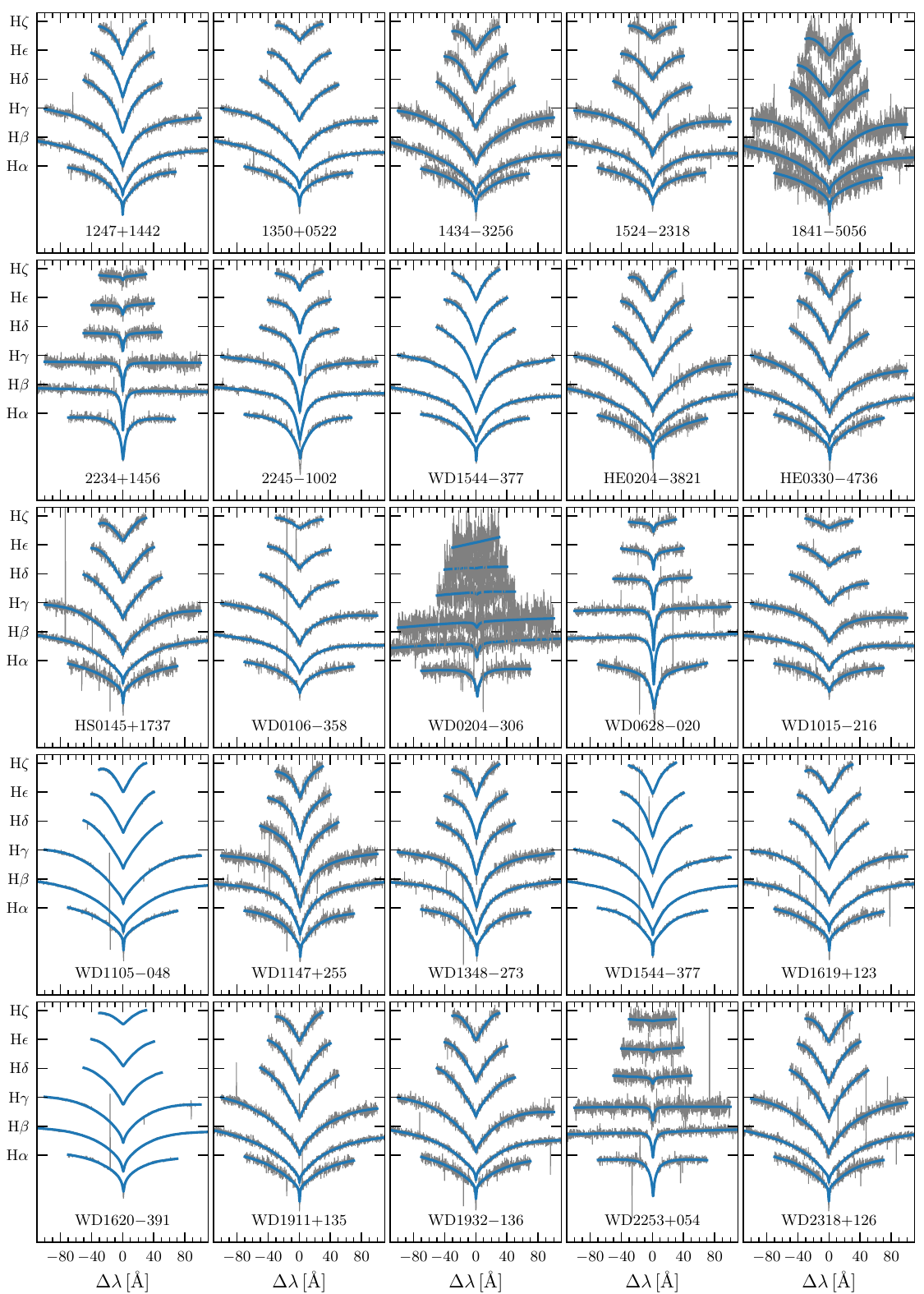}
    \caption{As in Fig.\,\ref{fig:fig01}. We display first the most recent spectrum for WD\,1544--377, and next its SPY spectrum.}
    \label{fig:fig02}
\end{figure}
\begin{figure}
    \centering
    \includegraphics[height=22cm]{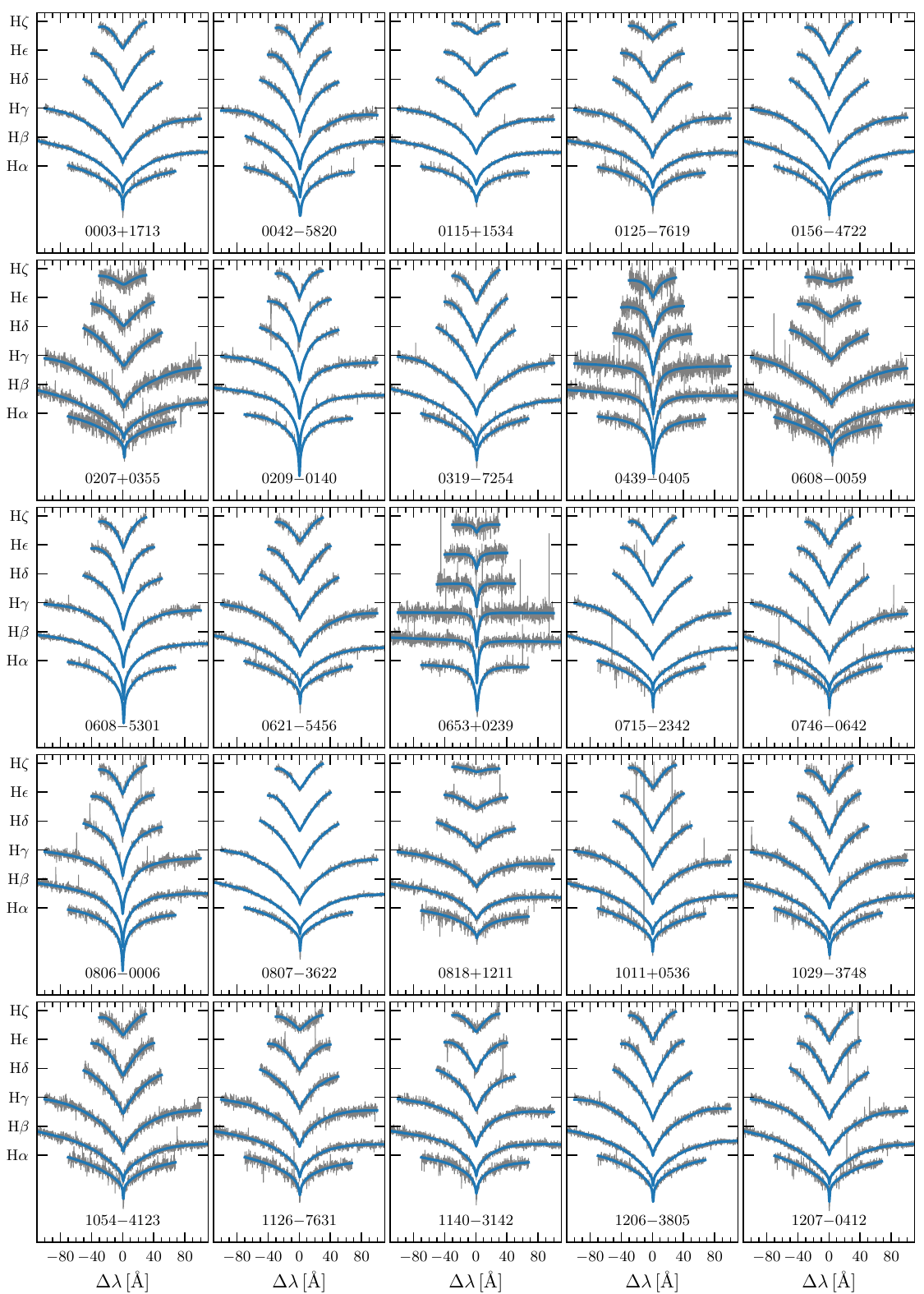}
    \caption{As in Fig.\,\ref{fig:fig01}, but showing the best fit-models from the Tremblay grid of synthetic spectra.}
    \label{fig:fig03}
\end{figure}
\begin{figure}
    \centering
    \includegraphics[height=22cm]{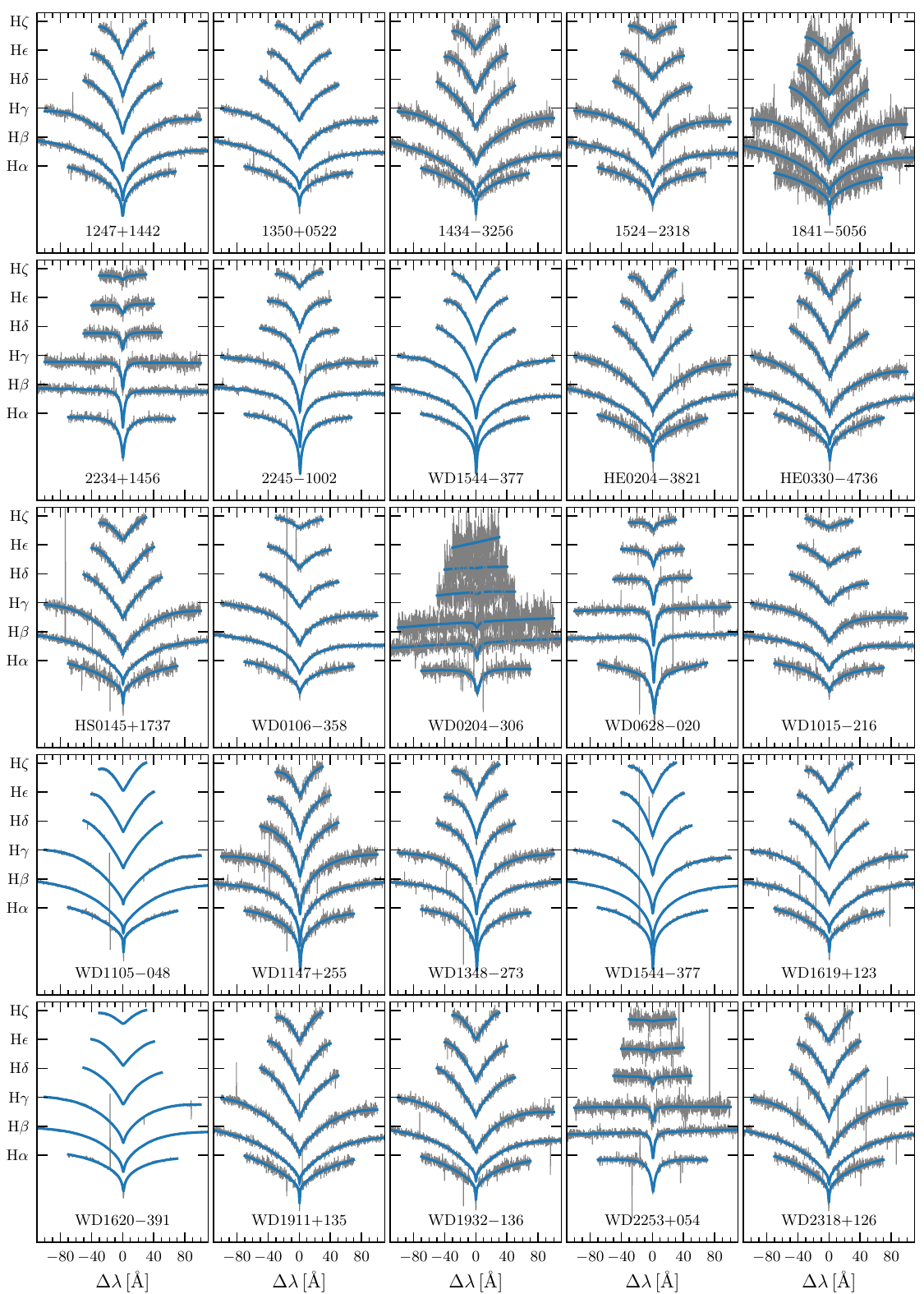}
    \caption{As in Fig.\,\ref{fig:fig02}, but showing the best fit-models from the Tremblay grid of synthetic spectra.}
    \label{fig:fig04}
\end{figure}
\twocolumn
\newpage
\onecolumn
\section{Gallery of radial-velocity best fits}
\begin{figure}[h!]
    \centering
    \includegraphics[height=22cm]{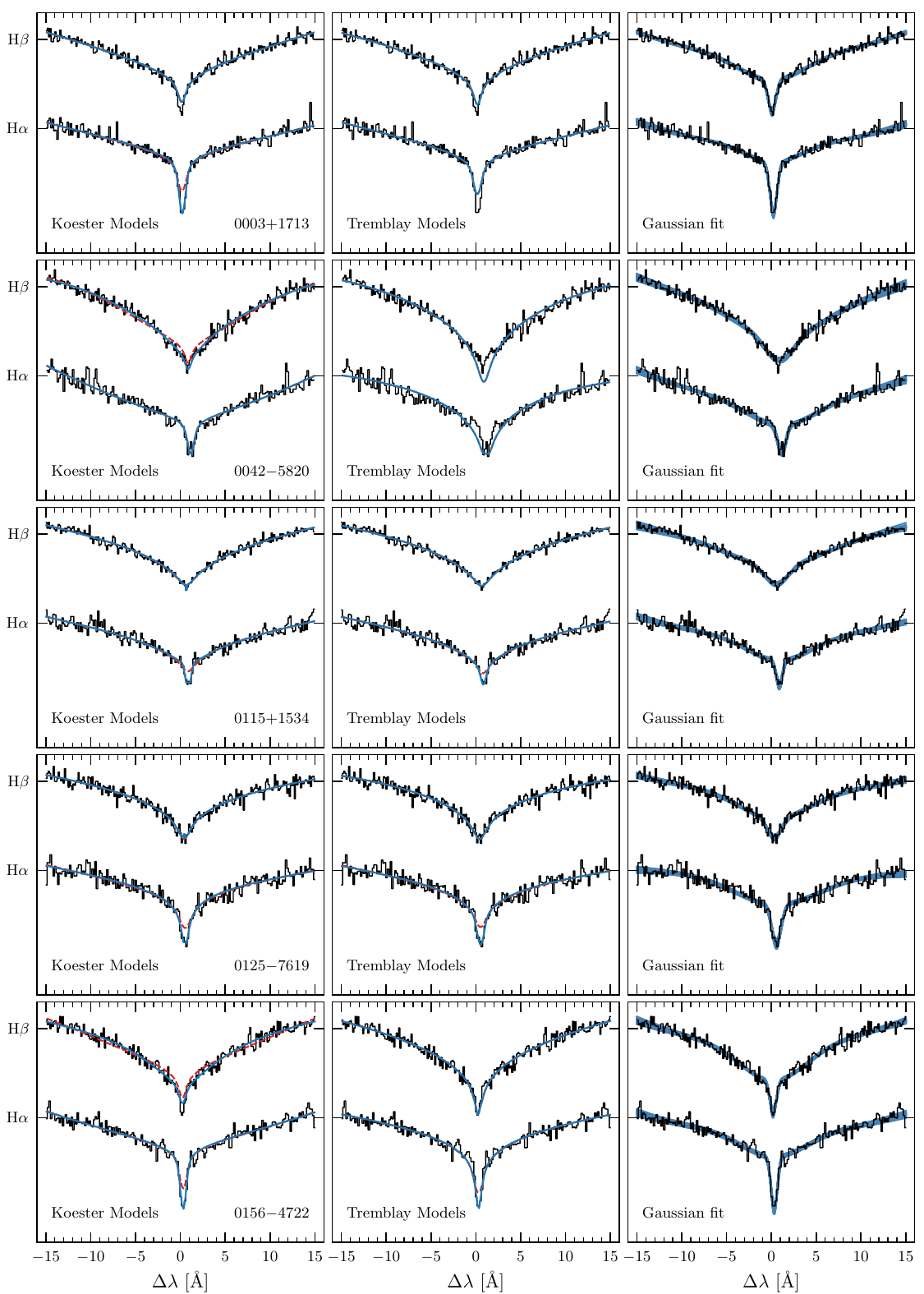}
    \caption{Radial-velocity fitting. The red-dashed curves represent the model-fit without the additional Gaussian functions, which are included in the blue-solid curves. The shaded region in the Gaussian fit panels represent 50 random draws of the best-fit distributions.}
    \label{fig:fig01rv}
\end{figure}
\begin{figure}
    \centering
    \includegraphics[height=22cm]{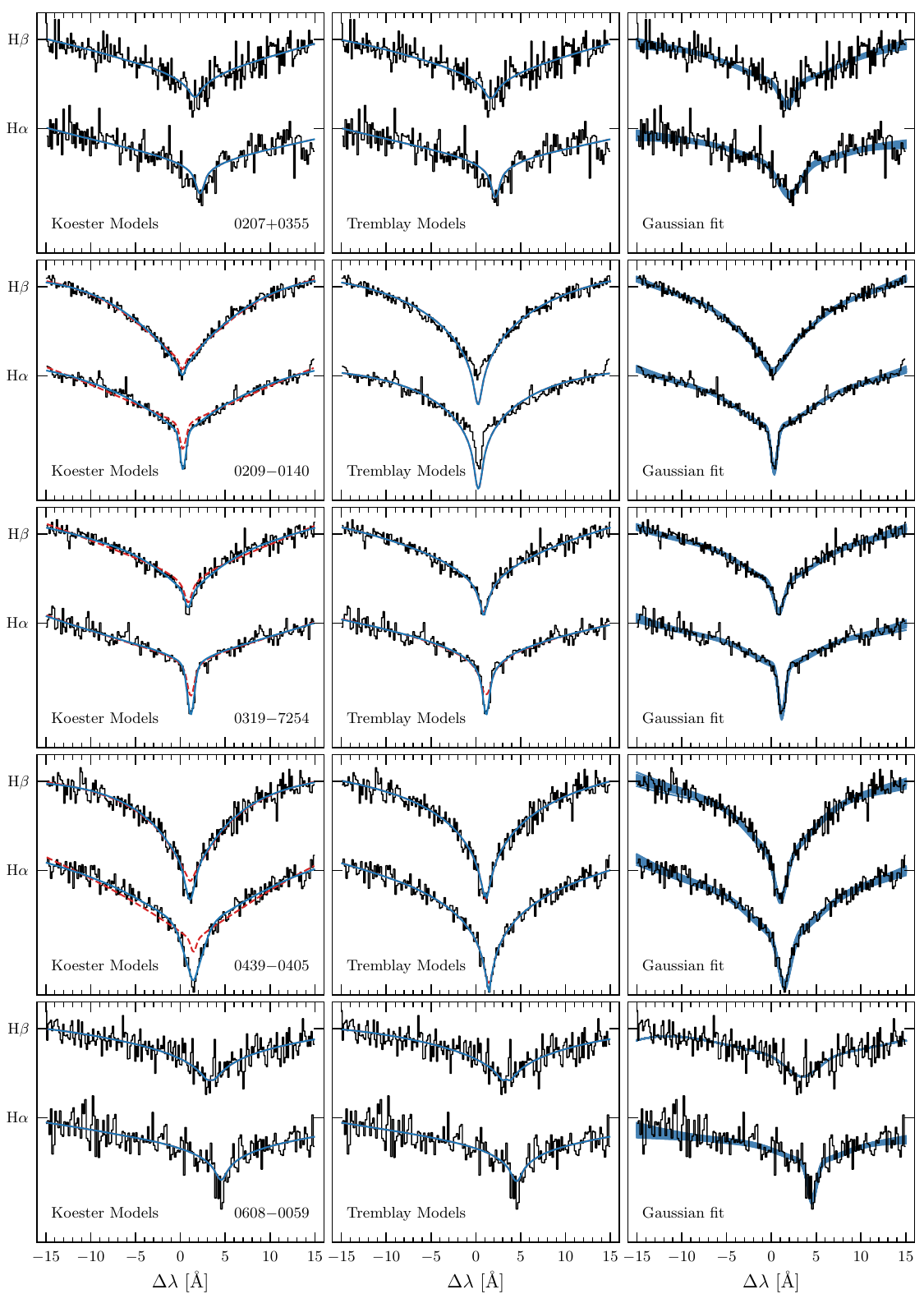}
    \caption{Radial-velocity fitting as in Fig.\,\ref{fig:fig01rv}.}
    \label{fig:fig02rv}
\end{figure}    
\begin{figure*}
    \centering
    \includegraphics[height=22cm]{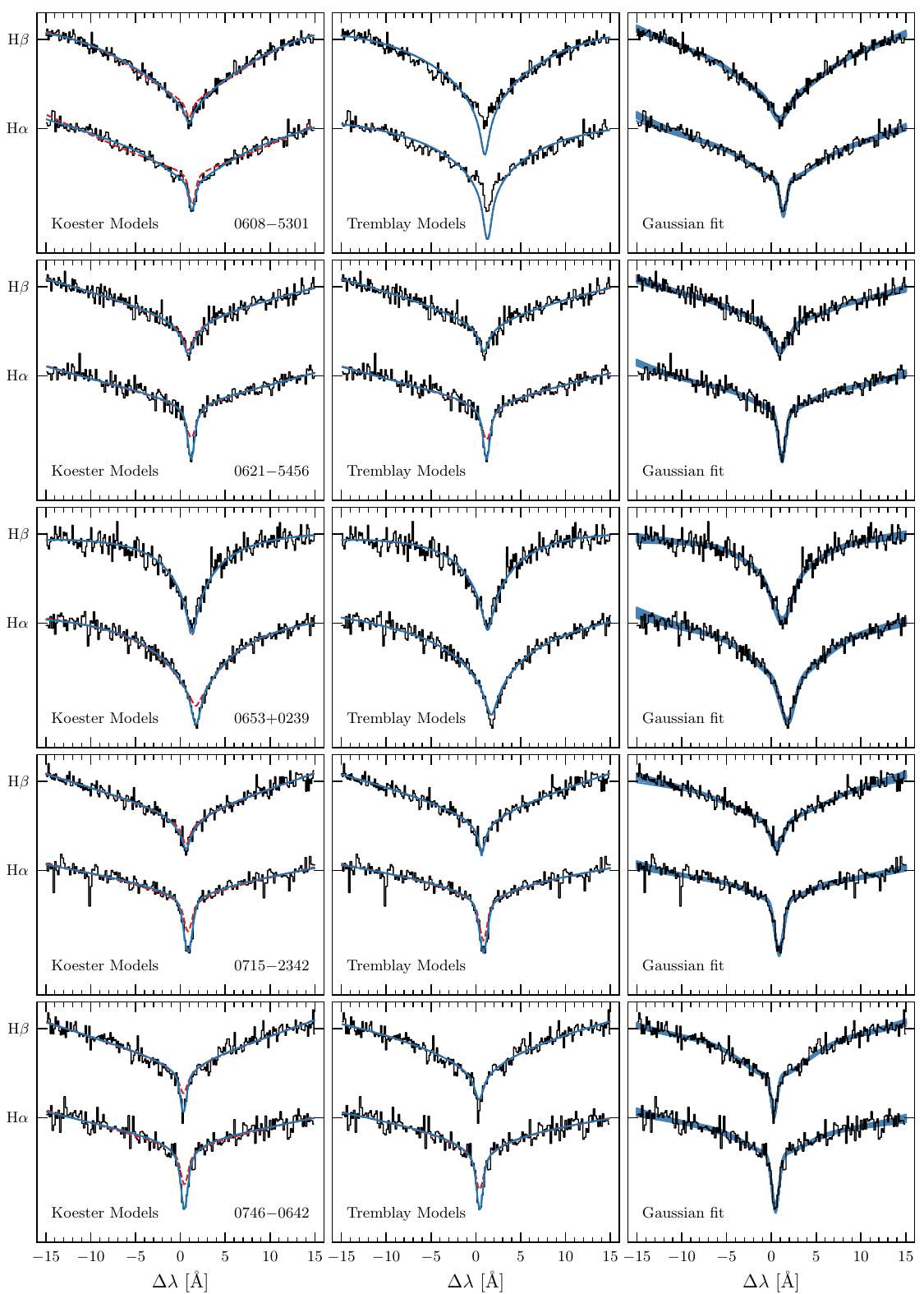}
    \caption{Radial-velocity fitting as in Fig.\,\ref{fig:fig01rv}.}
    \label{fig:fig03rv}
\end{figure*}    
\begin{figure}
    \centering
    \includegraphics[height=22cm]{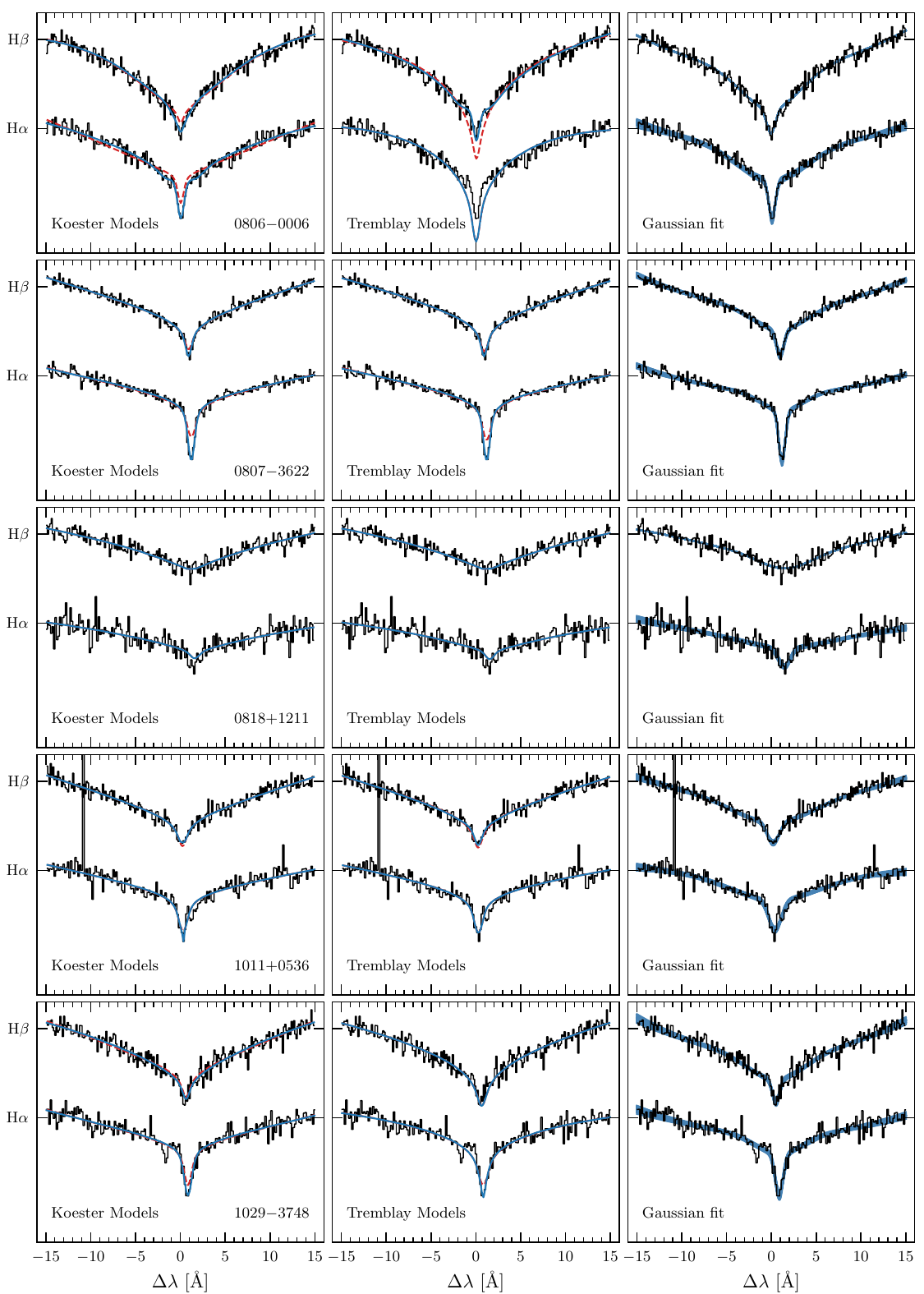}
    \caption{Radial-velocity fitting as in Fig.\,\ref{fig:fig01rv}.}
    \label{fig:fig04rv}
\end{figure}    
\begin{figure}
    \centering
    \includegraphics[height=22cm]{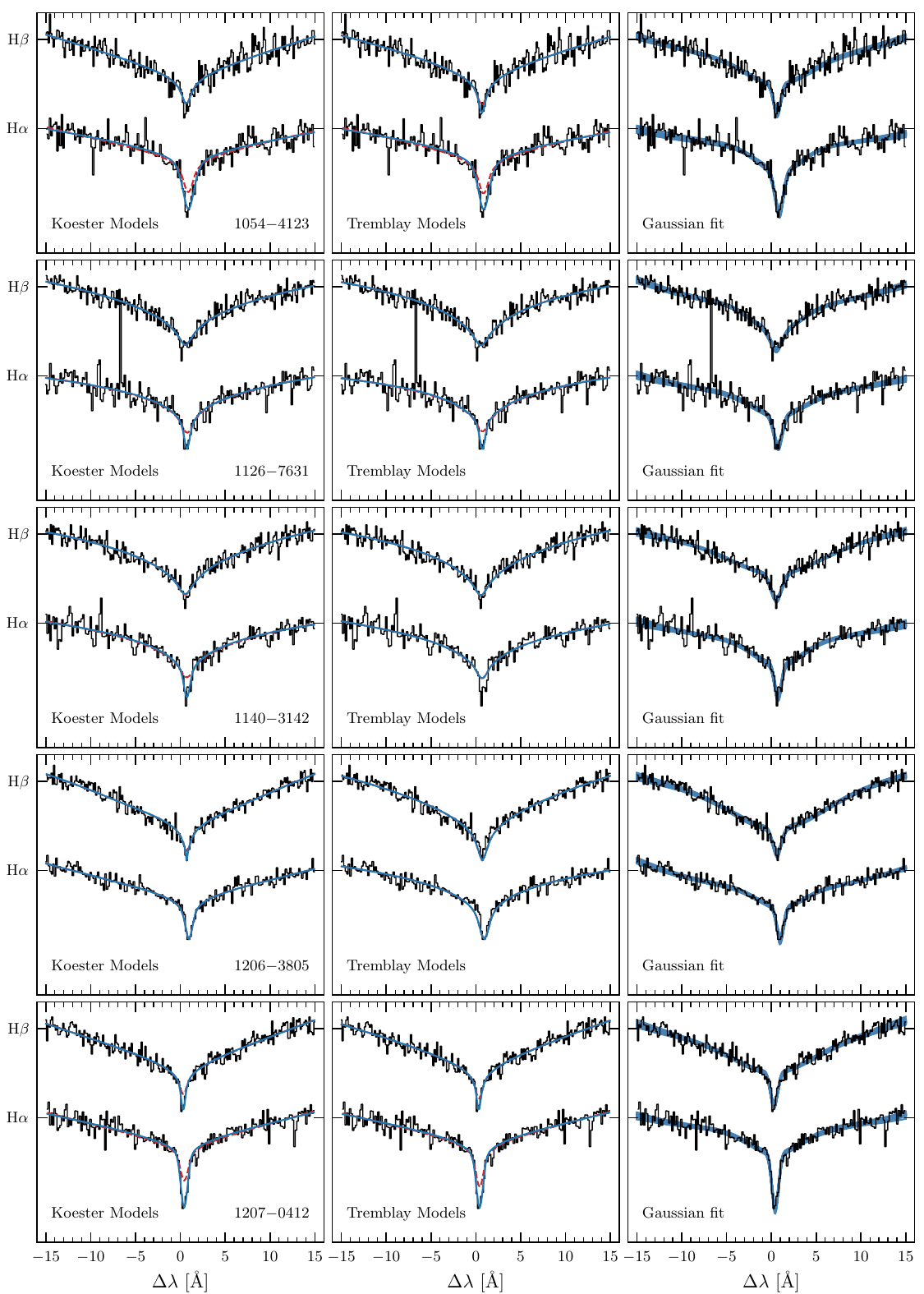}
    \caption{Radial-velocity fitting as in Fig.\,\ref{fig:fig01rv}.}
    \label{fig:fig05rv}
\end{figure}    
\begin{figure}
    \centering
    \includegraphics[height=22cm]{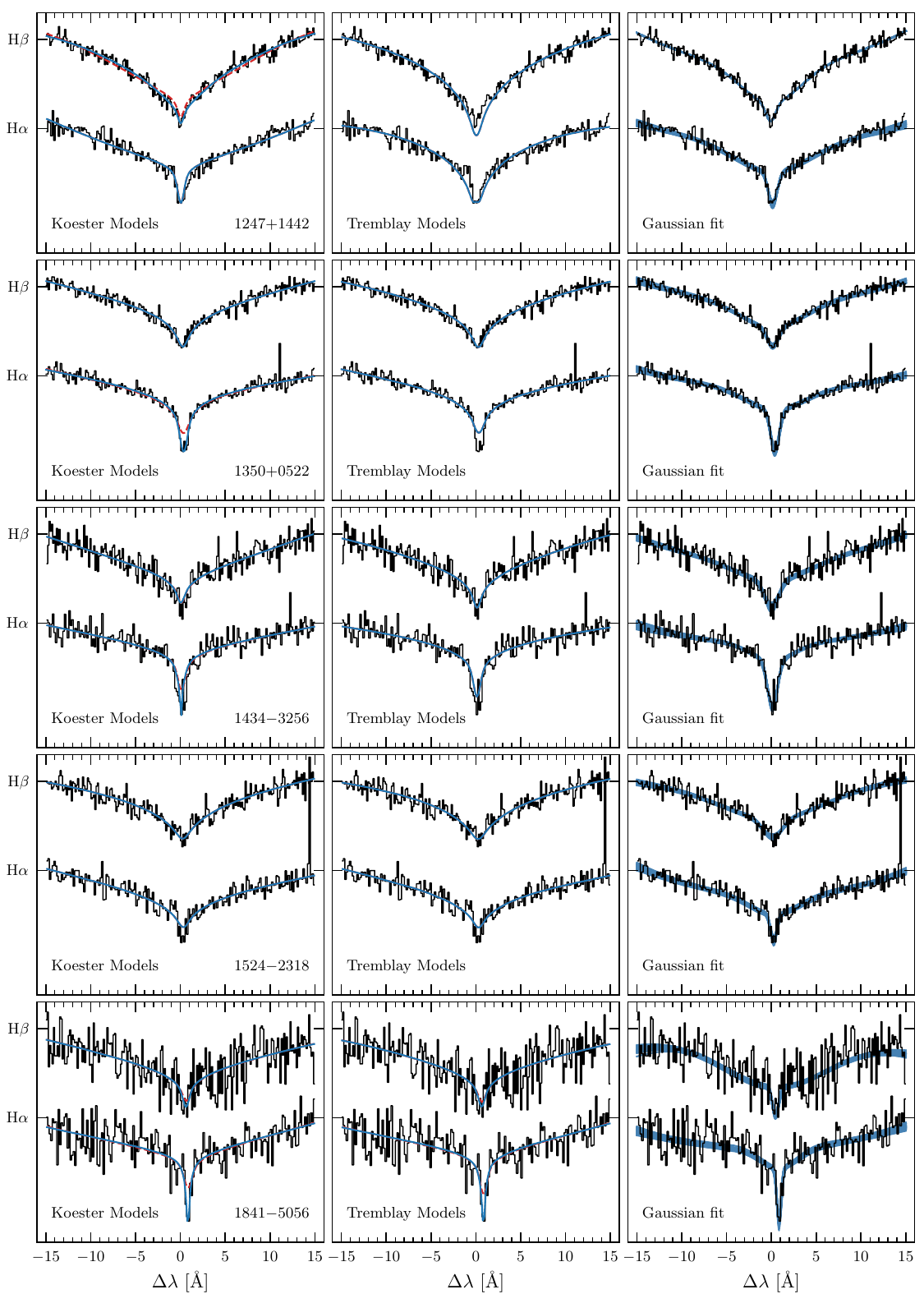}
    \caption{Radial-velocity fitting as in Fig.\,\ref{fig:fig01rv}.}
    \label{fig:fig06rv}
\end{figure}    
\begin{figure}
    \centering
    \includegraphics[height=22cm]{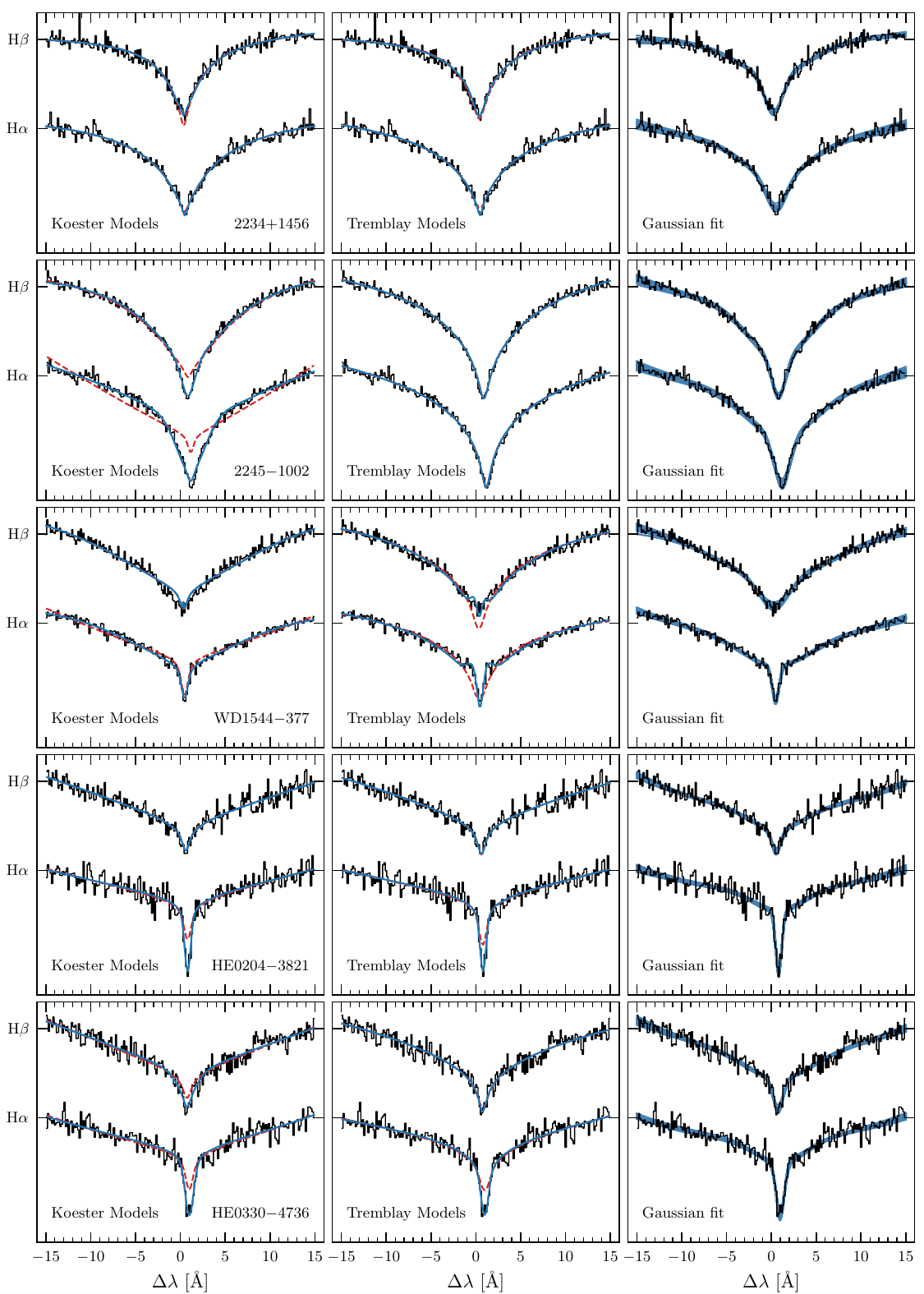}
    \caption{Radial-velocity fitting as in Fig.\,\ref{fig:fig01rv}.}
    \label{fig:fig07rv}
\end{figure}    
\begin{figure}
    \centering
    \includegraphics[height=22cm]{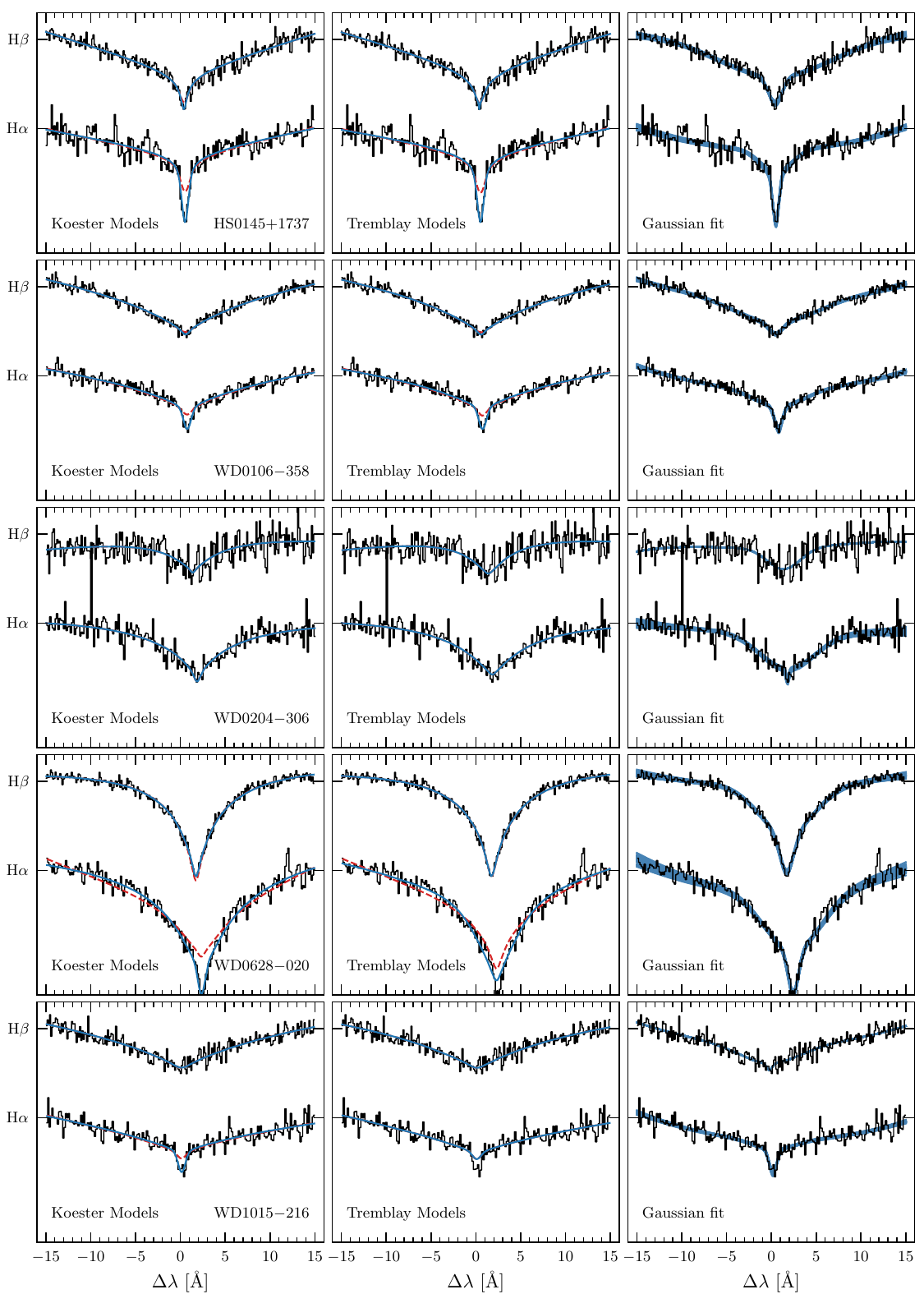}
    \caption{Radial-velocity fitting as in Fig.\,\ref{fig:fig01rv}.}
    \label{fig:fig08rv}
\end{figure}    
\begin{figure}
    \centering
    \includegraphics[height=22cm]{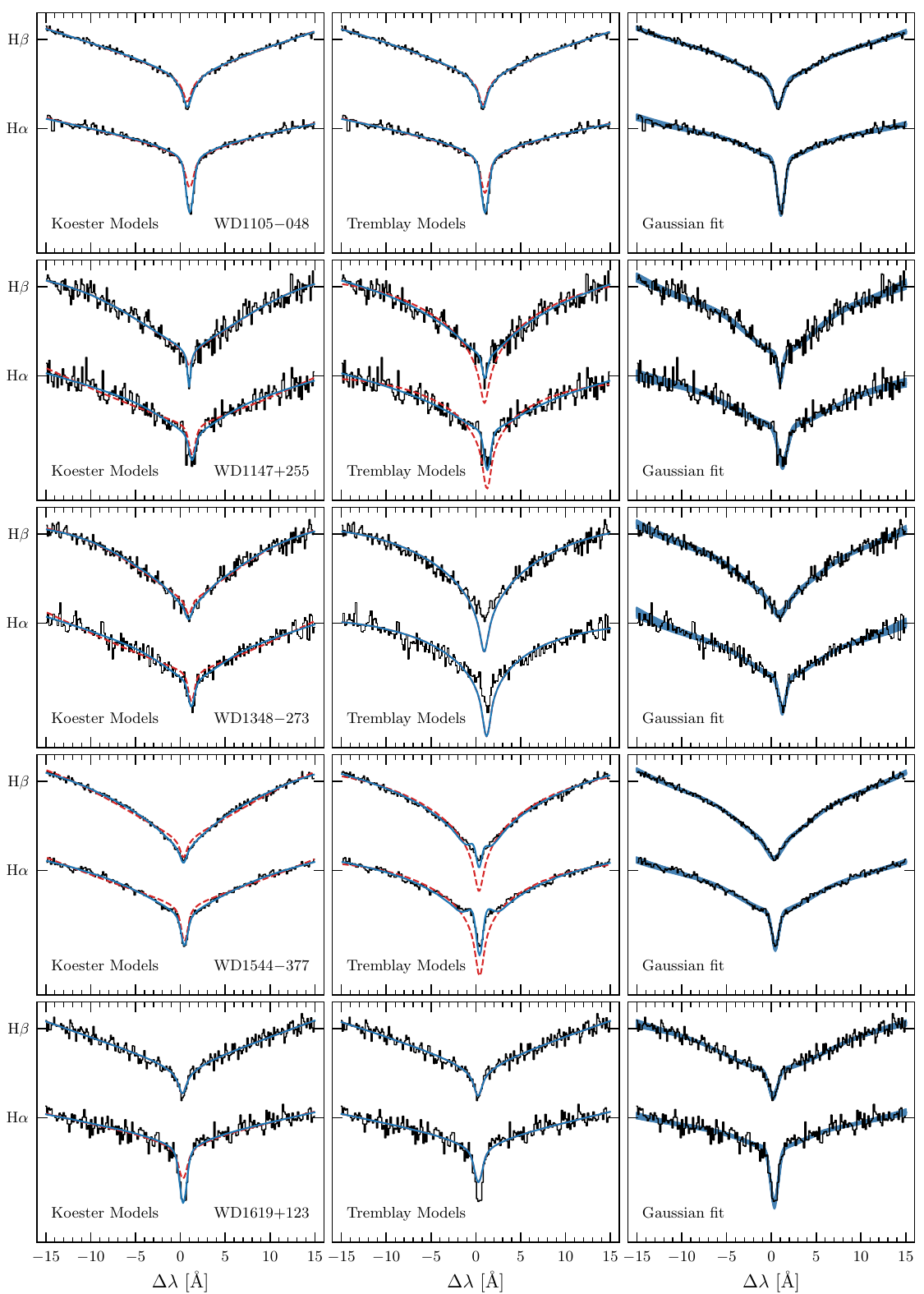}
    \caption{Radial-velocity fitting as in Fig.\,\ref{fig:fig01rv}.}
    \label{fig:fig9rv}
\end{figure}    
\begin{figure}
    \centering
    \includegraphics[height=22cm]{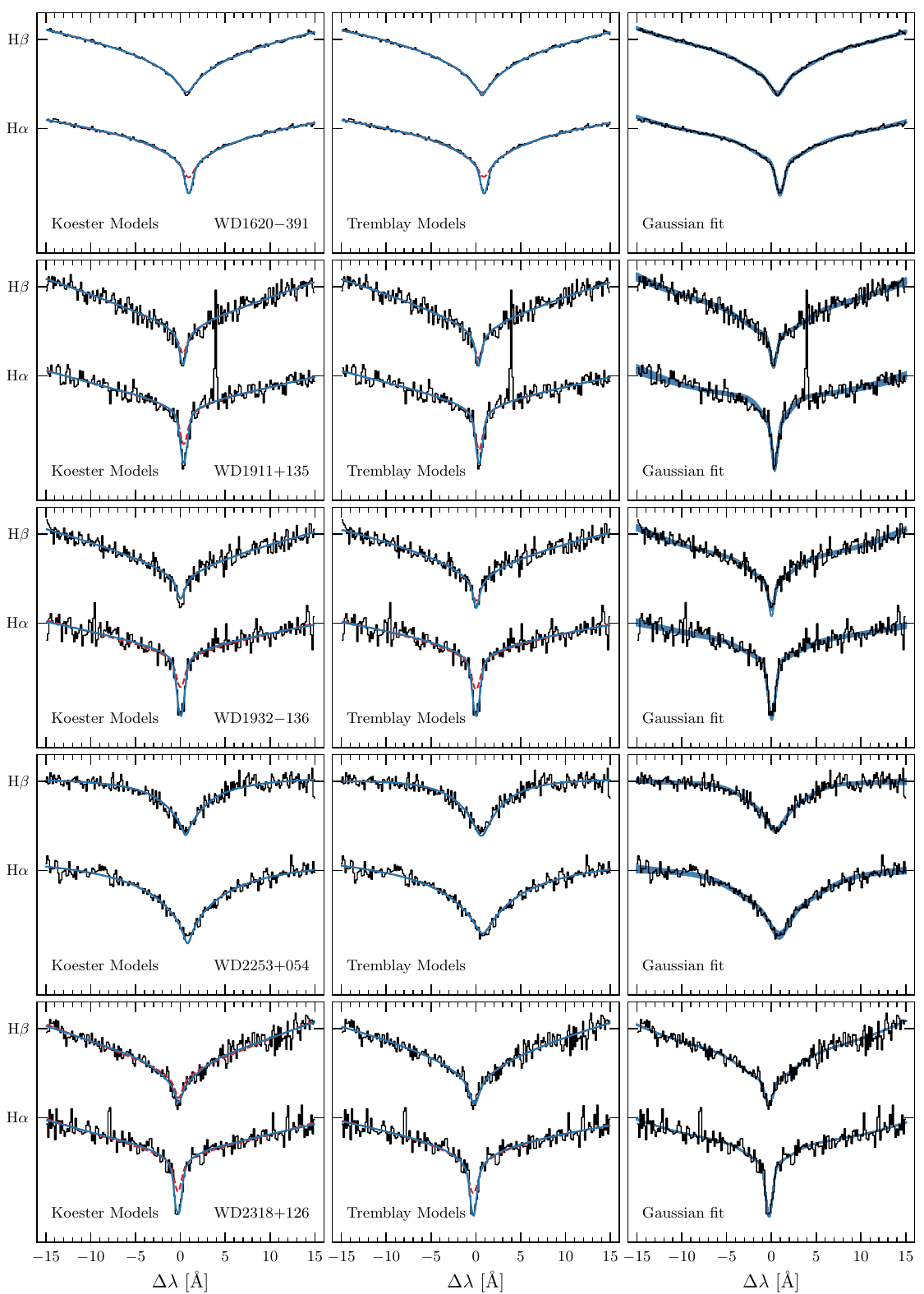}
    \caption{Radial-velocity fitting as in Fig.\,\ref{fig:fig01rv}.}
    \label{fig:fig10rv}
\end{figure}    
\twocolumn
\newpage
\onecolumn
\section{Photometric parameters}

\begin{table}[h!]
        \caption{Photometric parameters of the studied white dwarfs.}
    \centering
    \scriptsize
    \begin{tabular}{@{}l m{3.4} c c c | m{3.2} m{4.0} m{1.2} m{4.1} m{4.1} | m{3.2} m{4.0} m{1.2} m{4.1} m{4.1}@{}}
    \hline
    \hline
    \noalign{\smallskip} \multicolumn{5}{c}{} & \multicolumn{5}{c}{Using the averaged $\varv({\rm H}\alpha+{\rm H}\beta)_{\rm WD}$} & \multicolumn{5}{c}{Using the averaged $\varv({\rm H}\alpha)_{\rm WD}$} \\
        \noalign{\smallskip}
Short Name &  \multicolumn{1}{c}{$T_{\rm eff}$} & \multicolumn{1}{c}{$R$}& \multicolumn{1}{c}{$d$} & \multicolumn{1}{c}{$A(55)$} & \multicolumn{1}{c}{$M$} &   \multicolumn{1}{l}{$\rho$(M,\,R)}  &  \multicolumn{1}{c}{$\varv_{\rm gr}$} &  \multicolumn{1}{c}{$\Delta R$} &  \multicolumn{1}{c}{$\Delta M$} &  
\multicolumn{1}{c}{$M$} &   \multicolumn{1}{l}{$\rho$(M,\,R)}  &  \multicolumn{1}{c}{$\varv_{\rm gr}$} &  \multicolumn{1}{c}{$\Delta R$} &  \multicolumn{1}{c}{$\Delta M$}\\
 &  \multicolumn{1}{c}{(K)} & \multicolumn{1}{c}{(R$_\odot$)}& \multicolumn{1}{c}{(pc)}  & \multicolumn{1}{c}{(mag)} 
&  \multicolumn{1}{c}{(M$_\odot$)} &  & \multicolumn{1}{c}{(km/s)} & \multicolumn{1}{c}{(\%)}  & \multicolumn{1}{c}{(\%)} &
 \multicolumn{1}{c}{(M$_\odot$)} &  & \multicolumn{1}{c}{(km/s)} & \multicolumn{1}{c}{(\%)}  & \multicolumn{1}{c}{(\%)} \\
    \noalign{\smallskip}
    \hline  
    \noalign{\smallskip}    
0003+1713 & 16485,575  & 0.0137 & 107 & 0.007 &  0.60, 0.05 &  0.30  & 28.1,2.1 & 1.9 & 1.1 &  0.65, 0.03 &  0.53  & 30.5,1.3 & 5.8 & 5.5 \\
0042$-$5820 & 10305,160  & 0.0123 & 60 & 0.003 &  0.69, 0.05 &  0.29  & 35.8,2.2 & 2.9 & 3.4 &  0.68, 0.04 &  0.34  & 35.7,1.8 & 2.7 & 2.4 \\
0115+1534 & 25075,815  & 0.0129 & 105 & 0.007 &  0.74, 0.05 &  0.40  & 36.8,2.2 & 6.3 & 5.7 &  0.79, 0.05 &  0.47  & 39.3,2.1 & 9.3 & 9.0 \\
0125$-$7619 & 23395,925  & 0.0137 & 176 & 0.011 &  0.57, 0.05 &  0.34  & 26.6,2.3 & -2.5 & -3.4 &  0.61, 0.04 &  0.49  & 28.6,1.7 & 1.9 & 1.6 \\
0156$-$4722 & 12410,310  & 0.0137 & 84 & 0.004 &  0.54, 0.03 &  0.25  & 25.3,1.4 & -3.0 & -3.8 &  0.56, 0.03 &  0.34  & 26.3,1.3 & -0.4 & -0.7 \\
0207+0355 & 16820,840  & 0.0076 & 191 & 0.013 &  1.09, 0.10 &  0.70  & 90.9,5.5 & 4.1 & 4.5 &  1.05, 0.08 &  0.79  & 88.5,4.0 & 2.4 & 1.9 \\
0209$-$0140 & 9350,225  & 0.0127 & 56 & 0.004 &  0.62, 0.04 &  0.65  & 30.9,1.3 & 0.9 & 0.8 &  0.64, 0.03 &  0.69  & 32.4,1.2 & 2.3 & 1.9 \\
0319$-$7254 & 13270,435  & 0.0133 & 97 & 0.005 &  0.54, 0.04 &  0.32  & 25.4,1.6 & -5.4 & -5.2 &  0.55, 0.03 &  0.43  & 26.3,1.3 & -3.3 & -3.3 \\
0439$-$0405 & 8060,140  & 0.0132 & 53 & 0.003 &  0.57, 0.04 &  0.35  & 27.6,1.8 & 0.2 & -0.4 &  0.60, 0.04 &  0.45  & 29.1,1.5 & 2.7 & 2.3 \\
0608$-$0059 & 17155,775  & 0.0046 & 62 & 0.004 &  1.31, 0.06 &  0.75  & 179.6,7.0 & 2.2 & 4.4 &  1.31, 0.07 &  0.74  & 179.5,7.1 & 2.2 & 4.5 \\
0608$-$5301 & 9795,160  & 0.0133 & 58 & 0.003 &  0.57, 0.04 &  0.34  & 27.3,1.8 & -0.2 & -0.7 &  0.60, 0.04 &  0.43  & 28.8,1.3 & 2.3 & 1.7 \\
0621$-$5456 & 16550,570  & 0.0135 & 93 & 0.005 &  0.65, 0.05 &  0.28  & 30.5,2.3 & 4.7 & 5.4 &  0.62, 0.03 &  0.47  & 29.3,1.4 & 3.1 & 3.3 \\
0653+0239 & 6410,130  & 0.0147 & 51 & 0.003 &  0.46, 0.07 &  0.23  & 20.3,2.6 & -2.0 & -3.2 &  0.54, 0.05 &  0.38  & 23.4,1.7 & 4.5 & 4.1 \\
0715$-$2342 & 14455,360  & 0.0136 & 112 & 0.006 &  0.61, 0.04 &  0.28  & 28.8,1.7 & 3.0 & 2.6 &  0.63, 0.03 &  0.39  & 29.7,1.2 & 4.8 & 4.5 \\
0746$-$0642 & 13735,425  & 0.0126 & 101 & 0.006 &  0.63, 0.03 &  0.44  & 32.1,1.6 & 0.0 & -0.1 &  0.65, 0.03 &  0.47  & 33.0,1.5 & 0.9 & 0.5 \\
0806$-$0006 & 9040,215  & 0.0137 & 61 & 0.004 &  0.69, 0.05 &  0.49  & 32.2,2.0 & 9.7 & 9.0 &  0.74, 0.04 &  0.69  & 34.6,1.4 & 12.2 & 12.0 \\
0807$-$3622 & 16540,285  & 0.0133 & 49 & 0.003 &  0.64, 0.04 &  0.11  & 30.5,1.8 & 2.9 & 3.7 &  0.60, 0.03 &  0.26  & 28.9,1.1 & 0.6 & 0.5 \\
0818+1211 & 31225,1440  & 0.0116 & 284 & 0.021 &  0.78, 0.16 &  0.24  & 42.0,8.2 & 1.4 & 4.3 &  0.61, 0.15 &  0.34  & 33.5,7.7 & -10.3 & -10.0 \\
1011+0536 & 12985,540  & 0.0204 & 191 & 0.009 &  0.41, 0.07 &  0.16  & 12.2,2.1 & 1.9 & 6.2 &  0.47, 0.06 &  0.26  & 13.9,1.8 & 8.5 & 13.0 \\
1029$-$3748 & 12240,195  & 0.0123 & 63 & 0.004 &  0.67, 0.05 &  0.02  & 35.3,2.4 & 2.2 & 0.3 &  0.72, 0.04 &  0.15  & 37.3,1.8 & 4.5 & 3.9 \\
1054$-$4123 & 16665,450  & 0.0125 & 124 & 0.007 &  0.81, 0.05 &  0.26  & 41.4,2.7 & 10.0 & 9.4 &  0.82, 0.05 &  0.30  & 42.2,2.5 & 10.7 & 9.7 \\
1126$-$7631 & 21990,835  & 0.0140 & 145 & 0.013 &  0.77, 0.06 &  0.38  & 35.0,2.6 & 11.2 & 11.0 &  0.78, 0.06 &  0.42  & 35.7,2.4 & 12.2 & 12.1 \\
1140$-$3142 & 21565,875  & 0.0200 & 268 & 0.025 &  0.58, 0.09 &  0.24  & 18.5,3.1 & 15.4 & 15.1 &  0.59, 0.11 &  0.26  & 19.2,3.2 & 17.4 & 15.5 \\
1206$-$3805 & 11845,185  & 0.0123 & 59 & 0.004 &  0.64, 0.04 &  0.12  & 33.5,2.1 & -1.2 & -2.3 &  0.69, 0.04 &  0.22  & 36.3,1.8 & 3.0 & 2.0 \\
1207$-$0412 & 14705,450  & 0.0143 & 127 & 0.007 &  0.54, 0.04 &  0.32  & 24.2,1.7 & -1.3 & -1.4 &  0.56, 0.04 &  0.29  & 25.1,1.5 & -0.2 & -0.8 \\
1247+1442 & 10490,175  & 0.0132 & 61 & 0.004 &  0.56, 0.07 &  0.09  & 27.0,3.3 & -1.6 & -1.3 &  0.67, 0.05 &  0.26  & 32.4,2.1 & 6.1 & 5.5 \\
1350+0522 & 20870,795  & 0.0133 & 184 & 0.012 &  0.56, 0.06 &  0.30  & 27.6,2.5 & -2.7 & -5.0 &  0.64, 0.05 &  0.54  & 30.8,1.9 & 2.7 & 2.5 \\
1426$-$5716 & 12590,315  & 0.0100 & 70 & 0.005 &   &    &   &   &   &  0.66, 0.03 &  0.39  & 41.9,1.5 & -10.5 & -10.5 \\
1426$-$5716$^{a}$ & 12405,315  & 0.0104 & 70 & 0.005 &   &    &   &   &   &  0.68, 0.03 &  0.44  & 41.9,1.7 & -7.2 & -7.0 \\
1434$-$3256 & 13680,515  & 0.0127 & 115 & 0.009 &  0.56, 0.07 &  0.19  & 28.3,3.5 & -4.2 & -4.4 &  0.58, 0.07 &  0.24  & 29.1,3.0 & -2.9 & -2.3 \\
1524$-$2318 & 19290,995  & 0.0160 & 156 & 0.092 &  0.74, 0.10 &  0.38  & 30.1,4.2 & 18.7 & 16.2 &  0.77, 0.12 &  0.43  & 31.4,4.7 & 20.1 & 17.7 \\
1841$-$5056 & 15870,665  & 0.0137 & 136 & 0.010 &  0.59, 0.08 &  0.20  & 27.5,3.2 & 1.6 & 1.1 &  0.60, 0.05 &  0.44  & 27.8,2.1 & 2.2 & 2.4 \\
2234+1456 & 6275,100  & 0.0126 & 34 & 0.003 &  0.60, 0.06 &  0.34  & 30.7,2.5 & 1.3 & 0.4 &  0.68, 0.04 &  0.51  & 34.1,1.6 & 4.6 & 5.0 \\
2245$-$1002 & 8335,275  & 0.0107 & 57 & 0.004 &  0.87, 0.05 &  0.87  & 51.5,1.4 & 7.5 & 7.5 &  0.91, 0.05 &  0.91  & 54.2,1.1 & 9.6 & 9.4 \\
WD\,1544$-$377$^b$ & 10470,70  & 0.0125 & 15 & 0.001 &  0.58, 0.04 & -0.04  & 29.2,1.8 & -3.7 & -2.6 &  0.59, 0.03 &  0.00  & 29.8,1.6 & -2.6 & -2.4 \\
HE\,0204$-$3821 & 13555,435  & 0.0136 & 95 & 0.005 &  0.65, 0.11 &  0.03  & 30.8,5.1 & 6.7 & 5.1 &  0.68, 0.11 &  0.02  & 32.0,4.5 & 7.9 & 7.1 \\
HE\,0330$-$4736 & 13180,315  & 0.0121 & 72 & 0.004 &  0.66, 0.03 &  0.30  & 34.8,1.8 & -0.6 & -0.8 &  0.65, 0.03 &  0.32  & 34.4,1.6 & -1.4 & -1.7 \\
HS\,0145+1737 & 18000,645  & 0.0125 & 88 & 0.007 &  0.59, 0.03 &  0.49  & 29.9,1.6 & -4.3 & -3.3 &  0.57, 0.04 &  0.43  & 28.9,1.8 & -5.9 & -6.1 \\
WD\,0106$-$358 & 28645,575  & 0.0146 & 94 & 0.005 &  0.51, 0.06 &  0.02  & 22.9,3.0 & -5.8 & -8.4 &  0.58, 0.07 &  0.05  & 25.7,2.8 & 0.3 & -1.6 \\
WD\,0204$-$306 & 5545,60  & 0.0126 & 29 & 0.002 &  0.67, 0.08 &  0.17  & 33.9,4.3 & 4.5 & 4.8 &  0.66, 0.08 &  0.15  & 33.5,4.4 & 4.1 & 3.5 \\
WD\,0628$-$020 & 6610,70  & 0.0131 & 21 & 0.002 &  0.55, 0.06 &  0.12  & 26.4,2.8 & -1.9 & -0.5 &  0.62, 0.06 &  0.12  & 29.8,2.8 & 3.4 & 4.1 \\
WD\,1015$-$216 & 30850,1210  & 0.0141 & 150 & 0.008 &  0.62, 0.17 &  0.07  & 30.0,8.0 & 4.0 & -1.6 &  0.78, 0.18 &  0.10  & 35.9,7.4 & 12.1 & 10.2 \\
WD\,1105$-$048 & 15700,275  & 0.0136 & 25 & 0.002 &  0.54, 0.01 &  0.53  & 25.5,0.7 & -4.3 & -4.5 &  0.57, 0.02 &  0.52  & 26.8,0.8 & -1.1 & -1.2 \\
WD\,1147+255 & 10210,230  & 0.0127 & 50 & 0.003 &  2.51, 0.13 &  0.61  & 125.6,5.0 & 53.5 & 54.1 &  2.48, 0.13 &  0.60  & 124.5,4.5 & 53.2 & 53.7 \\
WD\,1348$-$273 & 9720,155  & 0.0129 & 38 & 0.003 &  0.61, 0.04 &  0.37  & 30.0,1.7 & 1.4 & 1.2 &  0.63, 0.04 &  0.37  & 30.9,1.6 & 2.6 & 2.7 \\
WD\,1544$-$377 & 10460,70  & 0.0125 & 15 & 0.001 &  0.55, 0.02 &  0.04  & 27.7,1.3 & -6.1 & -5.5 &  0.55, 0.03 &  0.04  & 28.1,1.2 & -5.4 & -5.0 \\
WD\,1619+123 & 16445,350  & 0.0143 & 56 & 0.006 &  0.47, 0.04 &  0.19  & 21.1,1.8 & -7.6 & -7.9 &  0.50, 0.04 &  0.19  & 22.2,1.9 & -5.1 & -5.0 \\
WD\,1620$-$391 & 24335,475  & 0.0129 & 13 & 0.001 &  0.62, 0.01 &  0.68  & 30.6,0.6 & -1.6 & -1.0 &  0.65, 0.02 &  0.65  & 32.0,0.6 & 0.3 & 0.4 \\
WD\,1911+135 & 13795,175  & 0.0129 & 34 & 0.004 &  0.60, 0.03 &  0.06  & 29.6,1.6 & 0.3 & -0.2 &  0.62, 0.03 &  0.05  & 30.6,1.5 & 1.5 & 1.0 \\
WD\,1932$-$136 & 16200,540  & 0.0148 & 105 & 0.014 &  0.58, 0.04 &  0.43  & 24.8,1.7 & 3.1 & 2.7 &  0.59, 0.04 &  0.44  & 25.2,1.5 & 3.6 & 3.1 \\
WD\,2253+054 & 5860,100  & 0.0130 & 25 & 0.002 &  0.53, 0.04 &  0.42  & 26.4,1.8 & -2.8 & -3.8 &  0.62, 0.05 &  0.44  & 30.2,1.8 & 3.1 & 3.2 \\
WD\,2318+126 & 13440,500  & 0.0129 & 90 & 0.007 &  0.51, 0.11 & -0.05  & 25.8,5.8 & -8.2 & -10.2 &  0.55, 0.11 & -0.03  & 27.4,5.0 & -4.6 & -6.2 \\

    \noalign{\smallskip}
    \hline
        \noalign{\smallskip}
\multicolumn{15}{p{17.8cm}}{{\bf Notes.} The mean values and uncertainties of $d$ and $A(55)$ are constrained via their likelihoods in Sect.\,\ref{sec:likelihood} (less than 5\,\% and of the order of 10\,\%, respectively). The uncertainties on $R$ are typically of the order of 2--3\,\%. The listed physical parameters are averaged over the different radial estimates of $\varv($H$\alpha$+H$\beta)_{WD}$ and $\varv($H$\alpha)_{WD}$ provided in Table\,\ref{tab:spectra_params}, and averaging the results from both the Koester and Tremblay model grids. $\rho(M\,, R)$ is the correlation between mass and radius determined via the photometric analysis. The differences $\Delta R$ and $\Delta M$ are computed respect to theoretical estimates of $R$ and $M$ via interpolation of $T_{\rm eff}$ and $\varv_{\rm gr}$ onto the white dwarf evolutionary tracks \citep{althaus2013,camisassa2016,camisassa2019}. $a$:
the parameters are obtained with radiative synthetic spectra; $b$: the results are obtained from the most recent UVES observations. 
}
    \end{tabular}
    \label{tab:phot_params}
\end{table}

\end{appendix}
\end{document}